%% file: Wake_paper.tex
\documentclass[twocolumn]{aastex62}
\pdfoutput=1

\usepackage[bottom]{footmisc}
\received{February, 2019}
\revised{--}
\accepted{--}
\submitjournal{ApJ}

\shorttitle{Hunting for the DM wake Induced by the LMC}
\shortauthors{Garavito-Camargo et al.}
\begin{document}

\title{Hunting for the Dark Matter Wake Induced by the Large Magellanic Cloud}

\correspondingauthor{Nicol\'as Garavito-Camargo}
\email{jngaravitoc@email.arizona.edu}

\author[0000-0001-7107-1744]{Nicolas Garavito-Camargo}
\affil{Steward Observatory, University of Arizona, 933 North Cherry Avenue, Tucson, AZ 85721, USA.}

\author[0000-0003-0715-2173]{Gurtina Besla}
\affiliation{Steward Observatory, University of Arizona, 933 North Cherry Avenue, Tucson, AZ 85721, USA.}

\author{Chervin F.P Laporte}
\altaffiliation{CITA National Fellow}
\affiliation{Department of Physics and Astronomy, University of Victoria, 3800 Finnerty Road, Victoria, B.C., V8P 4HN, Canada}

\author[0000-0001-6244-6727]{Kathryn V. Johnston}
\affiliation{Department of Astronomy, Columbia University, New York, NY 10027, USA.}

\author{Facundo A. G\'omez}
\affiliation{Instituto de Investigaci\'on Multidisciplinar en Ciencia y Tecnolog\'ia, Universidad de La Serena, Ra\'ul Bitr\'an 1305, La Serena, Chile.}
\affiliation{Departamento de F\'isica y Astronom\'ia, Universidad de La Serena, Av. Juan Cisternas 1200 N, La Serena, Chile.}

\author[0000-0002-1343-134X]{Laura L. Watkins}
\affiliation{Space Telescope Science Institute, 3700 San Martin Drive, Baltimore, MD 21218, USA}
\affiliation{European Southern Observatory, Karl-Schwarzschild-Stra{\ss}e 2, D-85748 Garching bei M\"{u}nchen, Germany}
\nocollaboration



\begin{abstract}

Satellite galaxies are predicted to generate gravitational density wakes as they 
orbit within the dark matter (DM) halos of their hosts, causing their orbits
to decay over time. 
The recent infall of the Milky Way's (MW) most massive satellite galaxy, the 
Large Magellanic Cloud (LMC), affords us the unique opportunity 
to study this process in action. 
In this work, we present high-resolution
($m_{dm} = 4\times 10^4 \rm{M_{\odot}}$) $N$-body simulations of the MW-LMC 
interaction over the past 2 Gyr. 
We quantify the impact of the LMC's passage
on the density and kinematics of 
the MW's DM halo and the observability of these structures in the 
MW's stellar halo. 
  The LMC is found to generate a pronounced wake, which we decompose in
  \textit{Transient} and \textit{Collective} 
responses, in both the DM and stellar halos. The wake leads to 
overdensities and distinct kinematic patterns that should be observable
with ongoing and future surveys.
Specifically, the Collective response will result in redshifted radial velocities
of stars in the north and blueshifts in the south, at distances $>$45 kpc. 
The Transient response traces the 
orbital path of the LMC through the halo (50-200 kpc), resulting in a stellar
overdensity with a distinct, tangential kinematic pattern that 
persists to the present day.
The detection of the MW's
halo response will constrain the infall mass of the LMC and its orbital trajectory, 
the mass of the MW, and it may inform us about the 
nature of the DM particle itself.

\end{abstract}


\input{intro}
\input{LMC_mass}

\input{methods}

\input{results}
\input{observability}

\input{discussion}
\input{conclusions}

\section*{Acknowledgements}

It is a pleasure to thank Dennis Zaritsky, Ekta Patel, Jerry Sellwood, 
Hans-Walter Rix, Alis Deason, Robyn Sanderson, Adrian
Price-Whelan, and Jeff Carlin
for their valuable discussions. We specially thank Martin Weinberg for insightful 
discussions about the terminology used in this paper.
We thank the referee for valuable comments that helped to improve the quality of 
this work
We thank Volker Springel for giving access to the \verb+Gadget-3+ code.
We also thank Dennis Yurin for providing support
with \verb+GalIC+ and Andrew Cooper for providing us with the DESI footprint.
This work has been supported by the \textit{HST} grant AR 15004, NASA ATP grant
17-ATP17-0006, and the Vatican Observatory Stoeger-McCarthy fellowship.
All the simulations where run on \textit{El-Gato} super computer which 
was supported by the National Science Foundation under Grant No. 1228509. 

N.G.-C., acknowledges the support from the Writing Skills Improvement Program from 
the University of Arizona, in particular to Jen Glass, Rixin Li, Ben Lew, 
and Nadia Moraglio whose valuable input helped to improve the writing process of this work.

K.V.J.'s contributions were supported by NSF grant AST-1715582. 
F.A.G. acknowledges financial support from CONICYT through the project 
FONDECYT Regular Nr. 1181264. F.A.G. acknowledge financial support from 
the Max Planck Society through a Partner Group grant.
L.L.W. acknowledges funding from the European Research Council (ERC) under the 
European Union's Horizon 2020 research and innovation programme under grant 
agreement No 724857 (Consolidator Grant ArcheoDyn).

This work has made use of software developed by the 
\textit{Gaia} Project Scientist Support Team and the \textit{Gaia} Data Processing and Analysis
Consortium (DPAC, \href{https://www.cosmos.esa.int/web/gaia/dpac/consortium}
{https://www.cosmos.esa.int/web/gaia/dpac/consortium}). Funding
for the DPAC as been provided by national institutions, in particular the institutions
participating in the Gaia Multilateral Agreement.

\textit{Software:} Astropy \citep{astropy:2013, astropy:2018},  pygadgetreader
\cite{pygadgetreader}, matplotlib \cite{Hunter:2007}, numpy \cite{numpy}, scipy \cite{scipy},
ipython \cite{ipython}, scikit-learn \citep{scikit-learn}, jupyter
\cite{jupyter}, healpy \cite{healpy}, reproject 
\href{https://github.com/astrofrog/reproject}{https://github.com/astrofrog/reproject}, pyh5
\href{http://depsy.org/package/python/h5py}{http://depsy.org/package/python/h5py}.  ADS, Arxiv.

\bibliography{references} 

\appendix

\input{appendix}

%\begin{thebibliography}{}

%\end{thebibliography}



\end{document}

%% file: intro.tex
\section{Introduction}

Perturbations induced by orbiting satellite galaxies within the dark matter (DM) 
halos of their hosts have been studied since the seminal work of \cite{Chandrasekhar43}.
It has been recognized that satellite galaxies generate density wakes by direct 
gravitational scattering of particles that pull back on the satellite, causing 
the satellite to lose angular momentum and energy in a process referred to as 
dynamical friction \citep{BinneyTremaine}.

It was later discovered that such local scattering is a manifestation of 
the resonant nature of the system \citep{Tremaine84}. 
In particular, \cite{White83} and \cite{Weinberg98b} found that, on global
scales, resonances between orbital frequencies of the DM particles 
in the halo and the satellite's orbital frequency can effectively transfer 
angular momentum and energy from the satellite to the DM halo. These resonances 
produce overdensities and underdensities, which also consequently 
affect the kinematics of the DM halo.

During the first passage of a satellite around a host galaxy,
the frequency of the satellite's orbit is continuous, and it has a broad
range of frequencies that resonate with those of the DM
particles of the host galaxy. These resonances produce the classical `conic'
wake that trails the
satellite described in \cite{Chandrasekhar43}. However, as the satellite
continues orbiting around
the host galaxy, its orbit's frequencies range gets narrower and hence it
resonates with particular frequencies of the DM particles. As a consequence, the
classical `conic'
wake weakens and overdensities in other regions of the DM halo start to take
place.
In this paper, we will refer to these density and kinematic perturbations as
the \textit{Transient} response and \textit{Collective} response. 
The Transient response corresponds to the classical Chandrasekhar's wake, which
trails the satellite galaxy. The Collective response corresponds to those
overdensities and underdensities not trailing the satellite, generated by the
narrow range resonances after the first passage.

For a detailed and comprehensive review of these resonant processes, we refer the 
reader to \cite{Choi09}, where a theoretical framework using perturbation theory 
is derived and compared to $N$-body simulations to investigate the resonances 
induced by a satellite in the DM halo of its host. \cite{Choi09} found that the 
location of the resonances within the halo is dictated by the orbital frequency 
and trajectory of the satellite. As such, detailed studies of the DM halo wake produced by a 
particular satellite must accurately account for the satellite's exact orbit.

In addition to the DM halo responses, the gravitational acceleration induced by a satellite 
galaxy will also offset the DM halo cusp of the host galaxy 
from the original DM halo center of mass
(COM) \citep{Choi09, Ogiya16}. Consequently, the orbital barycenter 
of the host galaxy will move \citep{Gomez15}. 
Accounting for these effects 
is crucial to properly interpret astrometric data of observed satellites, 
streams and globular clusters. For example, 
\cite{Gomez15} showed that accounting for the Milky Way's (MW) barycenter motion due to the 
gravitational pull from the Large Magellanic Cloud (LMC) can reconcile the 
mismatch between observations and simulations of the morphology of the 
Sagittarius Dwarf spheroidal galaxy (Sgr dSph) stellar stream, without invoking a
triaxial DM halo model.

In reality, the MW's halo is embedded with multiple substructures,
such as satellite galaxies, globular clusters, and smaller DM subhalos, 
 that induce localized perturbations to the DM and stellar halo.
For example, \cite{Loebman17} find 
disturbances in the velocity distribution of stars along sightlines 
that pass through individual satellite galaxies. These velocity changes 
manifest as ``dips" in the
anisotropy parameter profile, $\beta(r)$ defined as

\begin{equation}\label{eq:beta}
  \beta = 1 - \frac{\sigma_t^2}{2\sigma_r^2},
\end{equation}

\noindent where $\sigma_r$ and $\sigma_t$ are the radial and tangential 
velocity dispersions, respectively. Positive values of $\beta$ correspond 
to radially biased orbits, while negative values correspond to tangentially
biased orbits.

In corroboration, \cite{Cunningham18b} used
the \textit{Latte} cosmological-zoom simulations of two MW-like galaxies \citep{Wetzel16}
to show that substructure can cause $\beta$ to vary locally from -1 to 1.
However, these perturbations do not fundamentally alter the kinematics
of the entire MW stellar halo or DM halo itself, but are instead localized perturbations 
associated with substructure that only span a few kpc. 
In addition to the variations in $\beta$ caused by substructure, \cite{Loebman17} 
found a long-lived tangential bias in the $\beta$ profile of galaxies that have 
undergone a recent major merger. Moreover, ``$\beta$ dips" can correspond 
to breaks in the stellar density profile resulting from the assembly 
history of a galaxy \citep{Rashkov13}. While none of these studies were specific to the 
MW system, they do strongly suggest that the LMC should cause a potentially 
significant observable kinematic signature in the stellar halo.
Indeed, the LMC is likely inducing significant perturbations to the MW's disk
\citep{Laporte17,Laporte18a} and 
stellar streams, such as Tucana III \citep{Erkal18a} and the 
Orphan Stream \citep{Erkal18b}.

The LMC is the most massive satellite galaxy of the MW and 
is most likely on a highly 
eccentric orbit, only just past its first pericentric approach to our Galaxy 
\citep{Besla07, Kallivayalil13}. \cite{Patel17a} found using the Illustris simulation
that is extremely rare to find a high-speed, massive 
($\sim$10$^{11}$ M$_\odot$) satellite in close proximity to a massive host at $z=0$ 
in cosmological simulations, see also \citep{Boylan-Kolchin11, Busha11,
Cautun19}. 

As such, studying the DM halo wake induced by the LMC
in a cosmological context 
remains a challenge and is beyond the scope of this paper .
Instead, we construct detailed $N$-body models of 
the LMC's recent orbit, from its first crossing of the MW's virial radius, 
$\sim$2 Gyr ago, to the present day. With controlled 
numerical experiments, we can predict the general form and locations of 
perturbations in the kinematics of the MW's stellar halo and ultimately link 
those perturbations to the passage of the LMC and its induced wake within the MW's DM 
halo. These constrained simulations allow us to 
match the LMC's current 6D phase-space properties within 2$\sigma$ of 
observations \citep[see also][]{Laporte18a}, which is not currently possible 
with cosmological simulations.

The amplitude of the DM wake induced by typical MW satellite galaxies 
($M_{halo} \sim 10^9-10^{10}$ M$_\odot$) is expected to be much smaller than that
of the LMC ($M_{halo} \sim 10^{11}$ M$_\odot$). We will illustrate this point by comparing 
the properties of the LMC's DM wake to that of the next most massive 
perturber, the Sagittarius dwarf galaxy \citep{Laporte18}. Indeed,
DM wakes induced by massive orbiting satellites are identifiable in 
cosmological-zoom-in simulations of MW-like galaxies \citep{Gomez16}, despite the 
presence of multiple smaller orbiting bodies. In such cases,
DM wakes are found to not only affect the DM and stellar 
halo density and kinematics but 
also the structure of the galactic disk \citep[e.g.][]{Weinberg06,Gomez16, Gomez17}.  
However, note that
perturbations in the halo from 
the combination of multiple subhalos can be coupled in nontrivial
ways \citep{Weinberg07a}.
 
Critically, in this study, we will
assess the ability of current and future surveys to identify the 
signatures of the DM wake generated by the LMC within the MW's stellar halo. 
Current and near-future observational studies of the kinematics and structure of
the stellar halo (\textit{Gaia}, RAVE, H3, DES, DESI, APOGEE, GALAH, LAMOST, LSST, 4MOST,
and WEAVE) will reveal the structure and the kinematic
state of the stellar halo of the MW. Soon, the phase-space
information, i.e., distances, proper motions and radial velocities, of millions
of stars out to at least $100$ kpc will be known, in addition to that of other
halo tracers, such as satellites and globular clusters.

Ultimately, the phase-space information of halo tracers can inform us about the underlying
DM potential, the total mass, and the accretion history of the
MW \citep{Jhonston08, Helmi08, Gomez10a, Carlin16}. 
However, the LMC is a major perturber
to the MW's halo that has not yet been properly accounted for in such studies.
This study of the kinematic and density perturbations induced by 
the LMC is essential
to properly compute the uncertainty in current MW mass estimates. 
Strong variations in the kinematics of the stellar and DM halos 
(i.e. in $\beta$) across the sky
will cause variations in estimates of the mass of the MW inferred through Jeans 
modeling \citep[e.g.,][]{Watkins10}. Furthermore, given the lack of 6D phase-space 
information in the outer regions of the stellar halo, it is common to extrapolate
DM profiles to large radii using constraints within the inner 50 kpc. 
However, the LMC can strongly modify the distribution of mass in the outer halo -  
the resulting asphericity will also affect MW mass estimates \citep{Wang18}.

Specifically, we seek to answer the following questions:
are the phase-space properties of the stellar halo conserved in the
presence of the LMC?  What are the kinematic signatures of the DM halo
wake induced by the LMC? Can we identify the LMC's DM wake and track
the past orbit of the LMC through the stellar halo? Addressing these questions
is essential to properly interpreting the data
from current and upcoming high-precision astrometric surveys
\citep{DESI, Sanderson19WP, MSE} and may provide new cosmological tests of the total DM mass of the LMC
and MW and the nature of the DM particle itself.

The structure of this paper is as follows: in \S \ref{sec:LMC_mass}, we discuss 
current estimates for the mass of the LMC. \S \ref{sec:methods} describes the
numerical methods and initial conditions. 
In \S \ref{sec:results}, we discuss the main results of our
simulations, focusing on the density and the kinematics of the Transient and
Collective response induced within the stellar and 
DM halos of the MW.
In \S \ref{sec:obs} we discuss the observability of our findings, given
current and upcoming surveys. In \S \ref{sec:discussion} we discuss: the 
convergence of our simulations; how our results scale as a 
function of the LMC mass; comparisons between the DM wake produced by the LMC against that of 
Sgr; how the Transient response can be distinguished from stellar debris associated with the Magellanic Stream; and the prospects for studying the nature of DM using the LMC's DM wake. 
We conclude in \S \ref{sec:conclusion}.

%% file: LMC_mass.tex
\section{The mass of the LMC}\label{sec:LMC_mass}

The response of the MW's DM halo and corresponding
perturbations to the kinematics of the MW's stellar halo will depend on the
total mass of the LMC.  Moreover, owing to dynamical friction,
the orbital history of the LMC also strongly depends on its mass
\citep{Kallivayalil13}.
However, the LMC's mass is uncertain within a factor of $\sim$10. Many
theoretical models of the Magellanic System have assumed
low halo masses for the LMC 
\citep[ $\sim 10^{10}$ M$_\odot$, e.g.][]{Gardiner96, Connors06, Yoshizawa03, Diaz11, Guglielmo14}.
However, massive LMC-halo models ($10^{11} $M$_\odot$) have also been
shown to reproduce several observations, such as the global properties of the Magellanic System
\citep{Besla10, Besla12, Besla13, Salem15, Pardy18}, the morphology of the MW's
HI disk and its resulting line of nodes \citep{Weinberg06, Laporte18a}, and in
the misalignment of the velocity vectors of the Orphan stream \citep{Erkal18b}. 
In addition, there are a mounting number of arguments
that together strongly support a high infall mass for the LMC,
$\rm{M_{vir}} > 8 \times 10^{10}$ M$_\odot$ as listed below.

\begin{enumerate}
\item \textit{Rotation Curve:}  \cite{Vdm14} derived
the rotation curve of the LMC using the \textit{HST} proper
motions of 22 stars with known line-of-sight velocities.
The derived rotation curve peaks at  $91.7 \pm 
18.8 $km s$^{-1}$ at $8.7$ kpc, which implies an enclosed
dynamical mass of $\rm{M(<8.7)} = 1.7 \times 10^{10} \rm{M_{\odot}}$.
This is a strict minimum mass for the LMC, which is already at odds
with many existing theoretical models.

\item \textit{Extent of the LMC's Stellar Disk:} The stellar disk of
the LMC has been observed out to a radius of $\sim 19$ kpc \citep{Saha10,
Mackey16, Nidever18} from the LMC's optical center.
This indicates that the LMC is not tidally truncated at $8.7$ kpc \citep{Besla16}.
As such, the mass of the LMC must be larger than the dynamical
mass estimate, within 8.7 kpc, of $1.7 \times 10^{10} \rm{M_{\odot}}$. 
If the rotation curve remains flat to $\sim 19$ kpc the enclosed 
mass is $\sim 3.7 \times 10^{10} \rm{M_{\odot}}$.
On the other hand, if one assumes that the tidal radius of the LMC
is $19$ kpc, one can back out the mass of the LMC. Assuming an enclosed MW halo mass
    within $50$ kpc of $\sim 5\times10^{11} \rm{M_{\odot}}$ \citep{Kochanek96},  
the minimum mass of the LMC must 
be $8.3 \times 10^{10}  \rm{M_{\odot}}$ at the present day.

But the LMC does not illustrate clear evidence for tidal truncation,
suggesting its infall mass could be much larger.
In this study, we assume a minimum mass of the LMC at infall of $8 \times 10^{10} 
\rm{M_{\odot}}$. 

\item \textit{Cosmological Expectations:}
The total stellar mass of the LMC is $3.2 \times
10^{9}$ M$_{\odot}$  \citep{vanderMarel09}. Using
abundance matching, a statistical technique used
to assign a DM halo mass to a galaxy of a given stellar mass,
the total mass of the LMC, \textit{prior to accretion} by the 
MW, should be $\sim 1.6 \times 10^{11}$ M$_{\odot}$  \citep{ Behroozi10, Guo10,
    Moster10}.

Similarly, if a typical baryon fraction of $3\%$ (appropriate for spiral galaxies) is
assumed,
the total mass of the LMC before accretion should be $\sim 10^{11}$ M$_{\odot}$.  
These calculations indicate that the LMC was likely
quite massive at infall.

\textit{HST} proper motions indicate
that the LMC was 
likely recently captured ($<2$ Gyr ago) \citep{Kallivayalil13}. 
This first infall scenario is the cosmologically preferred orbital 
history for massive satellites of MW-mass hosts at z=0 
\citep{Boylan-Kolchin11, Busha11, Gonzalez13, Patel17}. In such a scenario, the LMC should
retain a significant fraction of its infall mass at the present day 
\citep[e.g.,][]{Sales11}.

\item \textit{Satellites of the LMC:}. The presence of the SMC and potentially multiple smaller satellites
companions \citep{DOnghia08, Jethwa16, Kallivayalil18}, also indicates that the LMC
must have been relatively massive at infall. 
In particular, satellites with stellar masses similar to the LMC that also have an
SMC companion usually reside in DM halos with a mass of
$\rm{M}_{200}=3.4^{+1.8}_{-1.2} \times 10^{11} 
\rm{M_{\odot}}$ \citep{Shao18}. This high mass
is also supported by studies of cosmological dwarf galaxy pairs in the field
\citep{Besla18}. 

\item \textit{Timing argument:}
The mass of galaxies in the Local Group can be
derived using the \textit{timing argument}. This method compares
the galaxies' currently observed positions and velocities
to the solution of their equations of motion in an expanding
universe  \citep{Kahn59, Lynden-Bell81, Sandage86, Partridge13, Penarrubia14}.
These equations can be
solved if the potential, the rate at which the universe is expanding, and the time
since the galaxies separated ($\approx 13.7$ Gyr, assumed to be the age of the universe) are known.
\cite{Penarrubia16} applied
a Bayesian inference method to constrain the total mass of the LMC using the
\textit{timing argument}. They found that the
LMC's total virial mass before infall is most likely 
$\rm{M}_{LMC}=2.5 ^{+0.9} _{-0.8} \times 10^{11}$ M$_{\odot}$ \citep[see also,][]
{Peebles10}.
We use this estimate as an upper limit on the mass of the LMC.
Our team has recently illustrated that such a high LMC 
mass is able to induce a strong warp in the outer disk. However, it
does not cause significant kinematic perturbations to the MW's disk
in the solar neighborhood
that violate observational constraints
 \citep{Laporte18a}.

\item{\bf Perturbations to Stellar Streams:} Recently, \cite{Koposov18} 
identified prominent twists in the shape of the Orphan Stream on the sky and
 nonzero motion in the across-stream direction.
\cite{Erkal18b} then illustrated that the misalignment
between the debris track and the streaming velocity cannot be reproduced in a static gravitational potential, but is instead best explained by 
perturbations from the LMC, provided it had an infall mass of $1.3^{+0.27}_{-0.24} \times 10^{11}$ M$_\odot$.

\end{enumerate}

Given the above arguments, we choose a range of $8-25 \times
10^{10} 
\rm{M_{\odot}}$ for the LMC's virial mass at the time it first
crossed the virial radius of the MW, noting that it could have been larger. 
We then simulate the evolution of the LMC to its present location on an orbit
consistent with the latest \textit{HST} proper motions of
\cite{Kallivayalil13}.
This is the first study of the global impact of such high LMC masses
on the MW's stellar and DM halo, mapping its the full extent of the wake  out to 200 kpc.

%% file: methods.tex
\section{Numerical Methods}\label{sec:methods}

The $N$-body simulations were
carried out with the Tree Smoothed Particle Hydrodynamics
code \verb+Gadget-3+ \citep{Springel08}, which is a modified version of \verb+Gadget-2+
\citep{Springel05} with an improved gravity solver.
We use the publicly available code \verb+GalIC+
\citep{Yurin14} to generate the initial conditions for the MW and the LMC.

\subsection{Galaxy Models}

Table \ref{tab:MWmodels} summarizes the parameters of the adopted MW
model.  Our MW model has a virial mass
of $\rm{M}_{vir}=1.2 \times 10^{12}$ M$_{\odot}$ \citep{McMillan17}. 
This is defined as the mass enclosed within the virial radius $R_{vir}$, 
where $R_{vir}$
encloses an overdensity of $\Delta_{vir}=357$. That is,
$\rho_{vir} = \Delta_{vir} \Omega_m \rho$, where $\rho$ is the average density
of the universe. We do not vary the mass of the MW in this study, as the first infall orbits
for the LMC are not recovered for massive MW models ($\rm{M}_{vir} > 1.5 \times 10^{12}
$M$_{\odot}$) and the LMC orbit has behaved
similarly in lower-mass MW models over the past 1-2 Gyr \citep{Kallivayalil13, Gomez15}.

The DM halo of the MW is represented by a
Hernquist profile \citep{Hernquist90}, where the scale length $r_a$ was chosen in
order to guarantee that the enclosed mass at the virial
radius is the same as that of the equivalent
NFW profile (for details of this procedure, see the appendix of
\cite{Vdm12}). Note that \verb+GalIC+ uses quantities evaluated at $R_{200}$ as
input parameters, which is the radius at which the enclosed
density is 200 times the critical density of the universe.
We have changed these definitions to virial quantities
in \verb+GalIC+ in order to
ensure equivalence between the NFW density profile and the
Hernquist profile.

The halo spin parameter $\lambda=0.027$
and concentration are consistent with typical MW-like DM halos in
cosmological simulations \citep{Klypin11}.
We adopt two different internal kinematic profiles for the MW's DM halo,
represented by the anisotropy parameter $\beta$ (see \S
\ref{sec:initial_beta} for a detailed description).

The disk of the MW is represented by an exponential profile and
the stellar bulge of the MW is modeled using a Hernquist profile.
 The stellar and DM particle mass are both $m_p = 4 \times 10^4$ M$_\odot$.
 The adopted MW disk, bulge, and halo parameters are within $2 \sigma$ of the best-fitting
MW parameters in \cite{McMillan17}, such that the rotation curve reaches a peak
of $\sim$240 km/s. We have included a disk and bulge in order to ensure
that the potential is realistic in the inner regions of the halo ($<$ 30 kpc)
and to accurately track the COM of the system.

Note that our simulations have small discreteness and can accurately 
capture distortions to the LMC's DM distribution. While small-scale resonances
can be affected by discreteness noise, we are interested in structures over 
larger scales (several kpc).
In general, the number of particles used in our
simulations is six orders of magnitude greater than the regimes where these
effects take place, as discussed in \cite{vandenbosch18a} and \cite{vandenbosch18b}.

\begin{table}
  \centering
  \begin{tabular}{c c c  }
  \hline
  \hline
    \textbf{MW Component} & \textbf{Parameter} & \textbf{Value}  \\
    \hline
    DM halo & M$_{vir}$, M$_{200}$ $[\times 10^{12} \rm{M_{\odot}}]$ & $1.2, 1.03$ \\
    & $R_{vir}, R_{200}$ [kpc] & $279, 208$ \\
    & concentration $c_{vir}$ & 15 \\
    & scale length $a_{halo}$ [kpc] & 40.82 \\
    & DM halo particles & $10^{8}$ \\
    & Mass per DM particle [M$_\odot$] & $4 \times 10^{4}$ \\
    \hline
    Disk & M$_{disk} [\times 10^{10}$M$_{\odot}]$ &  5.78 \\
    & Disk scale length $r_a$ [kpc] &  3.5\\
    & Disk scale height $r_b$ [kpc] &  0.5\\
    & Disk particles & 1382310 \\
    \hline
    Bulge & M$_{bulge} [\times 10^{10} \rm{M_{\odot}}]$ &  \\
    & scale length $a_{bulge}$ [kpc] & 0.7 \\
    & Bulge particles & 335220 \\
  \hline
  \hline
  \end{tabular}
  \caption{\textbf{Milky Way Model Parameters} Parameters defining the simulated
  MW. Models 1 and 2 have the same parameters, but are initialized with
  different anisotropy profiles (\S \ref{sec:initial_beta}). The MW is
  modeled using a Hernquist profile to describe the halo and bulge, and an
  exponential profile for the stellar disk. }
  \label{tab:MWmodels}
\end{table}

\begin{figure}
\centering
\includegraphics[scale=0.5]{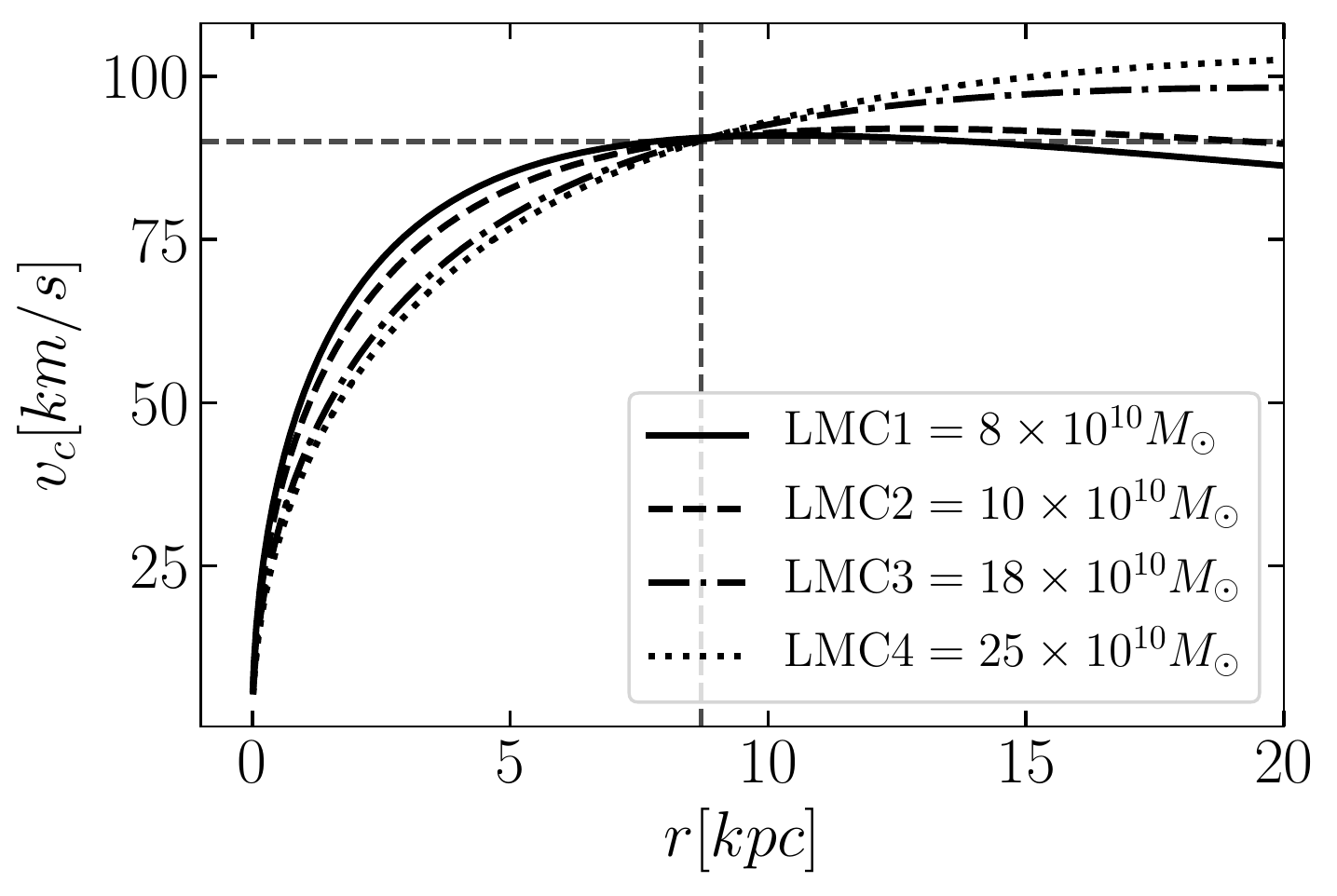}
\caption{Rotation curves of the four LMC models at infall. The LMC's
DM halo is modeled as a Hernquist profile with a scale factor chosen
to match the observed rotation curve. For visual
reference, the vertical and horizontal gray dashed lines are at 8.7 kpc
and at $91.7 \rm{km s^{-1}}$ respectively, illustrating that the models
agree with the measurements of \citet{Vdm14}.}
\label{fig:LMC_models}
\end{figure}

For the LMC, we construct four models with total halo masses of
$\rm{M}_{vir} = 0.8, 1.0, 1.8, 2.5 \times 10^{11}$ M$_\odot$.
The bulk of this study will focus on a fiducial LMC model, with a 
halo mass of $1.8 \times 10^{11}$ M$_\odot$, which is consistent with 
both models of the Magellanic System on a first infall \citep{Besla12,Besla13,Besla16} and 
the mean halo mass expected from abundance matching 
(see \S~\ref{sec:LMC_mass}).
The LMC is modeled using a Hernquist profile to represent
the DM halo. We do not include a disk,
as we are interested in the impact of the LMC on the MW halo
kinematics, where the dominant perturbations comes from the DM halo of
the LMC. We identify an adequate Hernquist scale length, $r_a$, to guarantee that
the circular velocity at $8.7$ kpc is $\sim 92$ km s$^{-1}$, as shown in
Figure \ref{fig:LMC_models}. The parameters of the LMC models are presented in
Table \ref{tab:LMC_models}. Note that the DM particle mass of each LMC model 
matches that of the MW ($m_{dm} = 4 \times 10^4$ M$_\odot$).

\begin{table}
  \centering
  \begin{tabular}{c | c | c | c | c }
  \hline
  \hline
  & LMC1 & LMC2 & LMC3 & LMC4  \\
  \hline
  $\rm{M}_{vir}, \rm{M}_{200} [\rm{M_{\odot}} \times 10^{10}]$ &  $ 8, 6.7$& $10, 8.35$ & $18,
    14.7$ & $25, 20.1$ \\
  $r_a [kpc]$ & 10.4& 12.7& 20 & 25.2\\
    $R_{vir}, R_{200}$ [kpc] & 113, 83& 121, 89& 148,108 & 165, 120\\ 
  \# DM particles [$10^6$] &  6.66 & 8.33 & 15 & 20.84 \\  
  \hline
  \hline
  \end{tabular}\par
  \caption{\textbf{LMC model parameters:} The virial mass 
  $\rm{M}_{vir}$ of the LMC halo for all models is
  consistent with the arguments given in \S \ref{sec:LMC_mass}.
  The values of the Hernquist scale length, $r_a$, are chosen to
  match the observed LMC rotation curve, as
  illustrated in figure \ref{fig:LMC_models}. The fiducial LMC model is LMC3. }\label{tab:LMC_models}
\end{table}

\subsection{The Milky Way's Anisotropy Profile, $\beta$}\label{sec:initial_beta}

One of the main advantages of using \verb+GalIC+ to generate galaxy initial conditions 
is that it allows us to specify an initial anisotropy profile, $\beta(r)$, for the DM
halo. We build two MW models with different forms for the radial anisotropic profile,
$\beta(r)$: (1) \textit{Model 1} assumes an isotropic 
DM halo ($\beta=0$); and (2) \textit{Model 2} assumes a radially
varying profile \citep{Hansen06}:

\begin{equation}
\beta(r)=-0.5-0.2\alpha(r) \,; \qquad \alpha(r)=\frac{ {\rm dln} \, \rho(r)}{{\rm dln}\, r}. 
\end{equation}

Model 1 allows us to study
perturbations from the LMC in the simplest case of an isotropic halo. Once the
halo response is understood in this idealized setting, we will use the gained
intuition to interpret the perturbations in the more realistic, radially
varying profile (Model 2).

Model 2 is radially biased,
where the radial dispersion is always larger than the tangential dispersion
(see the top right panel of Figure \ref{fig:beta_stability}).
Such a profile agrees with cosmological
numerical simulations where both the DM and the stellar halo anisotropy
profiles of MW type galaxies increase monotonically with increasing
Galactocentric radius \citep{Abadi06,Sales07}.

\subsubsection{Stability of the initial $\beta(r)$ profiles}\label{sec:betastable}

One of our main goals
is to study perturbations in the kinematics 
of the MW's stellar halo induced by the LMC. 
As such, we must first test the kinematic stability of the MW models 
generated with \verb+GalIC+. We use \verb+Gadget-3+ to evolve Models 1 and 
2 in isolation for 5 Gyr to test the
stability of the kinematic and density profile of the MW's DM halo.

\begin{figure*}
  \centering
  \includegraphics[scale=0.6]{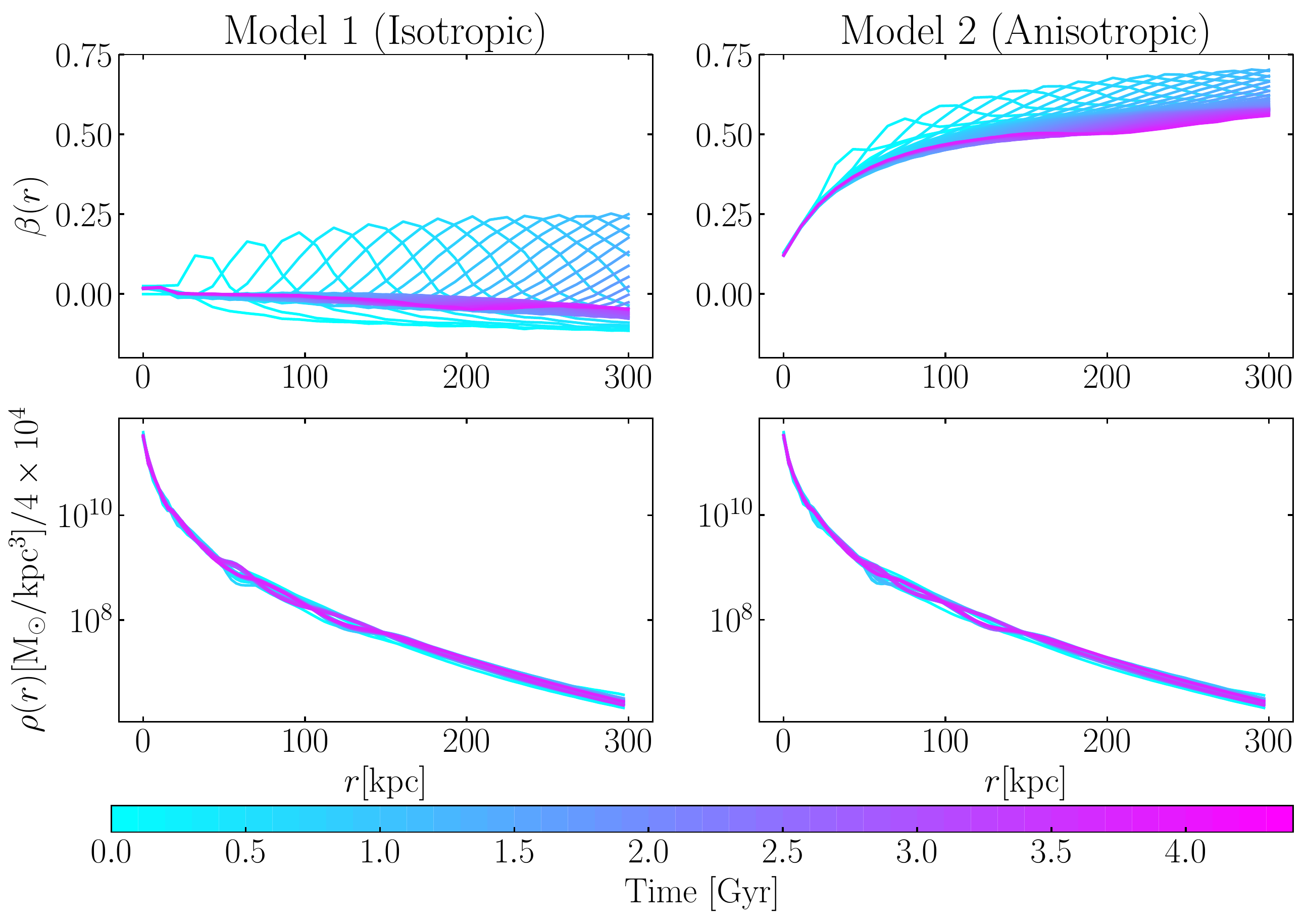}
  \caption{Time evolution of the anisotropy parameter profile, $\beta(r)$ (top), and
  density profile, $\rho(r)$ (bottom), for the DM halo of Model 1 ($\beta=0$; left panel), and Model 2 ($\beta=-0.5-0.2\alpha$; right panel).  The models are evolved 
    in isolation (without the LMC) over 5 Gyr.  The timeline is indicated
    by the color bar (in Gyr). The $\beta$ profile is initially unstable, but variations
    are minimal after $2.5$ Gyr. In contrast, the density profile remains stable for the duration of the run. 
    As such, we introduce the LMC to the simulation of the MW's DM halo after $2.5$ Gyr. 
    With these isolated MW halo stability tests, we can confidently isolate
    the impact of the LMC on the kinematics of the MW's stellar
    and DM halo from numerical artifacts. 
    }
 \label{fig:beta_stability}
\end{figure*}

The density profiles show minimum variation over
5 Gyr (bottom panel of Figure \ref{fig:beta_stability}).
On the other hand, $\beta(r)$ is not perfectly stable
for the first 2 Gyr (top panel of Figure
\ref{fig:beta_stability}). A ``bump" in the $\beta(r)$ profiles 
appears and evolves
with radius over $2$ Gyr. 
We have identified this to be a numerical artifact of \verb+GalIC+. 
However, for both models,
the variations are minimal
after 2.5 Gyr. As such, 
we introduce the LMC after the MW has been run in
isolation for $\sim$ 2.5 Gyr (purple colors in Figure
\ref{fig:beta_stability}).

\subsection{Orbit Reconstruction}

Using the described MW and LMC models, we set up a suite of $N$-body
simulations of the LMC's orbit within the MW's DM halo using \verb+Gadget-3+.  Table \ref{tab:sims}
summarizes the simulation suite.
The softening length is $\epsilon=0.08$ kpc, following the criteria of
\cite{Power03} (their equation $15$).

The initial 6D phase-space coordinates of the LMC, i.e.
when it first crossed the virial radius of the MW $\sim 2$ Gyr ago,
are identified by integrating the orbit of the LMC backwards in time from the present
observed position and velocity \citep{Kallivayalil13} following the same methodology
as in \cite{Gomez15}. We use the dynamical friction
equation derived by \cite{Chandrasekhar43},
where we adopt the following Coulomb Logarithm definition following \cite{Hashimoto03}.

\begin{equation}
\centering
    Ln(\Lambda) = \xi \left( \frac{b_{max}}{b_{min}} \right),
\end{equation}

\noindent where $b_{max}$ is the Galactocentric position of the LMC,
$b_{min}=1.4 \, r_s$, and $r_s$ is the scale radius in kpc. 
$\xi$ is a free parameter included as a fudge factor in
the dynamical friction acceleration adjusted to match the
$N$-body orbits.

We start with $\xi=1$ and integrate the orbit of the LMC backwards 
in time until it reaches the
MW's virial radius, storing the 3D position and velocity vector as a first 
guess for the initial starting point for the simulated LMC.
Then, we run a low-resolution ($m_{dm} = 1.2\times10^{6} \rm{M_{\odot}}$) $N$-body test
simulation using the identified first guess initial conditions.
We iterate the backwards orbital integration with lower values of $\xi$ until we
find good agreement between the analytic and the N-body orbit. This iterative procedure usually
requires between two and three iterations. Once an optimal value of $\xi$ is
identified for each LMC-MW simulation, the correct initial conditions are found by integrating the observed
present values 2 Gyr into the past.

As a result of this iterative procedure, our $N$-body simulations
reproduce the magnitude of the LMC's present-day position and velocity 
 within $2 \sigma$ of
the observed values ($\Delta r$ and $\Delta v$ in Table \ref{tab:sims}). Appendix \ref{appendix:sim}
contains further details of the exact position and velocity vector for each simulated
LMC model. 
In addition, both the velocity and the distance of the LMC at pericenter (50 Myr ago)
are within 1$\sigma$ of the analytic expectations
($r_{peri}= 48 \pm 2.5$ kpc, $v_{peri}= 340 \pm 19$ km/s) \citep{Salem15}.

\begin{figure}
\centering
\includegraphics[scale=0.5]{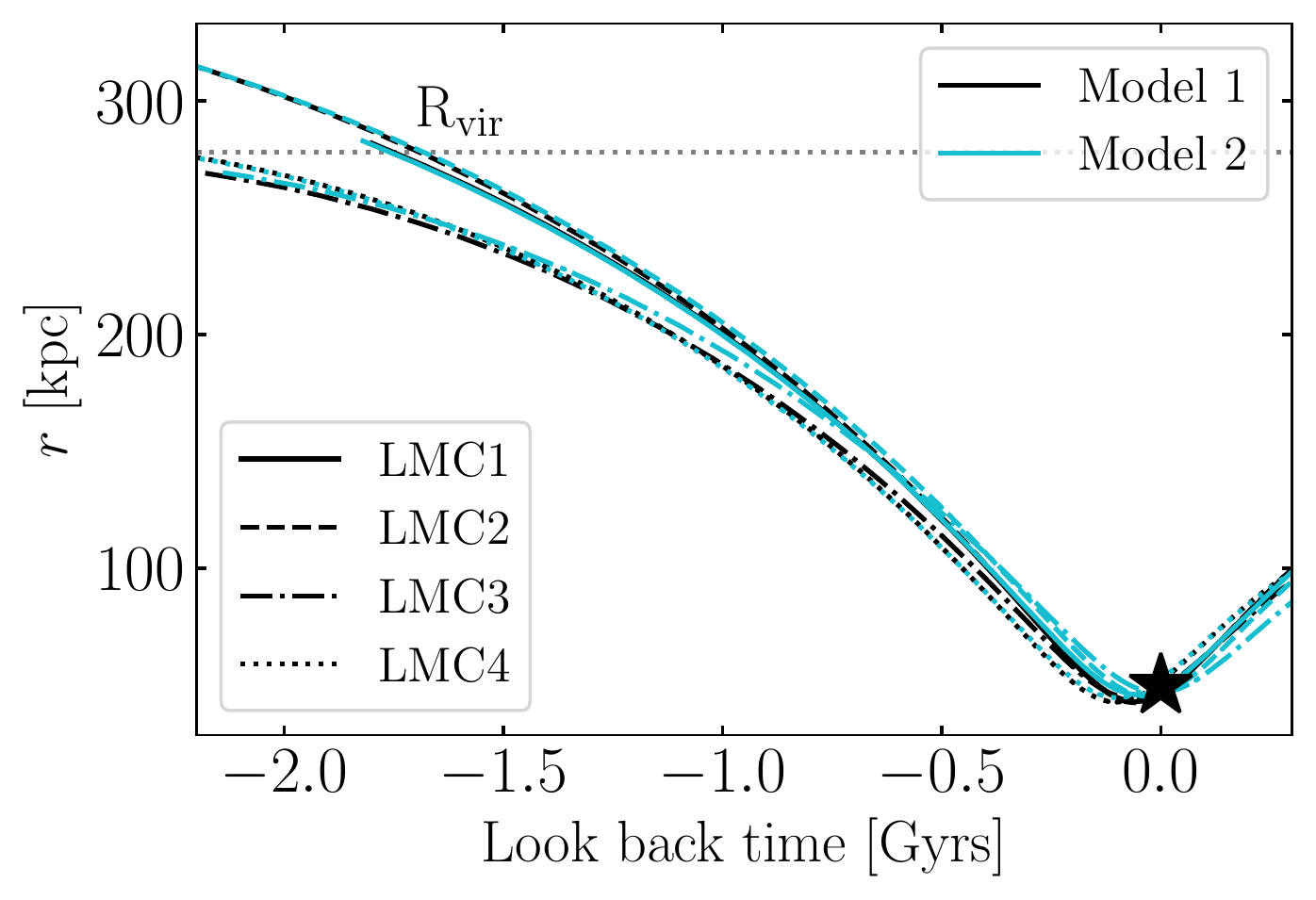}
\caption{The separation of the LMC's COM from that of the MW as a function of time
from the $N$-body simulations.
Results are illustrated for the four different LMC models using
 the isotropic (Model 1; black lines) and anisotropic (Model 2; blue lines) MW models. Note that
 the orbits are not arranged by mass. As such, the most massive LMC model does
 not reach the largest distance at infall (i.e. when its orbit crosses the
 virial radius of the MW). This is because the present-day position and
 velocity vectors are different in each simulation (see Appendix~\ref{appendix:sim}), and 
 the MW motion about the combined MW+LMC orbital barycenter increases with the 
 mass of the LMC, impacting the inferred separation of the MW-LMC system at early times
 \citep{Gomez15}.
 All orbits have pericenter distance and
 behavior consistent with backward analytic orbit integration using the observed proper
 motions \citep{Salem15}. The black star shows the current position of
 the LMC. Even though there are small deviations between the orbits, 
 all orbits are roughly consistent with each other over the
 last 1 Gyr. Therefore, the
 kinematics of the LMC's orbit are not a significant variable in this analysis
 \citep{Besla07, Kallivayalil13}.}
 \label{fig:LMC_orbits}
\end{figure}

The orbital separation of the LMC COM from 
that of the MW is 
illustrated in Figure \ref{fig:LMC_orbits} for all of the $N$-body LMC-MW simulations.
The COM position of the MW is computed using disk particles
within 2 kpc radius of the most bound particle. For the LMC, we use a 
shrinking sphere algorithm, following \cite{Power03}, to compute the COM of its DM halo.
We calculate the COM velocity
within a sphere of radius 10$\%$ of the virial radius, centered on the most bound
particles in each galaxy.
Regardless of LMC mass, all orbits agree with each other within the past 1 Gyr.
This is also true in higher-mass MW models \citep{Kallivayalil13}. As such,
the orbit of the LMC is not treated as a significant variable in this study.

\begin{table}
\begin{tabular}{c c c c c}
\hline
\hline
Sim. &  LMC model &  $\Delta r$ (kpc) & $\Delta v$ (km/s) & MW Model \\
\hline
$1$ & LMC1 & $2.51$ & $49.05$ & 1 \\
$2$ & LMC2 & $4.36$ & $64.43$ & 1 \\
{\bf $3$} & {\bf LMC3} & {\bf $2.49$} & {\bf $26.77$} & {\bf 1}\\
$4$ & LMC4 & $2.88 $& $31.99$ &  1 \\
$5$ & LMC1 & $1.83$ & $53.39$ & 2\\
$6$ & LMC2 & $2.98$ & $65.12$ & 2 \\
{\bf $7$} & {\bf LMC3} & {\bf $2.29$} & {\bf $44.89$} & {\bf 2} \\
$8$ & LMC4 & $3.54$ & $56.71$ &  2\\
\hline
\hline
\end{tabular}
\caption{Summary of the simulations. $\Delta r$ and $\Delta v$
denote the difference in the magnitude of the simulated present-day LMC 3-D position and
velocity vectors with respect to the observed values of
\citep{Kallivayalil13}. The MW kinematic profile is either isotropic ($\beta =0$; Model 1)
or radially anisotropic ( $\beta(r) = -0.15-0.2 \alpha(r)$; Model 2).
The fiducial simulations are Sim. 3 and 7, marked in bold.}
\label{tab:sims}
\end{table}

\subsection{Constructing the MW's Stellar Halo: Tagging DM Particles}
\label{sec:stellarhalo}

In this study, we will track perturbations in the density and kinematics of the
MW's DM halo induced by the LMC and aim to relate them to observations of the MW's stellar halo.
However, the $N$-body models of the MW created in this study 
do not explicitly include a live stellar halo due to its negligible self-gravity.
Instead, we build a mock smooth stellar halo using a weighting scheme implemented by
\cite{Laporte13a, Laporte13b} \citep[see also,][]{Bullock05, Penarrubia08}. 

In short, the technique works as follows.
We compute the fraction of stellar particles in
energy bins, $N_{\star}(E)$, from the distribution function and the density of states of
the DM particles of the MW's halo (we do this separately for both Model 1 and Model 2). 
The ratio of $N_{\star}(E)$ to the fraction of
DM particles in each energy bin, $N(E)$, provides the weight, $\omega$, that
each DM particle contributes to a stellar halo particle within that energy bin.
That is,

\begin{equation}
  \omega(E) = \frac{N_{\star}(E)}{N(E)}=\frac{f(E_{\star})}{f(E)},
\end{equation}

\noindent where the differential energy distribution is defined in terms of the density of
states, $g(E)$, and the distribution function, $f_{\star}(E)$, as

\begin{equation}
  N_{\star}(E) = g(E) f_{\star}(E).
\end{equation}

To compute the kinematics of the stellar halo particles, we utilize weights $\omega$, for
the DM particles, as follows:

\begin{equation}
  \sigma = \frac{\sum_i \omega_i(v_i - \bar{v})}{\sum_i \omega_i},
\end{equation}

where $v_i$ are the DM velocities of each particle, and $\bar{v}$ are the mean
velocities of all of the DM particles.
We assign the weights using the stabilized isolated MW models 
(i.e. after 2.5 Gyr of evolution in isolation; see \S~\ref{sec:betastable}). 
Because the  
DM halo is spherical and in equilibrium, 
the distribution function of the DM halo can be computed using Eddington's
equation \citep{Eddington16b}. We build two stellar halos for each MW model: {\bf MW-X} and
{\bf MW-H}. Both stellar halos 
have Einasto density profiles \citep{Einasto65},

\begin{equation}
  \nu(r) = \nu_e e^{-dn (r/r_{eff})^{(1/n)-1}},
\end{equation}

\noindent where $\nu_e$ is the normalization set by the total mass of the stellar halo. For our purposes, since we are measuring 
relative changes, the value of $\nu_e$ is set to 1.  For values of $n$ larger than 0.5,  $dn$ is defined as
in \cite{Merritt06}: 

\begin{equation}
    dn = 3n - 1/3 + 0.0079/n .
\end{equation}

The use of this profile is motivated by recent observations of K-Giants \citep[MW-X; ][]{Xue15}
and RR Lyrae \citep[MW-H;][]{Hernitschek18}. See Table \ref{tab:stellar_halo_models} for 
parameter details.

\begin{table}
  \centering
  \begin{tabular}{c c c c c c}
    \hline
    \hline
     & MW-X & MW-H \\
    \hline
    Density profile & Einasto & Einasto \\
    ($n$; $r_{eff}$) & (3.1; 15 kpc) & (9.53; 1.07 kpc)  \\
    Distances (kpc) & 10$-$80 & 20$-$131 \\
    Tracers &  K-giants & RRLyr \\
    Reference & \cite{Xue15} &  \cite{Hernitschek18} \\
    \hline
    \hline
  \end{tabular}\par
  \caption{MW stellar halo model parameters from observations of K-giants and RR Lyrae (RRLyr). }
  \label{tab:stellar_halo_models}
\end{table}

\begin{figure}
  \centering
  \includegraphics[scale=0.5]{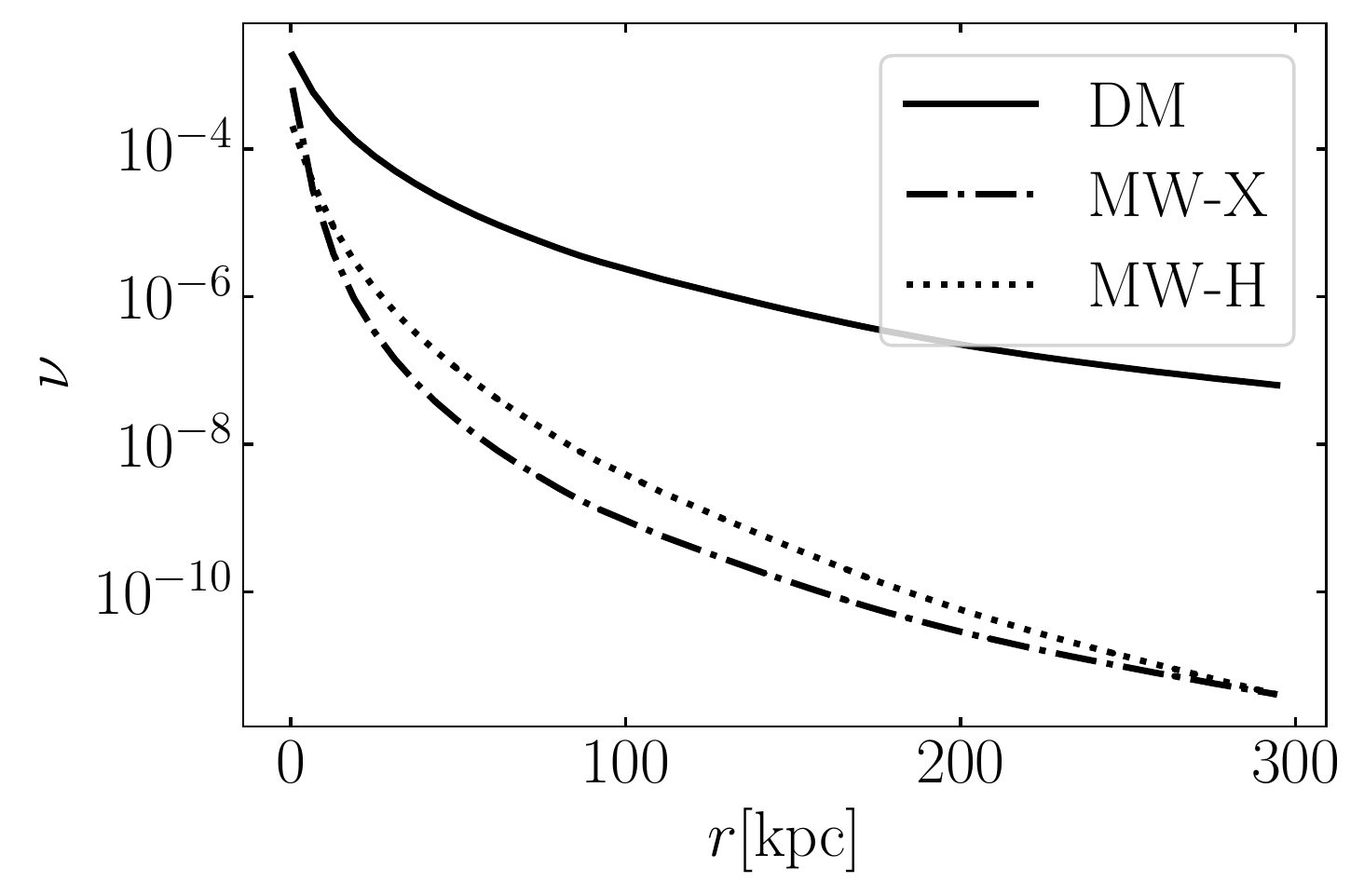}
  \caption{Initial number density profiles of the stellar halo (\#/kpc$^3$), built by
    applying the stellar tracer method, outlined in \cite{Laporte13a, Laporte13b}, and using the observed density 
    profiles for K-giants \citep[MW-X; dashed line][]{Xue15} and RR Lyrae \citep[MW-H; dotted line]{Hernitschek18}. 
    The DM density profile for the MW halo (solid line) is shown for
    comparison. This stellar halos are going to be used in both Models 1 and 2.}\label{fig:st_density}
\end{figure}

\begin{figure}
  \centering
  \includegraphics[scale=0.4]{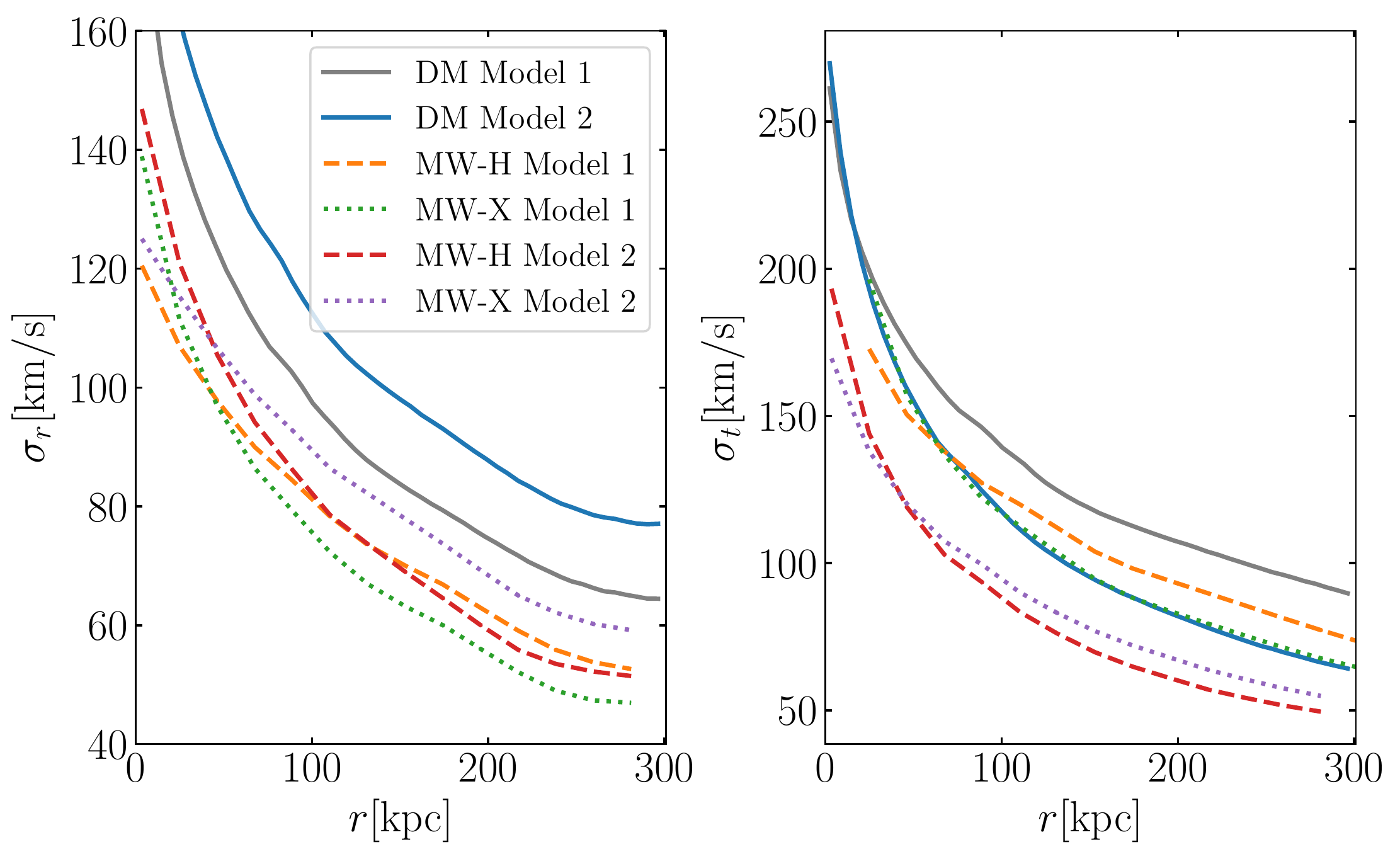}
  \caption{Radial ($\sigma_r$; left panel) and Tangential ($\sigma_t$; right panel) dispersion
    profiles corresponding to the stellar halo profiles from Figure~\ref{fig:st_density} (colored lines).  
    Results for the
    DM particles (solid lines) are shown for Model 1 (gray) and
    Model 2 (blue). Using this scheme, 
    the dispersion profiles for the stellar halo
    are not modeled to be the same as that of the 
    DM halo.
    }\label{fig:ST_kinematics}
\end{figure}

Figure \ref{fig:st_density} shows the resulting initial stellar halo number density (\#/kpc$^{3}$) 
and Figure \ref{fig:ST_kinematics} shows the velocity dispersion
profiles (tangential, $\sigma_t$, and radial, $\sigma_r$) using this outlined technique 
for MW Models 1 and 2. The results for the initial isolated MW DM halos are also shown for comparison.   

The MW-H density profile does not decrease as fast as MW-X with increasing 
Galactocentric radius. On the other hand, the
velocity dispersion profile is flatter for MW-X. Note that in this analysis, we extrapolate
the density and kinematic profiles of the stellar halo to distances larger than 100 kpc. 
This could be, in principle, an oversimplification since the outer halo likely
is not smooth; however, it lets us understand the simplest scenario as a first step. 

%% file: results.tex
%%%% ----------------------- RESULTS -----------------------------

\section{Results: The Response of the MW's DM and Stellar Halo to the LMC}\label{sec:results}

\begin{table}
\centering
\begin{tabular}{c c c c}
\hline
\hline
Octant \# & Longitude $(^{\circ})$ & Latitude$(^{\circ})$ & $t_{LB}$ (Gyr Ago) \\
\hline
$1$ & $[-180, -90]$  & $[-90, 0]$ & 0 - 0.12 \\
$2$ & $[-90, 0]$  & $[-90, 0]$ & 0.12 - 0.22 \\
$3$ & $[ 0, 90]$  & $[-90, 0]$ & 0.22 - 1.06 \\
$4$ & $[90, 180]$  & $[-90, 0]$ & $-$ \\
$5$ & $[-180, -90]$  & $[0, 90]$ & $-$ \\
$6$ & $[-90, 0]$  & $[0, 90]$ & $-$ \\
$7$ & $[0, 90]$  & $[0, 90]$ & 1.06 - 2.36 \\
$8$ & $[90, 180]$  & $[0, 90]$ & $-$ \\
\hline
\hline
\end{tabular}
\caption{Definition of the Octants illustrated in the Mollweide
projection in Figure \ref{fig:LMC_orbit} in Galactocentric
coordinates (Longitude and Latitude). The last column shows the look back time in Gyr, indicating the time since the 
LMC traveled through the given Octant. If no time is shown it is because the LMC never
passed through that Octant. The LMC spends the majority of its recent orbital history traveling in Octants 7 \& 3.
It is currently located in Octants 1 \& 2.}
\label{tab:octants}
\end{table}

Here, we study the perturbations induced by the LMC in the
properties of a smooth stellar halo (\S~\ref{sec:stellarhalo}) 
that is in equilibrium with the MW's DM halo, which is either 
initially isotropic (Model 1) or radially anisotropic (Model 2). 

\begin{figure}
\centering
\includegraphics[scale=0.5]{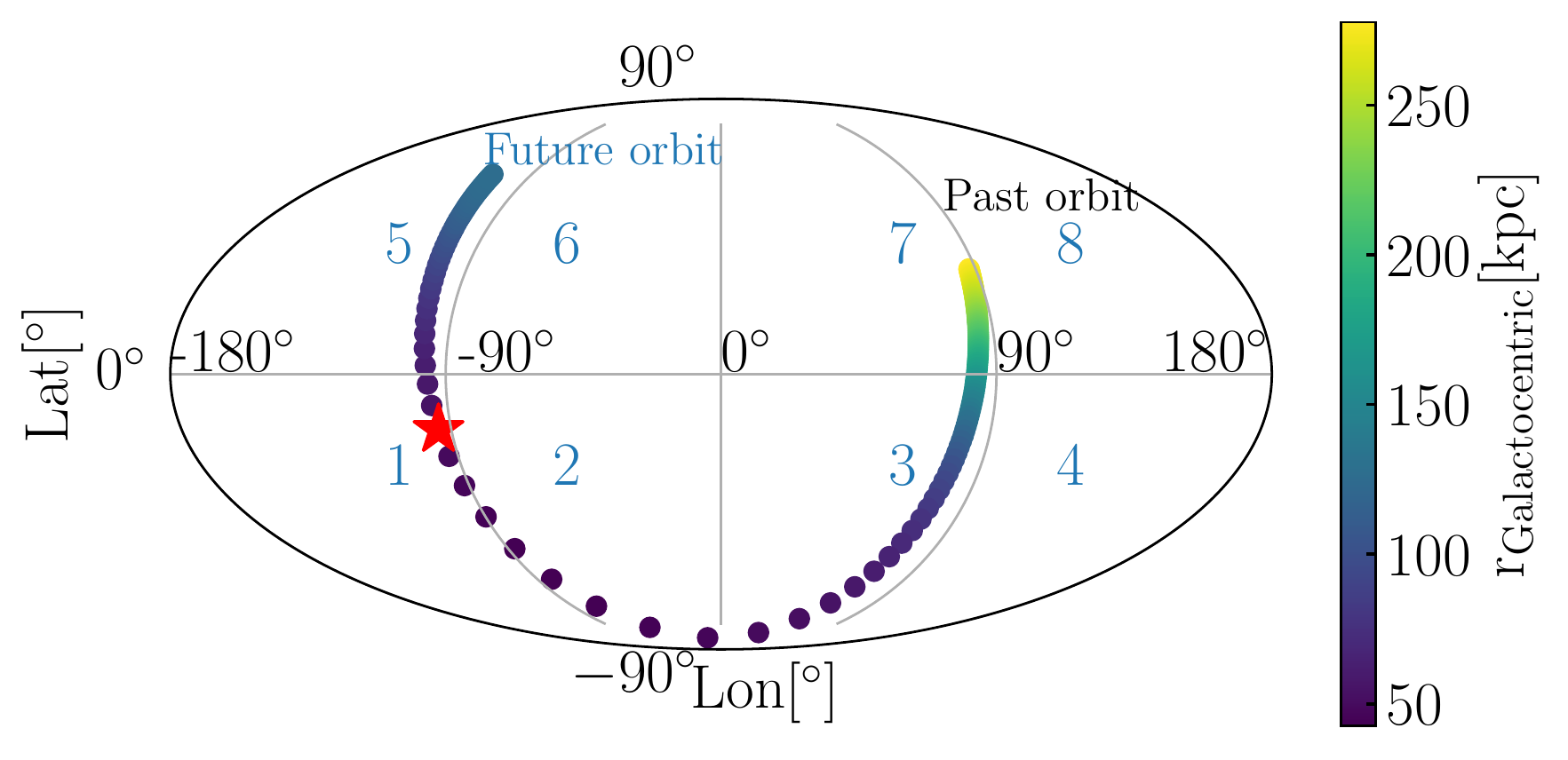}
\caption{Mollweide projection of the LMC's orbit from the fiducial isotropic simulation 
(Sim. 3), with
  Galactocentric longitude and latitude values marked in degrees. We use the
  convention for Galactocentric coordinates as defined in the Astropy library.
  Numbers denote Octants on the sky, defined in Table \ref{tab:octants}, which are used 
  as a reference for tracking the orbit of the LMC on the sky.
  The LMC enters the halo in Octant 7 ($\sim$ 2 Gyr ago) and is currently between Octants 1 and 2
  (marked by the red star).
  It reaches pericenter in Octant 2, $\sim$50 Myr ago at a distance of $\sim$45 kpc. 
  Octants 5 and 6 represent the predicted orbit
  of the LMC 0.5 Gyr into the future. The colorbar represents the Galactocentric
  distance of the LMC at each point in its orbit.
  The orbits from all 
  LMC-MW simulations are very similar; see Figure \ref{fig:LMC_orbits}.
  Note that this viewing perspective is flipped with respect to traditional
Mollweide projections ($-180^{\circ} \leq $ lon $ \leq 180^{\circ}$).}
\label{fig:LMC_orbit}
\end{figure}

We aim to identify regions of the
stellar halo that are responding to the passage of the LMC.
Figure \ref{fig:LMC_orbit}, shows the past and future orbit of the LMC in a Mollweide 
projection using Galactocentric coordinates following the 
convention of the Astropy library \footnote{\url{http://docs.astropy.org/en/stable/api/astropy.coordinates.Galactocentric.html}}.
The numbers in blue indicate Octants, which are defined in Table
\ref{tab:octants}. The decomposition of the sky into Octants
is used to guide the analysis, allowing 
us to relate the past location of the LMC to specific areas in the stellar halo. 
The LMC starts at the virial radius of the MW in Octant 7. The star
illustrates the present-day position of the LMC (which is located in Octants 1 and 2). 
The future orbit of the LMC
is depicted by the points to the left of the star, moving through Octants 1 and 5. The colorbar 
represents the Galactocentric distance of the LMC along its orbit.
The LMC reaches Octant 5 at a distance of $\sim$120 kpc. 
It takes $\sim 2$ Gyr for the LMC to travel from $R_{vir}$ to its current
location, which is marked by the red star.

In the following, we quantify the
perturbations in the density (\S \ref{sec:density}) and kinematics (\S \ref{sec:kinematics}) of the stellar
halo induced by the LMC at three different
Galactocentric radii: at 45 kpc, where the effects of the LMC on the halo are
the strongest, and at 70 kpc, to illustrate the extent to which the LMC's perturbations can be traced in the frontier
discovery space for LSST and future surveys. 

Note that we discuss only results for the fiducial LMC model of $\rm{M}_{vir}=1.8\times
10^{11}$ M$_{\odot}$ on the properties of the two different MW halo models (Sims 3 and 7 in Table~\ref{tab:sims}). 
Later, we will discuss how our results 
change as a function of the LMC's infall mass (\S \ref{sec:LMC_mass_wake}).

\subsection{Density Perturbations Induced by the LMC: The LMC's DM Wake}\label{sec:density}
In this section, we study the density perturbations induced within
the MW's DM and stellar halo owing to the recent orbit of a massive 
LMC. 

\subsubsection{The DM Halo Wake in Cartesian Coordinates}

\begin{figure*}
  \centering
  \includegraphics[scale=1]{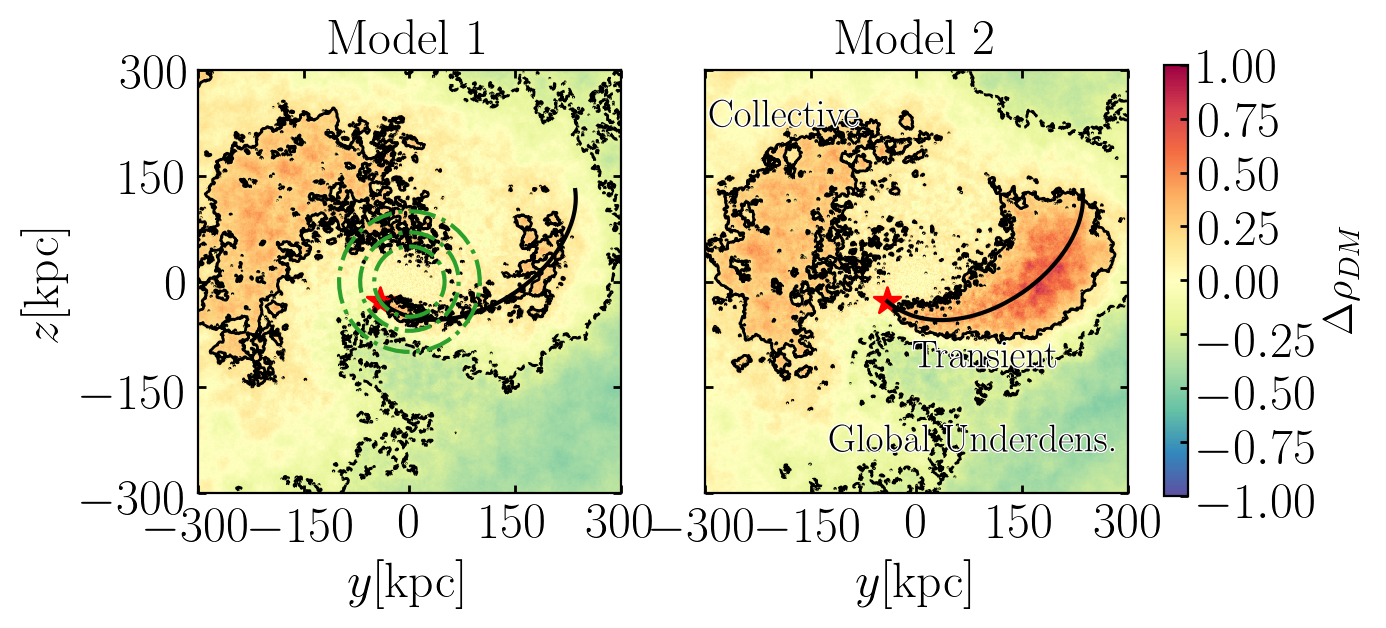}
  \caption{Density perturbations, revealing the DM halo wake induced by the LMC in the MW's DM halo in the
  $y-z$ Galactocentric plane, which is almost coplanar with the LMC's orbit. The
  MW's disk is in the $x-y$ plane, where the Sun is located at $x=-8.3$ kpc. Results are for a slice, 10 kpc in thickness, centered at $x=0$. 
  The green circles illustrate galactocentric distances of 45, 70 and 100 kpc.
  No LMC particles are included in this plot. 
  The color map illustrates the density
  ratio of the MW's DM distribution at the present day
  with respect to that of the MW in isolation (prior to the infall of the LMC) for Model 1 (left) and Model 2 (right). 
  Red colors indicate overdense regions and blue underdense regions.
  The current position of the LMC is represented by a red star and its orbital path is traced by the
  solid black line. Contours indicate underdensities of $\Delta \rho_{DM}=$ 0.8 (dashed) and overdensities
  of $\Delta \rho_{DM}=$1.2 (solid). Three main morphological features are
  identified. 1) {\bf The Transient response:} the DM overdensity
  trailing the LMC, tracing its past orbit. 2) {\bf The Collective response:} An overdensity that appears in the north
  (at $z>0$ \& $y<0$). 3) {\bf The Global Underdensity:} Underdense regions that
  surround the Transient response. 
  The morphology and the strength of these features are somewhat different 
  for both the isotropic (Model 1; left panel) and anisotropic (Model2; right
  panel) MW halos, reflecting the resonant nature of the DM halo. In particular, 
  the overdensity in the DM wake is stronger in Model 2.}
  \label{fig:wake_yz}
\end{figure*}

Figure \ref{fig:wake_yz} illustrates the density perturbations 
($\Delta \rho$) to the MW's DM halo (Models 1 and 2) by the LMC in Cartesian coordinates. 
Changes in the local MW 
DM density are measured with 
respect to the MW's halo in isolation (prior to the 
infall of the LMC; $MW_{iso}$). This is defined as:
%Perturbations are measured with respect to the MW halo 
%(Model 1 or 2) 
%excluding the LMC and computed as:

\begin{equation}\label{eq:deltarho}
  \Delta \rho_{DM} = \frac{\rho_{MW}}{\rho_{MW_{iso}}} - 1.
\end{equation} 

Note that LMC particles are not included in this 
calculation. Figure \ref{fig:wake_yz} shows a slice, 10 kpc in thickness,
of the present-day simulated MW DM halo 
in the Galactocentric $y-z$ plane, which is roughly co-planar
with the LMC's orbital plane. The MW's disk lies in the $x-y$ plane, and the Sun 
is located at $x=-8.3$ kpc.

Figure \ref{fig:wake_yz} reveals the extended and anisotropic
nature of the DM halo wake. We identify three main components:
\begin{enumerate}
    \item A {\bf Transient response}, seen as a DM overdensity trailing the LMC along 
    its orbit. This feature is marked by solid contours (red regions at positive $y$). 
    This is analogous to the classical Chandrasekhar wake.
    \item A large underdense region shown with dashed contours
    (blue regions) south of the Transient response. We call this the {\bf Global Underdensity}.
    \item An extended overdensity in the Galactic north, 
    shown with solid black contours (red regions at positive $z$ and negative $y$). 
    This {\bf Collective response} covers roughly one quarter of the sky.
\end{enumerate}

These maps indicate that the perturbations in the density 
field of the MW's DM halo are stronger at larger Galactocentric distances, $R_{GC}>45$ kpc. This likely 
reflects the 
longer duration that the LMC spend in those regions.
Within 45 kpc, the halo wake is the weakest since the LMC has not yet passed 
through the inner halo. Yet, the LMC does 
impact the structure of the MW's outer disk \citep{Laporte18a}. 
According to \cite{Weinberg89, Weinberg95, Weinberg98a, Choi09}, an inner 
DM wake should also be created due
to inner resonances, but this is not apparent in these simulations, likely due to the 
LMC's high speed. 
The resonant modes induced in the halo will be quantified in upcoming work 
(Garavito-Camargo et al in prep). 

Overall, the density maps agree for both Models 1 and 2. However, we note some
differences: 1) the Transient response is stronger in Model 2 (left panel in
Figure \ref{fig:wake_yz}); 2) the Collective response is stronger for Model 1; 
3) the morphology of the Transient and Collective DM responses, and the
Global Underdensity vary slightly. These differences are a consequence of the
internal kinematics of the two DM halo models as also found by Amorisco
in prep.

\subsubsection{The Wake in the Stellar Halo: Mollweide Projections.}

\begin{figure*}
  \centering
  \includegraphics[scale=0.9]{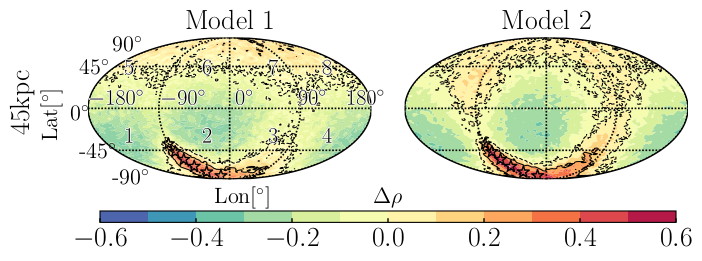}
  \includegraphics[scale=0.9]{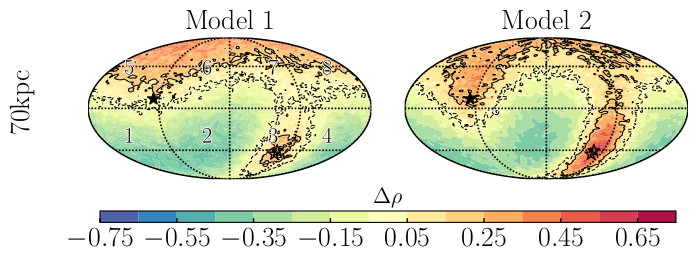}
  \includegraphics[scale=0.9]{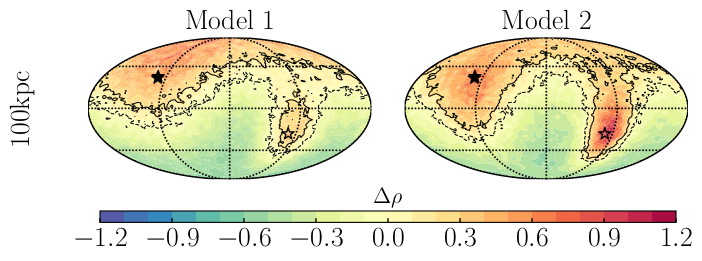}
  \caption{Mollweide maps of the MW stellar halo, illustrating local enhancements 
  and decrements in stellar density within a spherical shell, 5 kpc in thickness, 
  at three illustrative Galactocentric radii: 45 kpc (top), 70 kpc (middle), and 100 kpc 
  (bottom). Results are shown for the MW-X halo that follows the K-giant density 
  profile of \cite{Xue15} constructed in equilibrium with the
  the MW halo Model 1 (isotropic; left panel) and Model 2 (anisotropic; right
  panel); 
  results are similar for the MW-H model. 
  The color bar represents the ratio between the local density at a
  given grid point, $\rho(r)_{local}$, with respect to the spherical average
  in the shell at that distance, $\overline{\rho(r)}$. 
  The open black stars represent the past orbital path of
  the LMC at each Galactocentric distance, while the solid black stars indicate its future
  orbit. Contours mark stellar
  overdensities of $\Delta \rho=$0.1 (dashed) and $\Delta \rho=$0.2 (solid) relative to the mean
  in all panels except at 100 kpc, where contours mark overdensities of 
  $\Delta \rho=$0.2 (dashed) and $\Delta \rho=$0.6 (solid). 
  The stellar halo exhibits a stellar wake that tracks that of the DM
  halo and is similarly composed of three main features: (1) the Transient
  response tracing the LMC's past orbit
  in the south (red region marked by open stars); (2) the Collective response, seen as a large red overdensity
  in the north that persists at all distances; and (3) the Global Underdensity, seen as blue regions
  surrounding the stellar Transient response in the south.
  The contours illustrate that the initial halo kinematics impact the amplitude
  of the Transient response, 
  which is more pronounced in Model 2 than Model 1. 
  The Collective response is very extended and its structure and amplitude depend only
  mildly on the initial kinematic structure of the halo.}
  \label{fig:dm_wake_mollweide}
\end{figure*}

In this section, we explore how the DM wake induced by the 
LMC manifests within the stellar halo. 
In Figure~\ref{fig:dm_wake_mollweide}, we use the 
methodology outlined in \S \ref{sec:stellarhalo} to identify the
corresponding density 
perturbations seen in Figure~\ref{fig:wake_yz} in a
Mollweide map of the MW's stellar halo. 

The density perturbations are 
computed as follows.
We build a grid, with cell size of projected area $(1.6^{\circ})^2$,
on a spherical shell, 5 kpc in thickness, at a given Galactocentric radius.
%centered at a given radius. 
We define grid points as the corners of each cell. At each grid point,
we compute the local density, $\rho(r)$, using the 1000 nearest 
particles. At 200 kpc, the outskirts of the halo, grid cells
correspond to a volume of a cell of 13 kpc length and 5 kpc thickness.

The color scale in Figure~\ref{fig:dm_wake_mollweide} represents $\Delta \rho$, 
which we define as the ratio between the local 
stellar density,
$\rho(r)$, and the mean stellar density across the all-sky 
spherical shell, $\overline{\rho(r)}$:

\begin{equation}
\Delta \rho(r) = \frac{\rho(r)}{\overline{\rho(r)}}-1.
\end{equation}

Results are shown for spherical 
shells at 45, 70, and 100 kpc, centered on the Galactic center (see \url{https://bit.ly/2S25YzC} 
for additional plots at distances of 25, 150, and 200
kpc).

Given that the stellar halo is modeled in equilibrium with the DM halo, the 
same three features of the DM halo wake seen in Figure~\ref{fig:wake_yz} 
are also apparent in Figure~\ref{fig:dm_wake_mollweide}.
Furthermore, using the Mollweide projection, we can now identify the locations of these 
features on the sky. 

\begin{enumerate}
\item {\it The Transient response.} In the south, there exists a local stellar overdensity, 
coincident with the Transient DM wake, tracing the past orbit of the LMC 
(red region tracked by open black stars). The stellar Transient response persists over 
distances from 45 to 100 kpc (Figure~\ref{fig:dm_wake_mollweide})
and even at distances as large as 200 kpc.

\item {\it The Collective response.} An extended overdensity is apparent in the
  north, 
between Octants 5 and 8, at all distances. This coincides with the Collective DM
response that is generated by both resonances and the displacement of the orbital barycenter.

\item {\it The Global Underdensity.} In the south, primarily on
either side of the Transient response, underdense regions (blue) are apparent,
reflecting the removal of stellar mass and DM from these regions 
to form the higher-density Transient and Global responses.
\end{enumerate}

Perturbations in the stellar halo at distances smaller than 45 kpc do exist, but the amplitude is significantly 
lower. Instead, we focus our analysis on
the strongest wake amplitude in the hopes
of devising a viable observational strategy (see \S
\ref{sec:dens_observability}) to capture signatures of the wake in the stellar
halo.

We again see differences between the density perturbations in Model 1 vs. Model 2, indicating that
the internal kinematics of the DM halo affect the morphology and amplitude of the wake within the stellar halo.
The dashed and solid contours are
at the same density enhancement in both models, illustrating that the wake is
consistently stronger in Model 2. Regardless of the detailed
internal kinematics of the DM halo, we find that perturbations to the stellar 
halo caused by the LMC persist for 2 Gyr and will cover a very large volume of 
the stellar halo.

\subsection{Kinematic Perturbations in the Stellar Halo: The Kinematics of the LMC's Wake}\label{sec:kinematics}

As seen in the previous section, the DM and stellar halos are perturbed by the passage of the 
LMC, resulting in regions of over- and underdensities. This requires the displacement
of mass in the halo \citep[e.g.][]{Buschmann18}. 
Here, we identify the kinematic signatures of this motion. These signatures
complement the density perturbations studied in the previous section 
and provide key observables for the identification of the wake.

We compute the local mean velocities
and velocity dispersions of the stellar halo 
using the nearest 1000 particles at each grid point in the Mollweide projections,
as in Figure~\ref{fig:dm_wake_mollweide}. 

All kinematic quantities are computed in Galactocentric coordinates.
As in the previous section, we present results at three illustrative Galactocentric 
distances: 45, 70 and 100 (for $\Delta \sigma_r$) kpc (in \url{https://bit.ly/2S25YzC} 
we include the corresponding plots at 25, 100, 150 and 200 kpc). We first show our results 
for the radial velocities (\S \ref{sec:radial}), followed by the tangential
motion (\S \ref{sec:tangential}).

\subsubsection{Radial Motions in the Stellar Halo.}\label{sec:radial}

\begin{figure*}
  \centering
  \includegraphics[scale=1]{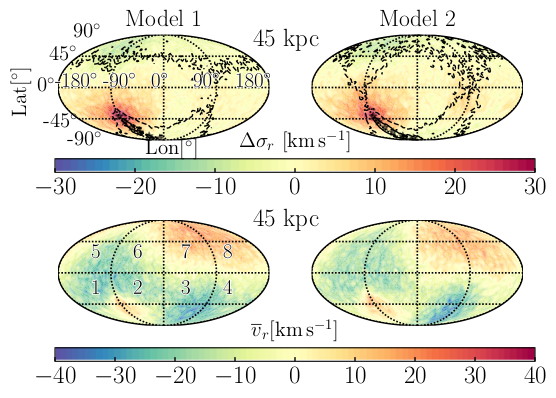}
  \caption{Mollweide projection in Galactocentric coordinates illustrating 
  radial motions of the stellar halo. 
  The top panel shows the radial-velocity dispersion relative to the average computed
  over a spherical shell, 5 kpc in thickness, at 45 kpc, $\Delta \sigma_r = \sigma_r/\overline{\sigma_r}$. 
  The bottom panel illustrates the local mean radial velocities, $\overline{v_r}$, across the 
  same spherical shell. Results are similar for both Model 1 (left) and Model 2 (right). 
  Contours illustrate the location of overdensities in the north (Collective
  response) and the south (Transient response),
  representing density enhancements, $\Delta \rho$, of 0.0 and 0.4 relative to the mean density
  (see Figure~\ref{fig:dm_wake_mollweide}). The gray stars show the past orbit of the
  LMC at this distance.
  Local values of both $\sigma_r$ and $\overline{v_r}$ are measured using the 1000 nearest stellar particles within
  a grid cell of $1.6^{\circ}$ squared. 
  For the top panels, the color scale indicates increases (red) or decreases (blue) in $\sigma(r)$
  relative to the shell average. In the bottom panels the color bar indicates the direction 
  of average radial motions, $\overline{v_r}$, on the sky (blueshift or redshift).
  $\sigma_r$ increases by $\sim$25 km/s near the LMC (Octants 1 and 2), 
  leading the Transient response, and decreases by $\sim$14 km/s to the north of the LMC (Octants 4 and 5),
  forming a kinematically ``cold region". 
  The latter coincides
  with the Collective response and the future orbital trajectory of the LMC. 
  In contrast, the velocity dispersion in Octants 7 and 8 in the north is largely unaffected. 
  The radial motions, $\overline{v_r}$, illustrate that the region of the
  Transient response closest to the 
  LMC (Octant 2) is moving away from the galactic center (redshifted; as is the LMC COM motion), but the 
  Transient response further away from the LMC (Octants 3 and 4) 
  follows the past orbital motion of the LMC toward the MW (blueshifted). On
  average, the northern Sky
  appears to be currently moving away from us.}
  \label{fig:radial_45}
\end{figure*}

\begin{figure*}
 \centering
 \includegraphics[scale=1]{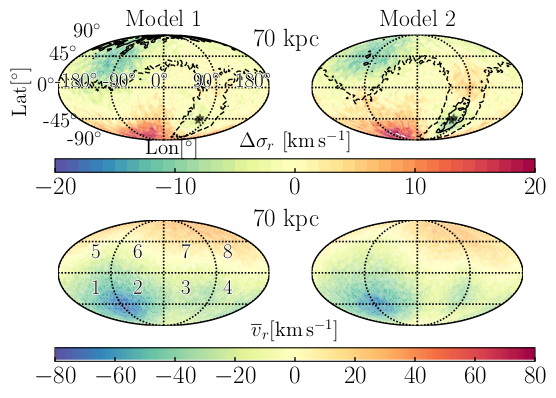}
 \caption{Same as Figure \ref{fig:radial_45} but at 70 kpc. The gray tars mark the past location 
 of the LMC on the stellar halo at this distance. The decrease in $\sigma_r$ 
 still persists in the north (Octants 4 and 5) and appears more clearly associated 
 with the Collective response. The increase in $\sigma_r$ in the south has 
 moved, with respect to the 45 kpc case, on the stellar halo to reflect the 
 past motion of the LMC, but still leads the 
 Transient response. The mean radial velocities in the north indicate that the
 Collective response is still moving away from the MW disk, but now there is a comparably 
 blueshifted region in the southern sky, forming a dipole pattern that may 
 reflect the COM motion of the disk about the LMC-MW orbital barycenter.
}
  \label{fig:radial_70}
\end{figure*}

\begin{figure*}
 \centering
 \includegraphics[scale=1]{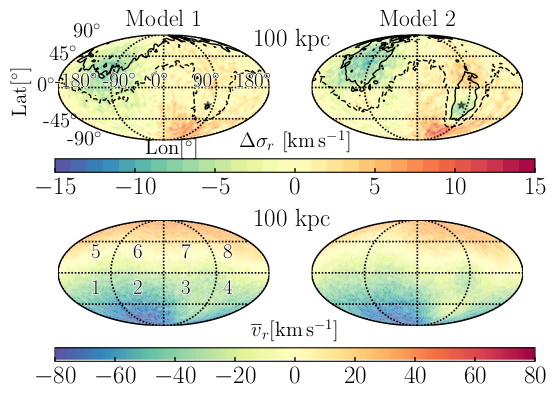}
 \caption{Same as Figure \ref{fig:radial_45} but at 100 kpc. 
 Results are similar to that of 
 Figure \ref{fig:radial_70}, but the dipole pattern in the $\overline{v_r}$ is even more 
 pronounced and the effect is very similar in both Model 1 and 2. This supports the idea
 that this velocity pattern is generated from the motion of the MW disk about
 a new orbital barycenter, as we expect the halo outskirts to exhibit the largest offsets
 with respect to the motion of the halo cusp.}
 \label{fig:radial_100}
\end{figure*}

Radial velocities are computed with respect to the Galactic center.
Figures~\ref{fig:radial_45},\ref{fig:radial_70}, and \ref{fig:radial_100} show 
the change in the local radial Galactocentric velocity
dispersion ($\sigma_r$) relative to the all-sky average dispersion ($\overline{\sigma_r}$):
\begin{equation}
\Delta \sigma_r (r) = \sigma_{r}-\overline{\sigma_r}
\end{equation}
\noindent and the local mean radial velocity, 
$\overline{v_r}$ of the stellar halo at 45, 70, and 100 kpc, 
respectively. Again, local quantities are computed at each grid point,
as in Figure~\ref{fig:dm_wake_mollweide}.

At 45 kpc, the radial-velocity dispersion maps show two main features: (1) an 
increase in $\sigma_r$ of $\sim 25$ km/s near the LMC (Octants 1 and 2); and (2) a
decrease in $\sigma_r$ by $\sim 14$ km/s, forming a ``cold region", in the north (Octant 5) 
that coincides with both the Collective response and the future of the orbit of the LMC. 
In the northern Octants 7 and 8, $\sigma_r$ appears unaffected by the LMC. 

The increase in $\sigma_r$ in Octants 1 and 2 correlates with 
positive motions in $\overline{v_r}$ of $20$ km/s in the same region. The stars
in the region of the Transient response closest to the LMC are thus 
tracing the COM motion the LMC, which is currently moving away 
from us (redshift). Further away from the LMC, 
stars in the Transient response trace the past orbital motion of the LMC toward the MW.

These kinematic perturbations persist at larger Galactocentric distances. 
In particular, the `cold region' associated with the Collective response is more apparent at 70 kpc
(Figure \ref{fig:radial_70}).
Also, $\overline{v_r}$ is consistently negative (blueshifted) along the
Transient response,
following the orbit of the LMC (black stars) to large distances.
There is thus a strong spatial correlation between 
$\overline{v_r}$ and the Transient response. 
 
Overall, the results are consistent for Models 1 and 2, but the amplitude of the 
blueshift in $\overline{v_r}$ within the Transient response is larger in Model 2. This likely
explains the increased strength of the Transient response in Model 2, seen in Figure~\ref{fig:dm_wake_mollweide}. 
The kinematic profile of the halo in Model 2 is radially anisotropic, naturally 
boosting the Transient response signature, which follows the radially infalling orbit of the LMC. 

The Collective response in the northern sky
exhibits positive radial velocities, indicating that this region
is moving away from the MW's disk. 
Moreover, at larger distances (Figures \ref{fig:radial_70} and \ref{fig:radial_100}), 
radial motions increase, approaching the COM motion of the MW's disk ($\sim$56 km/s in Sim. 3). 
In the south, we see 
negative radial velocities, particularly in Octants 1 and 2. Furthermore, the velocity 
pattern at 100 kpc is very similar in both Model 1 and Model 2. Together, these results 
support the idea that this pattern results from the 
motion of the MW's disk about the new LMC-MW orbital barycenter. 
The disk is moving toward the location of the LMC at its pericentric approach (Octant 3)
and so the radial-velocity pattern seen in
the outer halo is thus the reflex of this motion: {\it the 
northern sky displays redshifted motions and the south blueshifts}.

\subsubsection{Tangential Motions in the Stellar Halo}\label{sec:tangential}

We compute the tangential motions of the stellar halo particles 
with respect to the Galactic center. We consider the tangential motions 
of stars in the stellar halo separately in both 
the latitudinal ($Lat$) and longitudinal components ($Lon$) in Galactocentric coordinates.
\newline

\noindent {\bf Latitudinal Motions:}

\begin{figure*}
 \centering
 \includegraphics[scale=1]{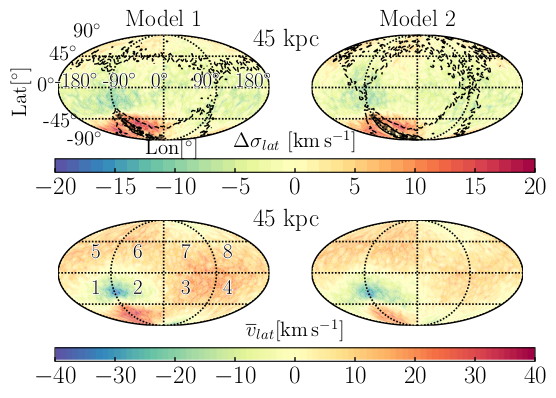}
 \caption{Mollweide projection in Galactocentric coordinates illustrating
 the ratio of the local latitudinal tangential velocity dispersion 
  with respect to the all-sky average ($\Delta \sigma_{lat}$; top panels) 
  and the local mean latitudinal tangential velocity ($\overline{v_{lat}}$; bottom panels). 
 $\sigma_{lat}$ increases by 20 km/s within the Transient response in Octants 1 and 2.
 For $\overline{v_{lat}}$, the color scale also indicates the 
 direction of motion in latitude (positive is toward the north and
  negative is toward the south).
north of the the LMC, between $-45<Lat<0$, $\sigma_{lat}$ 
decreases by $\sim 10$ km/s and increases 
up to $20$ km/s between $-90<Lat<-45$; the latter traces the Transient response. 
This bipolar behavior in $\overline{v_{Lat}}$ illustrates that stars both north 
and south of the LMC are being gravitationally pulled toward the LMC.}
  \label{fig:theta_45}
\end{figure*}

\begin{figure*}
 \centering
 \includegraphics[scale=1]{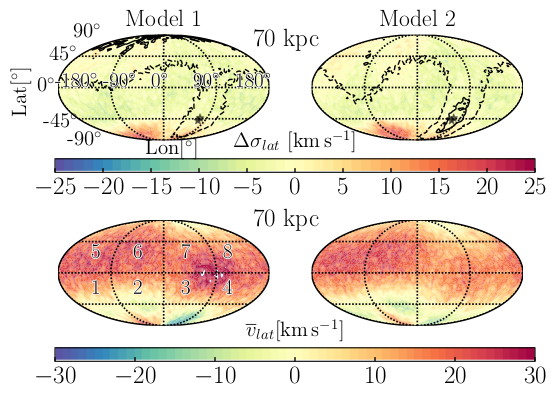}
 \caption{Same as Figure \ref{fig:theta_45} but at 70 kpc. An increase in $\sigma_{lat}$ leads 
 to the Transient response. The behavior of $\overline{v_{lat}}$ is more complex, but there is still evidence for 
 negative motions
 along the Transient response (following the LMC's orbit). }
  \label{fig:theta_70}
\end{figure*}

Figures \ref{fig:theta_45} and \ref{fig:theta_70} illustrate the
change of the local latitudinal tangential velocity dispersion with respect to to the all-sky average:

\begin{equation}
\Delta \sigma_{lat} = \sigma_{lat} - \overline{\sigma_{lat}}
\end{equation} 

\noindent and the mean local latitudinal tangential velocity ($\overline v_{lat}$) of 
stars in the stellar halo 
at 45 and 70 kpc, respectively. 

At 45 kpc, Octants 1 and 2 show the strongest changes in $\sigma_{lat}$, increasing by 
20 km/s in the Transient response. $\overline{v_{lat}}$ exhibits a 
bipolar behavior north and south of the LMC, indicating that stars are being gravitationally 
focused toward the LMC's COM.

At larger distances, Figures \ref{fig:theta_70} illustrate that an increase
in $\sigma_{lat}$ consistently leads to the Transient response. The behavior of $\overline{v_{lat}}$ shows more complex patterns at 
distances beyond 45 kpc. However, the kinematics still illustrate motion along
the Transient response
(negative values, indicating motion toward the south in Octants 3 and 4). We find this behavior at the outer 
regions of the halo too, but here we just show the results at 70 kpc for illustrative purposes.
\newline

%\begin{figure*}
% \centering
% \includegraphics[scale=1]{../code/tracers/theta_100_galactic.pdf}
% \caption{Same as Figure \ref{fig:theta_45} but at 100 kpc. An increase in $\Delta \sigma_{lat}$ still leads
% the Local Wake and $\overline{v_{lat}}$ illustrates motions towards the South Galactic Pole.}
%  \label{fig:theta_100}
%\end{figure*}

\noindent {\bf Longitudinal Motions:}

\begin{figure*}
 \centering
 \includegraphics[scale=1]{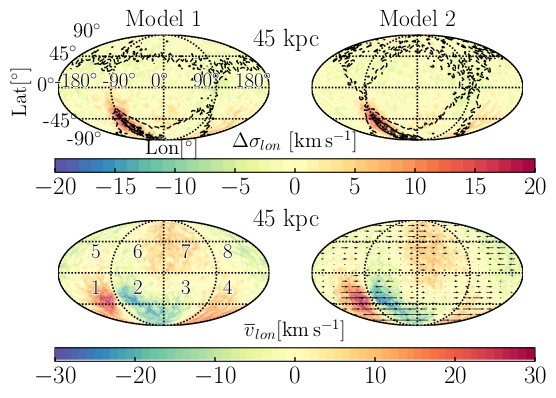}
 \caption{Same as Figure~\ref{fig:theta_45} except for the tangential motion in Galactocentric longitude. 
  The color bar indicates the magnitude of local changes in the dispersion, $\sigma_{lon}$, relative to the all-sky average and 
  both the magnitude and the direction of the local mean velocity, $\overline{v_{lon}}$, where negative indicates toward more negative longitudes and 
  positive toward more positive longitudes (see arrows on the bottom right panel for guidance). 
  $\sigma_{lon}$ increases at the location of the LMC by $\sim20$ km/s. 
  Surrounding the LMC, the direction of $\overline{v_{lon}}$ indicates a
  corresponding converging flow of particles moving toward the LMC.
  Further along the Transient response, the flow of stars is diverging (Octants 3 and 4). This flow results from the motions of stars that were once converging toward the LMC as it passed through 
  that location of the sky.}
  \label{fig:phi_45}
\end{figure*}

\begin{figure*}
 \centering
 \includegraphics[scale=1]{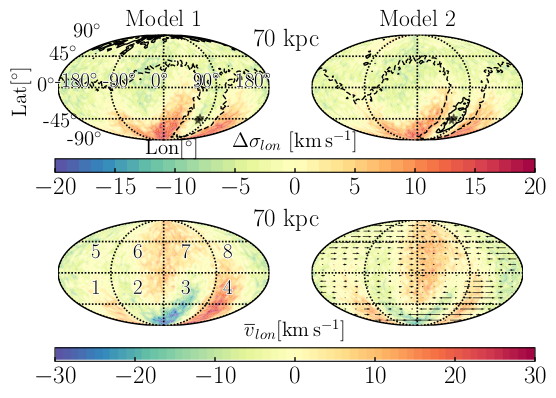}
 \caption{Same as Figure \ref{fig:phi_45} but at 70 kpc. $\sigma_{lon}$ increases by $\sim$20 km/s 
 surrounding the stellar wake. This has a corresponding signature in $\overline{v_{lon}}$ as 
 diverging motions, moving away from the Transient response. This is in contrast to the results in the 
 latitudinal tangential component, where motions appeared to follow the
  direction of the LMC orbit Transient response
 (Figure~\ref{fig:theta_45}). 
 }
  \label{fig:phi_70}
\end{figure*}

Figures~\ref{fig:phi_45} and ~\ref{fig:phi_70} illustrate the ratio of the local stellar longitudinal tangential
velocity dispersion with respect to the shell average:
\begin{equation}
\Delta \sigma_{lon} = \sigma_{lon} - \overline{\sigma_{lon}}
\end{equation}

\noindent and the local mean longitudinal tangential velocity ($\overline v_{lon}$) at 45 and 70 kpc, respectively. 

At 45 kpc, the velocity dispersion, $\sigma_{lon}$, increases by $\sim20$ km/s in the vicinity of the LMC.
Surrounding the LMC, $\overline{v_{lon}}$ indicates motions converging toward the LMC. 
Together with Figure~\ref{fig:theta_45} and Figure~\ref{fig:phi_45}, these results indicate that the motions of 
particles in the vicinity of the LMC are being gravitationally attracted to it.

At 70 kpc, Figures \ref{fig:phi_70} illustrates kinematics directly related to the 
motion of particles in and surrounding the Transient response.

On the other hand, in Octants 3 and 4, there is a divergent flow of stars in extended regions around $l=90^{\circ}$
in $\sigma_{lon}$.
This flow is the result of stars that were converging toward the LMC when it passed through 
that location of the sky $\sim 0.4$ Gyr ago, corresponding to the formation of
the Transient response at that time.
At the present day, those converging stars continued 
in their motion through the Transient response, and now appear to be diverging.
The $\overline{v_{lon}}$ maps reveal that particles in those regions are moving with opposite 
directions in longitudes, again exhibiting diverging motions.
The Transient response is located between these two regions, where the velocity dispersion 
is lower. The kinematic imprint of the Transient response beyond 45 kpc is thus stronger in the 
longitudinal component than in either the latitudinal or the radial-velocity components.
We find this effect to increase at distances larger than 70 kpc.

\subsubsection{Assessment of the Kinematic State of the Halo}

Our results indicate that the MW's stellar halo should hold kinematic signatures
of the Transient and Collective responses induced by the passage of the LMC. These 
effects manifest themselves as global, correlated kinematic patterns across the sky that 
persist over large ranges of Galactocentric distances, making them distinct from thin substructures
that are characteristic of disrupting satellites or globular clusters.  

The Transient response is most apparent kinematically in the tangential motions of the stellar halo, resulting in anticorrelated
responses in Galactocentric longitudes and latitudes. Specifically, the velocity 
dispersion in $\sigma_{lat}$ increases in the Transient response itself whereas $\sigma_{lon}$ increases
in the regions surrounding the Transient response. 

In contrast, the Collective response is best tracked by radial velocities, which likely 
reflects the reflex motion of the MW's disk about the orbital barycenter of the LMC-MW system, 
resulting in global redshifts in the north, vs. blueshifts in the south.

Overall, the impact on the MW's stellar halo kinematics across the sky in the tangential component 
is very similar for both Models 1 and 2, although the amplitude of the
perturbation is typically stronger in Model 2. 
Given that these two MW models have very different kinematics, we conclude that our conclusions are robust 
to any initial anisotropy in the kinematic structure of the halo. 

Furthermore, the measured increase in both the radial and tangential velocity dispersions by as much 
as 20-30 km/s, suggests that the stellar halo is being `heated' and is not in equilibrium locally.

We compute the change in the anisotropy parameter as the difference between the local anisotropy 
parameter, computed over the nearest 1000 neighbors within a grid cell ($\Delta \beta$), relative to the all-sky average ($\overline{\beta}$): 

\begin{equation}
\label{eq:deltabeta}
    \Delta \beta = \beta - \overline{\beta}.
\end{equation}
 
\begin{figure}
\includegraphics[scale=0.5]{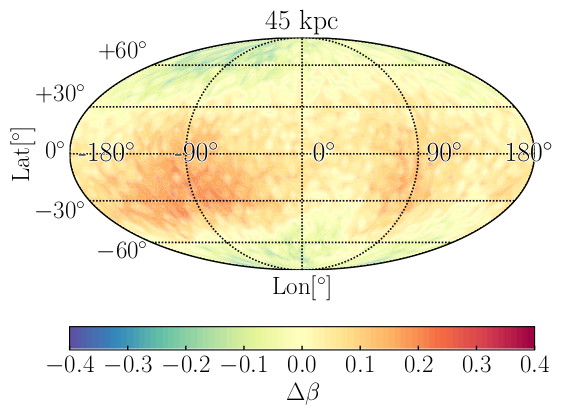}
\includegraphics[scale=0.5]{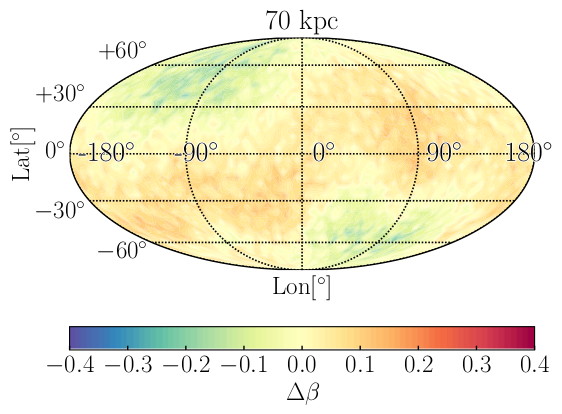}
\caption{Mollweide plots of the difference in the local value of the anisotropy parameter
relative to the average ($\Delta \beta$, Equation \ref{eq:deltabeta})
computed within a spherical shell,
  5 kpc in thickness at 45 (top) and 70 (bottom) kpc using Sim. 7 (Model 2).  
  $\Delta\beta$ ranges from -0.25 to
  0.25. Increases in $\beta$ are found preferentially near the plane of the MW while
  decreases are found toward the poles of the MW. Results at different distances and for Model 1
  can be found here \url{http://jngaravitoc.github.io/Garavito-Camargo/research/lmc_wake/}. These results suggest that the LMC
  has a nonnegligible effect on $\beta$. At 70 kpc, the effects of both the 
  Transient and Collective response can be seen as decreases in $\beta$.
  }\label{fig:beta_mollweide}
\end{figure}

The resulting all-sky map of $\Delta \beta$ for the simulated stellar halo in Sim. 7 (Model 2) 
within a spherical shell (5 kpc in thickness) at
a Galactocentric distance of 45 kpc is 
shown in Figure \ref{fig:beta_mollweide}. Versions of this map are also created for 
Model 1 and for shells at different distances in both models, see \url{https://bit.ly/2S25Yz}. 
We find that $\beta$ varies over large regions of the sky and distances. 
The largest positive values of $\beta$ (as high as 0.3) are induced at 45 kpc, 
as shown in Figure \ref{fig:beta_mollweide}. 

The Collective response in the north, which exhibited a 
`cold spot' region in the radial-velocity dispersion (Figure~\ref{fig:radial_45}), 
corresponds to a decrease in $\beta$ of -0.25. This structure persists out to the virial radius.
Decreases in $\beta$ are also seen corresponding to the Transient response at distances greater
than 70 kpc. This corresponds to changes in the
tangential velocity dispersion, which are discussed in $\S$ \ref{sec:tangential}.

Recently, \cite{Cunningham18b} studied the impact of substructure on $\beta$ using two of the \textit{Latte} 
FIRE-2 \citep{Wetzel16} cosmological simulations. The stellar particles from 
substructures in the MW halo have different kinematics than the stellar halo and thus produce local changes in $\beta$ from -1 to 1. 
Note that these substructures are smaller than the LMC and are not perturbing 
the stellar halo itself. However, distinguishing between small substructures and the
perturbations on the stellar halo due to the LMC could potentially be done using $\beta$. Thus, compared
to our results, we note that the effect of the LMC, without including the LMC stellar particles, on $\beta$ is $\sim$ 20\% (30\%
for the isotropic MW models) of 
that from substructure. In addition, the LMC should perturb  
some of the substructure in the stellar halo e.g., \cite{Erkal18b}. Future work examining this 
scenario in a cosmological setting, i.e. including 
both the LMC and substructure, will be done to properly capture the global
perturbations in $\beta$. In the mean time, the results presented in Figure 
\ref{fig:beta_mollweide} show that the effect of the LMC on $\beta$ is not
negligible when compared to perturbations expected from local substructure.

%% file: observability.tex
\section{Observability of the Wake}\label{sec:obs} 
In the previous sections, we characterized the density and the kinematic imprint                                                                       
of the interaction between the MW and the LMC in the stellar halo. 
In this section, we assess the                                                                          
observability of our results, including observational errors in both                                                                              
distances, and velocities. We also select regions of the sky within
 current                                                                           
or upcoming survey footprints to outline example observing strategies.
                                                                                  
We start our analysis by exploring the observability of the density enhancements
in the Transient and Collective responses induced by the LMC (\S \ref{sec:dens_observability}).
In \S \ref{sec:obs_kinematics}, we focus on the kinematic signature                                                                             
of these structures in the angular and radial components of the velocity dispersion. 

Finally, we estimate the number of                                                                                 
particles/stars needed in order to measure the predicted perturbations
(\S \ref{sec:ndensity}).                                                                             

The aim of this section is to provide the reader with a sense of how to sample the global patterns 
induced by the passage of the LMC on the smooth component of the stellar halo, 
rather than to make concrete predictions for specific surveys. 
A standing problem to observe these global patterns is the presence of
substructure in the stellar halo \citep[e.g.,][]{Bullock05}. Distinguishing substructure 
from the global patterns of the LMC wake might be possible
by using combinations of the predicted signatures in both density and kinematics.
We anticipate that the global patterns corresponding to the Transient and
Collective responses should be
present in the halo despite substructure and will persist over significantly larger 
distances than expected for tidal streams or individual satellites.

\begin{table*}
\centering
\begin{tabular}{c c c c c c}
\hline
\hline
  Region &  Quantity & Longitude $[^{\circ}]$ & Latitude $[^{\circ}]$ \\
\hline
  Regions 1-7 & $\sigma_r$ & [-129, -122, -118, -118, -100, -100, -100, -90, -90] & [67, 67, 66, 60, 55, 55, 50, 45, 45] \\
  Region 8 & $\sigma_r$ & 45 & 45 \\
  Region 9 & $\sigma_{lon, lat}$ & 130 & -55\\
  Region 10 & $\sigma_{lon, lat}$ & 80 & -45\\
  Regions 11-14 & $\rho_{\textit{transient}}$ overdensities & [-90, 90, 80, 90] & [-45, -45, -25, 15]\\
  Region 15 & $\rho_{\textit{transient}}$ underdensities & 145 & -50\\
  Region 16 & $\rho_{\textit{collective}}$ overdensities & -90 & 45\\
  Region 17 & $\rho_{\textit{collective}}$ underdensities & 145 & -50 \\
\hline
\hline
\end{tabular}
  \caption{Galactocentric coordinates of the centers of the target regions of the sky for the proposed observations. The regions are
           $20^{\circ}$ in width in both longitude and latitude. Regions 1-7 have coordinates centers that correspond
           to distances of [20, 30, 40, 50, 60, 70, 80, 90, 100] kpc respectively. For regions 8-10 the same
           coordinates centers are used at all the distances. Regions 11-14 have coordinate 
           centers that correspond to distances of [45, 70, 100, 200] kpc
           respectively, whereas Regions 15-17 have the same coordinates centers at every distance
           (see Figure \ref{fig:observations_1}). Transient and Collective refers to regions
           assigned to observe the Transient and Collective responses respectively. \label{table:densityRegions}}
\end{table*}

\subsection{Observing Density Enhancements Associated with the Transient and
Collective responses}

\label{sec:dens_observability}

\begin{figure*}
  \centering
  \includegraphics[scale=0.7]{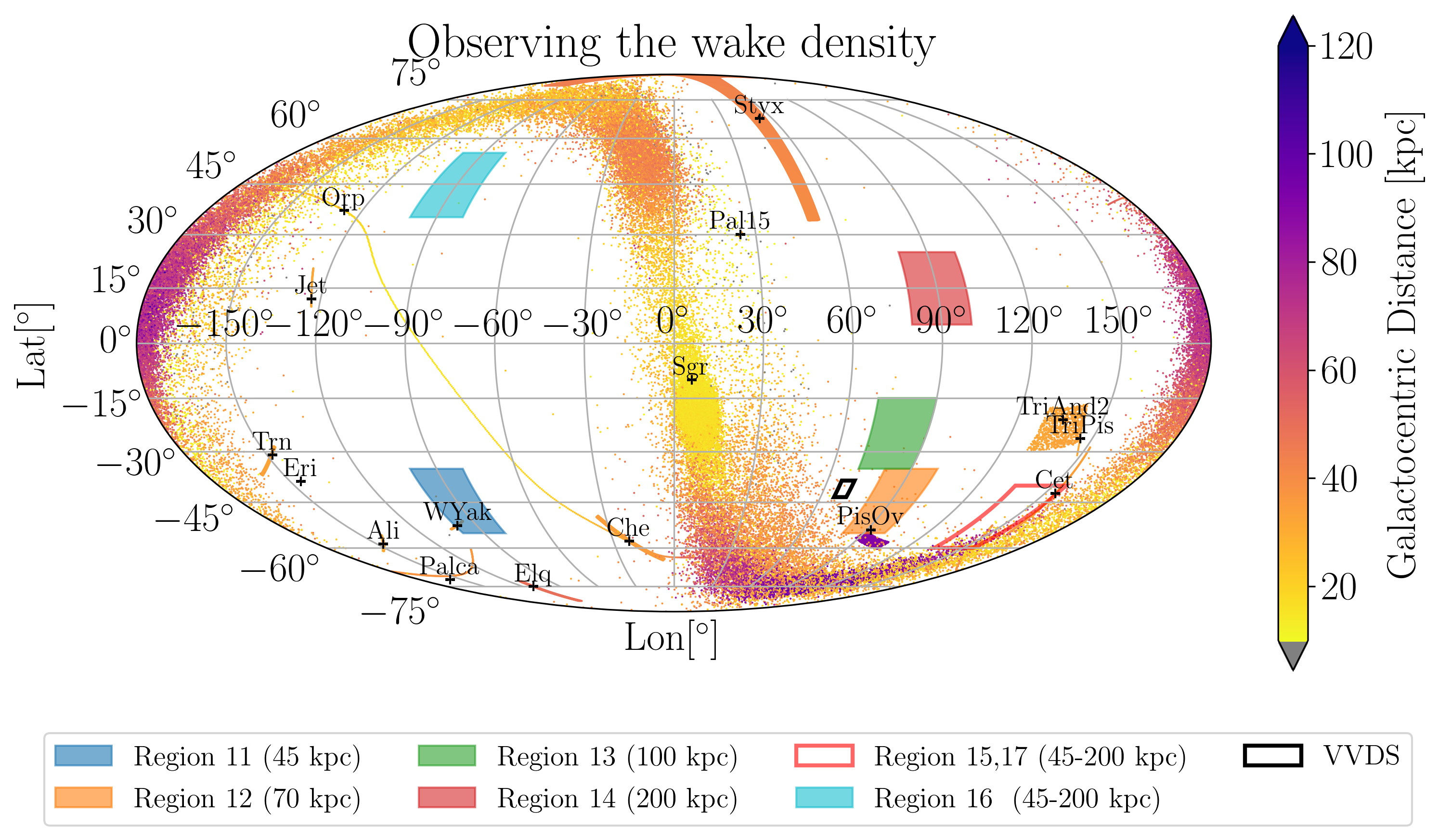}
  \caption{\textbf{Observing strategies for identifying the wake using stellar
  densities:} illustration of observing
  strategies to identify the predicted Transient and Collective responses induced by the LMC within 
  the stellar halo. 
  The figure shows a Mollweide projection in Galactocentric coordinates marking 
  the location of the
  currently known stellar streams that extend or are at distances greater than 50 kpc. The color bar indicates the
  Galactocentric distance of the streams. The most prominent substructure is the stellar
  stream from the Sag dSph. Also marked is the location of the VVDS survey
  \citep{Deason18} with a black empty square. 
  We select regions in which less substructure is
  present, marked by colored squares, in order to illustrate the observability
  of the wake.
  Filled boxes indicate overdense regions tracking the Transient response (blue,
  orange, green, and red) and Collective response (cyan) 
  at different Galactocentric radii. 
  The empty red box marks an example of an underdense region that will be used compute the density contrast.
  The image was made using the
  \textit{GALSTREAMS} library \citep{Mateu18}.}
  \label{fig:dm_regions}
\end{figure*}

Here, we study the observability of stellar density enhancements 
corresponding to the Transient and Collective responses induced by the LMC in the stellar halo.                                       
Our proposed strategy consists of measuring                                                                               
density ratios across the stellar halo, focusing on regions with few known substructures
and where the relative change in density is predicted to be largest. 
We illustrate this strategy in Figure \ref{fig:dm_regions},
which shows a map of the current known stellar streams                                                                  
beyond 30 kpc in a Mollweide projection in Galactocentric coordinates. 
The most prominent stream is that of the Sgr. dSph. 
We pick eight regions of 20 square degree at distances of 50, 70, 100, and 200 kpc that avoid 
the Sgr stream (see Table 
\ref{table:densityRegions} for exact coordinates of the proposed regions).
The cyan box indicates the overdensity on the Collective response. The 
blue, orange, green, and red boxes indicate regions centered 
on overdensities induced by the Transient response.                                                                             
The green box is centered on a region within the Collective response underdensity 
that will be used to compute the density contrast with the overdense regions. Note that there 
are multiple regions that could be chosen; here, we pick example regions that will 
be within the LSST footprint.

\begin{figure*}
    \centering
    \includegraphics[scale=0.5]{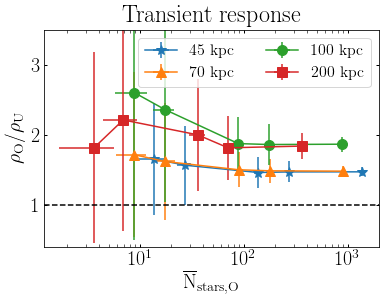}
    \includegraphics[scale=0.5]{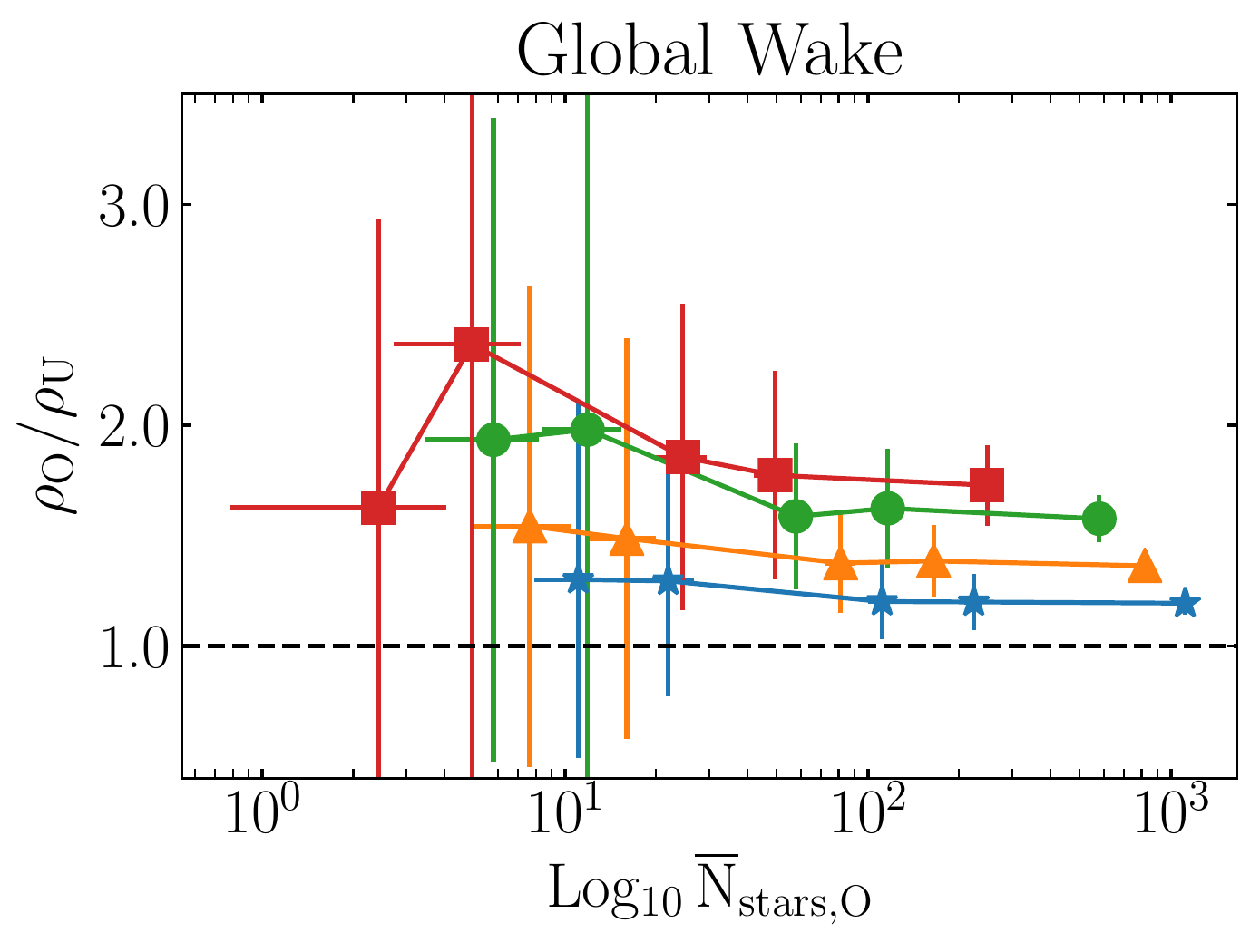}
    \caption{Observability of overdensities associated with the 
    Transient and Collective responses as a function of the number of sampled
    DM particles (see table ~\ref{table:densityRegions} for details). 
    To study how our results we change as a function of number of stars, we
    randomly sample the DM halo with  [$5, 10, 50, 100, 500$] $\times 10^4$ DM
    particles. With each of those samples, we proceed to measure the densities in
    the Regions 11 and 15 for the Transient response and 16 and 17 for the
    Collective response (see Table~\ref{table:densityRegions} for the
    coordinates of the proposed regions). Plotted are density ratios between 
    overdense (O) and underdense (U) regions
    (marked in Figure~\ref{fig:dm_regions}) at different Galactocentric distances, including 
    distance errors of 10\% and an assumed uncertainty in the Sun's Galactocentric distance of $\pm$0.09 kpc. Ratios 
    are plotted as a function of the number of particles sampled in each overdense region
    ($N_{\rm part,O}$). 
    The coordinates of the center of the O and U regions chosen for this experiment
    are listed in Table \ref{table:densityRegions}.
    Each region comprises a volume of $20$ square degree and 5 kpc in thickness. 
    The errors in the distances and the number of particles used do not
    have a strong effect on the density ratio. The strength of both the
    Transient and Collective responses is 
    stronger at larger distances, as discussed in \S ~\ref{sec:results}. We
    conclude that the Transient response induced by the LMC
    should be measurable even if there are only 20 stars in each 20 degree squared region.
    The corresponding minimum number density needed at each distance is listed in Table \ref{tab:ndensities}. 
    See Figure~\ref{fig:number_density} for an assessment of our choice of sampling 
    of stars in the stellar halo.}
    \label{fig:obs_wake}
\end{figure*}

The results of this observing strategy to hunt for the Transient and Collective  
responses are shown in Figure \ref{fig:obs_wake}. The average 
number density of stars located 
in the overdense regions ($\rho_O$) is divided by the 
average number density of stars 
in the underdense region ($\rho_U$). The 
resulting ratios are plotted as a function of the number of stars sampled 
inside the volume (a box of 20 square degrees and a thickness of 5 kpc). 
                                                                                    
In all cases, we have included assumed distance errors of 10\% which account for
the observational errors for typical current surveys. 
Furthermore, to account for observations being transformed to a Galactocentric
frame, we also include a distance error of 0.09 kpc for the Sun's Galactocentric position \citep{McMillan17}.
The error bars are computed using the bootstrapping                                                                         
technique and increase as the number of particles in each box decrease. 

We find that the predicted density contrast is unaffected by the number
of particles used or by the distance errors. These results suggest that 
measurements of 20-30 stars within each volume are sufficient to identify the
Transient and Collective responses. Table \ref{tab:ndensities} summarizes the corresponding number densities of stars 
within the selected regions (listed in Table \ref{table:densityRegions}) at different distances.
In \S \ref{sec:ndensity}, 
we discuss our sampling relative to realistic expectations for the number 
density of stars at these distances.         

\begin{table}
\centering
\begin{tabular}{c c }
\hline
\hline
$r $ (kpc) & $\nu$ [\# of stars kpc$^{-3}$]  \\   
\hline
45 kpc & $6.2\times 10^{-3}$ $\pm$ $1\times 10^{-4}$ \\ 
70 kpc & $3.6\times 10^{-3}$ $\pm$ $2\times 10^{-4}$ \\ 
100 kpc & $1.6\times 10^{-3}$ $\pm$ $1\times 10^{-4}$ \\ 
200 kpc & $4\times 10^{-4}$ $\pm$ $5\times 10^{-5}$ \\ 
\hline
\hline
\end{tabular}
\caption{Stellar number densities ($\nu$) corresponding to 20 stars in a 20 square degree 
region of 5 kpc thickness (i.e. boxes in Figure~\ref{fig:dm_regions} and \ref{fig:observations_1}) 
at the listed Galactocentric radius ($r$). This is the minimum density needed to observe 
the Transient and Collective responses as stellar overdensities.} 
\label{tab:ndensities}
\end{table}

Interestingly, \cite{Deason18} recently reported the discovery of an extended overdensity of 17
stars along the orbit of the LMC at distances of 50-100 kpc (disappearing at smaller radii). 
The region is marked as VVDS in Figure~\ref{fig:dm_regions}.
The authors attributed this material to stellar debris associated with Magellanic Stream. 
However, the spatial coincidence of these observations with expectations 
for the general location of the LMC's Transient response are suggestive (see $\S$ \ref{sec:MS_wake}).  
It is possible that other existing surveys may already have data to identify these 
proposed structures. We caution, however, that confirmation of the association 
of such overdensities with the LMC Transient and Collective responses must 
also involve matches with kinematic predictions, as outlined 
in \S \ref{sec:kinematics} and discussed in the next section.

\subsection{Observing the Kinematic Signature of the LMC's Wake}\label{sec:obs_kinematics}

\begin{figure*}
  \centering
  \includegraphics[scale=0.7]{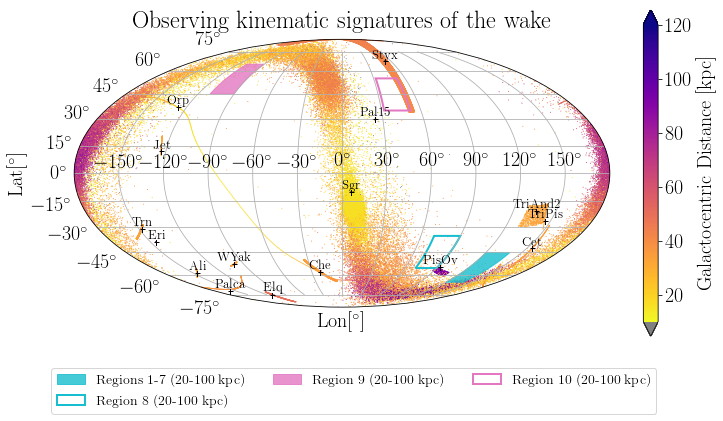}
  \caption{\textbf{Observing strategies for kinematic signatures:} 
  same as Figure~\ref{fig:dm_regions} but illustrating observing
  strategies designed to detect the kinematic signatures of the Transient and
  Collective responses 
  due to the passage of the LMC in the stellar halo.
  Each colored square shows the regions in which a given component of the
  velocity dispersion is measured: $\sigma_r$ (magenta box; Collective response); 
  $\sigma_{lat}$ and $\sigma_{lon}$ (cyan box; Transient response). 
   Since we focus on relative changes in the velocity dispersion,
  each component is measured in two distinct regions, selected to both maximize the 
  difference in the velocity dispersion and avoid known substructures.
  The empty and solid
  cyan boxes are adjacent to the density enhancement along the Transient
  response, which illustrates
  strong increases in $\sigma_{lon}$ (Figure~\ref{fig:phi_70})
  The solid magenta box is coincident with the `cold region' of lower radial-velocity dispersion in the Collective response whereas the empty box is a relatively unperturbed
  region of the halo (Figure~\ref{fig:radial_70}). 
  For illustrative purposes, the test regions 
  are selected within the DESI (for $\sigma_r$) and LSST (for $\sigma_{lat}$,
  $\sigma_{lon}$) footprints. }
  \label{fig:observations_1}
\end{figure*}

We discuss the observability of the kinematic signatures                                                                       
associated with the Transient and Collective responses induced in the stellar halo 
owing to the LMC's passage, as studied in \S
\ref{sec:kinematics}.                                                                   
We select example regions where the MW's stellar halo is predicted to have
the strongest kinematic response and that also                                                                            
avoid known substructures, as illustrated                                                                             
in Figure \ref{fig:observations_1}.                                                                                                                  
We select the marked regions to measure the relative change                                                                         
in the average radial-velocity dispersion (tracing the Collective response)
and tangential velocity dispersions (tracing the Transient response). 
These regions are also within the DESI, H3, LSST and                                                       
\textit{Gaia} footprints. We focus on changes in the velocity dispersion rather
than the mean velocities, 
as the velocity and distance errors have less of an impact on the measured dispersion.

We include 10\% and 20\% Gaussian errors in the distances. For the
radial velocities, we assume accuracies of 10 km/s and 20 km/s, which 
are similar to or greater than expectations for current surveys 
such as DESI and H3. Assumed tangential velocity accuracies of 50 km/s and 100 km/s 
are based on \textit{Gaia} and LSST proper motion accuracies, as 
discussed in Appendix \ref{app:errors}. We also account for errors
associated with the motion of the Local Standard of Rest, which we take as 
$\pm$5km$/$s in the $\hat{y}$ direction in Galactocentric coordinates, 
which is larger that that reported by \cite{McMillan17} ($\pm 3.0$ km/s).

\begin{figure*}
  \centering
  \includegraphics[scale=0.6]{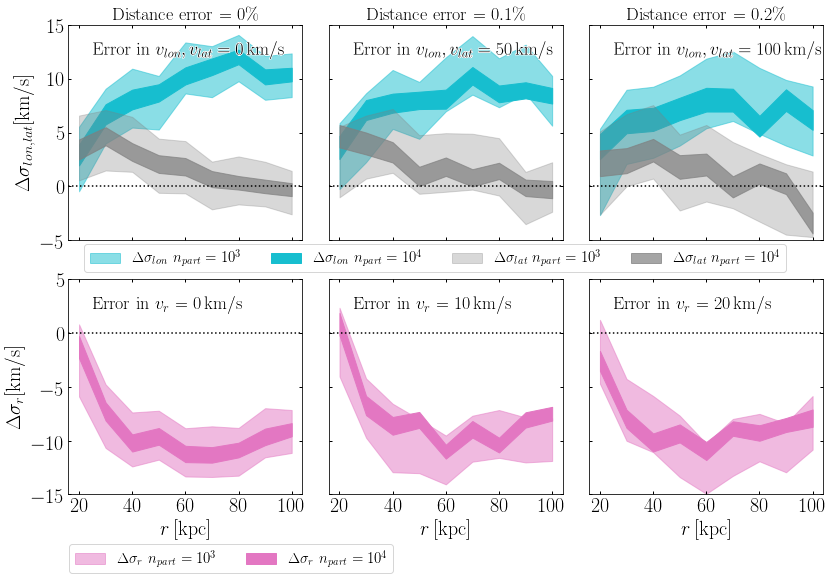}
  \caption{Relative difference of the velocity dispersion between different
  regions in the sky in a given velocity component, $\Delta \sigma$, 
  as a function of Galactocentric distance. Regions 1-8 are used for measuring 
  $\Delta \sigma_r = \sigma_{r | region 1-7} - \sigma_{r | region 8}$ and 9 to 10 for measuring
  $\Delta \sigma_{lon, lat} = \sigma_{lon, lat | region 9} -  \sigma_{lon, lat |
  region 10}$
  as defined in Table ~\ref{table:densityRegions}. We use these regions for distances from 20 to 100 kpc. These regions 
  are associated with the location of the Transient response (top panels: cyan, $\Delta \sigma_{lon}$; and
  gray, $\Delta \sigma_{lat}$) or the Collective response (bottom panels: magenta, $\Delta \sigma_r$) 
  relative to an unassociated region, as marked in Figure \ref{fig:observations_1} 
  (solid and empty boxes; 20 square degrees, 5 kpc in thickness). Each region is sampled 
  using $10^3$ and $10^4$ particles. The width of each line
  corresponds to the 1$\sigma$ deviation about the mean using $10^3$ and $10^4$
  particles,
  computed using the bootstrap technique.
  Errors in distances and velocities are included as marked above each column.
  Results for Model 1 and Model 2 are similar; here, we show only results for Model 1.
  $\Delta \sigma_{lon}$ illustrates that $\sigma_{lon}$ is boosted in the
  regions surrounding the Transient response. But, in the same region of the sky, $\Delta \sigma_{lat}$ is expected to 
  display distinctly different behavior, remaining similar to the average dispersion at all distances. 
  $\Delta \sigma_r$ is measured in the expected ``cold region" associated with
  the Collective response, which demonstrates velocity dispersions
  10 km/s lower than average. This ratio is expected to persist over a huge distance range of 50 kpc
  and should be largely unaffected by velocity and distance errors. We conclude that the
  kinematic impact of the LMC on the \textit{smooth} component of the stellar
  halo should be observable even when large uncertainties are included.}
  \label{fig:observability}
\end{figure*}

Figure \ref{fig:observability}, illustrates the ratio of the average 
velocity dispersion between the two selected regions (empty and solid boxes 
marked in Figure~\ref{fig:observations_1}) for each 
velocity component, as a function of Galactocentric distance. 
The line widths show the standard deviation from the mean 
ratio when the regions are sampled using a different number of particles, as marked in the legend.
In all cases, Models 1 and 2 show similar behavior, and so only the results for Model 1 are illustrated here.

The top panels of Figure \ref{fig:observability} show the results for the tangential dispersions, 
$\Delta \sigma_{lon}$ and $\Delta \sigma_{lat}$ in regions adjacent to the density enhancement 
corresponding to the Transient response. 
$\Delta \sigma_{lon}$ shows a median increase of $\sim 12$ km/s, when no errors are included (left panel). This increase persists over 30 kpc. 
In the same regions, $\Delta \sigma_{lat}$ does not change. This behavior is 
expected based on
the global maps presented in Figure~\ref{fig:theta_70} and \ref{fig:phi_70}, which 
illustrate that $\sigma_{lon}$ increases in the region surrounding the Transient
response, whereas $\sigma_{lat}$ 
changes along the Transient response itself.
The opposite behavior in these two components of the tangential velocity dispersion
is a characteristic signature of the Transient response. 
As we include larger distance and velocity errors, the strength of the mean ratio 
decreases, but the signal should be observable, provided the error in the tangential velocities 
is not larger than 100 km/s.

The bottom panels of Figure \ref{fig:observability}
illustrate the behavior of the ``cold region" associated with the Collective
response in the 
northern sky, which displays a lower-than-average radial-velocity dispersion over a significant distance range ($\sim$50 kpc).  
This ratio ($\Delta \sigma_r$) 
presents a clear predicted trend, where the ratio decreases with increasing 
Galactocentric distance from 20 to 50 kpc.

From Figure \ref{fig:observability}, it is clear that the number of stars sampled
is a crucial factor in reliably detecting the kinematic signal of the LMC's wake. 
When $10^4$ particles are used, the signal in any velocity component
is expected to be detectable even when large
distance or velocity errors are taken into account. On the other hand, when $10^3$ particles are 
chosen, the signal in 
$\Delta \sigma_{lon}$ will be difficult to detect if tangential velocity errors are 100 km/s (upper right plot).
In contrast, even when sampling $10^3$ particles and including large
velocity errors of $20$ km/s, the predicted decrease in the radial-velocity dispersion is expected to be observable. 
Therefore, it is essential to compare our sampled number densities within these regions
with those expected for the MW's stellar halo at these distances.

\subsection{Sampling Stars in the Stellar Halo}\label{sec:ndensity}

Observing the MW's DM halo wake in the stellar halo is not an easy
measurement. There is likely substructure and 
the stellar density is expected to decrease in the outer halo. 

Here, we estimate the expected number of stars in the stellar halo from recent
measurements of the stellar halo number density profile from RR Lyrae and K-giants.
Note that the K-giant density profile \citep{Xue15} drops faster
than the RR Lyrae profile \citep{Hernitschek18}, see Figure~\ref{fig:st_density}. 
This trend is also in agreement with BHB stars
\citep{Deason14, Deason18}.

\begin{figure}
  \centering
  \includegraphics[scale=0.6]{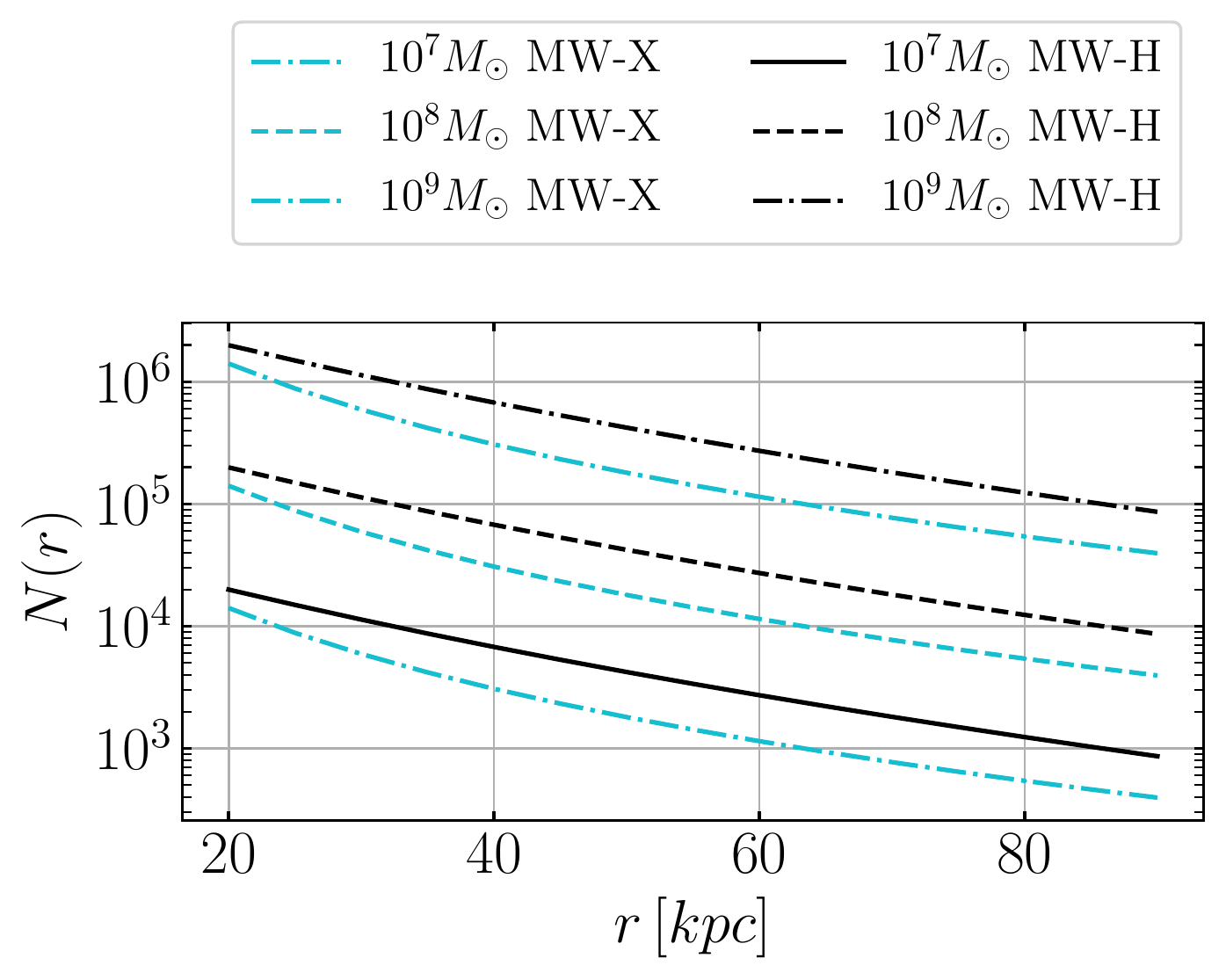}
  \caption{Number of stars, $N(r)$, within regions with the same volumes as the
  observational fields in Figure~\ref{fig:dm_regions} and \ref{fig:observations_1}
  (20 square degrees, and thickness of 5 kpc). 
  The cyan and black lines indicate the expected number of stars in such a volume
  using the observed 
  number density profiles for the stellar halo for K-giants \citep[MW-X][]{Xue15} 
  and RR Lyrae \citep[MW-H][]{Hernitschek18}, assuming different normalizations for the 
  total mass of the stellar halo, as marked. 
  In Figure \ref{fig:obs_wake}, we sample the observational fields using 100 and 1000
  particles, which is well below the expected total number of stars in 
  the stellar halo in the same volume at the same distances. In
  Figure \ref{fig:observability}, we
  increased the sampling to $10^4$ stars, which was found to be reasonable if the mass of the 
  MW's stellar halo was in excess of $10^8$  M$_{\odot}$.}
  \label{fig:number_density}
\end{figure}

We use these profiles to 
estimate the number of stars at a given distance assuming that the stellar
halo is homogeneous and is either entirely made up of RR Lyrae or K-giants. With such 
assumptions, the number of stars $N_{star}$ in a spherical shell of thickness $dr$
is

\begin{equation}\label{eq:n_stars}
  N_{\star}(r,r+dr) = \frac{\rm{M}_{\star halo} \int_r^{r+dr}\nu(r)}{ \rm{M}_{\star} \int_0^{R_{obs}} \nu(r)r^2dr}, 
\end{equation}

\noindent where $\nu(r)$ is the observed density profile, $R_{obs}$ is the radius to
which the stellar halo extends (here we assume 90 kpc), and  $\rm{M}_{\star halo}$ 
is the total mass of the stellar halo. Note that the normalization
factor in Equation \ref{eq:n_stars} is $M_{\star halo}/M_{\star}
\int_{0}^{R_{vir}}\nu(r)r^2 dr$. Where $\rm{M}_{\star}$ is the mass of a K-giant.
Figure \ref{fig:number_density} shows the number of stars inside a 20 square
degree field of 5 kpc thickness, 
as marked by boxes in Figure \ref{fig:dm_regions} and \ref{fig:observations_1}, as a function of
distance for the RR Lyrae (black lines) and the K-giants (cyan
lines) using three stellar halo masses $10^7$, $10^8$, and $10^9$ M$_{\odot}$. 
Figure \ref{fig:number_density} illustrates that, assuming that  
finding 100 or 1000 particles in a volume of $20$ square
degree and 5 kpc thickness in the stellar halo is consistent with current observations
of the number density profile, finding $10^4$ stars could be possible if the 
total mass of the stellar halo is larger than $10^8$ M$_{\odot}$.

%% file: discussion.tex
\section{Discussion}\label{sec:discussion}

Here, we discuss the details of our simulation suite
and place our results in a broader context. 
In \S \ref{sec:convergence}, we study the convergence of our simulations
and predicted observable signatures.
We then explore how our results change as a function of LMC mass 
in \S \ref{sec:LMC_mass_wake}. In \S \ref{sec:sag+lmc}, we compare 
the strength of the perturbations of the MW's DM halo from the
Sgr. dSph to those from 
the LMC. We consider how the Transient response can be distinguished from the stellar 
counterpart of the Magellanic Stream in \S \ref{sec:MS_wake}
and discuss how the properties of the wake can be used to constrain the 
mass of the MW in \S \ref{sec:MW_mass_wake}.
Finally, in \S \ref{sec:DM_nature}, we discuss how our results 
can be used to constrain the nature of the DM particle.

\subsection{Convergence of the Simulations}\label{sec:convergence}

In this section, we discuss the convergence of our simulations. Specifically,
we demonstrate that the chosen number of particles in our simulations is sufficient
to capture the DM halo response and to measure relative changes in the 
velocity dispersion. 

The DM wake is governed by resonances induced by the LMC, as
discussed in detail in \cite{Weinberg98b} and \cite{Choi09}. In particular, 
\cite{Weinberg07a} discuss that capturing these resonances in an $N$-body
simulation primarily depends on the number of particles used. For a satellite-host
interaction with mass ratios of 1:10, \cite{Choi09} showed that $10^6$ particles with equal mass for the
host satellite is sufficient to capture the resonances in MW-like DM halo.
However, to fully capture the innermost and low-order-resonances a $10^8$ halo is needed.
Therefore, our $10^8$ equal-mass particles should also capture the resonant
nature of the MW-LMC interaction; we test this statement in detail below using
the fiducial LMC model with $\rm{M}_{vir} = 1.8\times 10^{11} \rm{M_{\odot}}$  orbiting within an isotropic MW halo (Model 1).

\begin{figure*}
    \centering
    \includegraphics[scale=0.65]{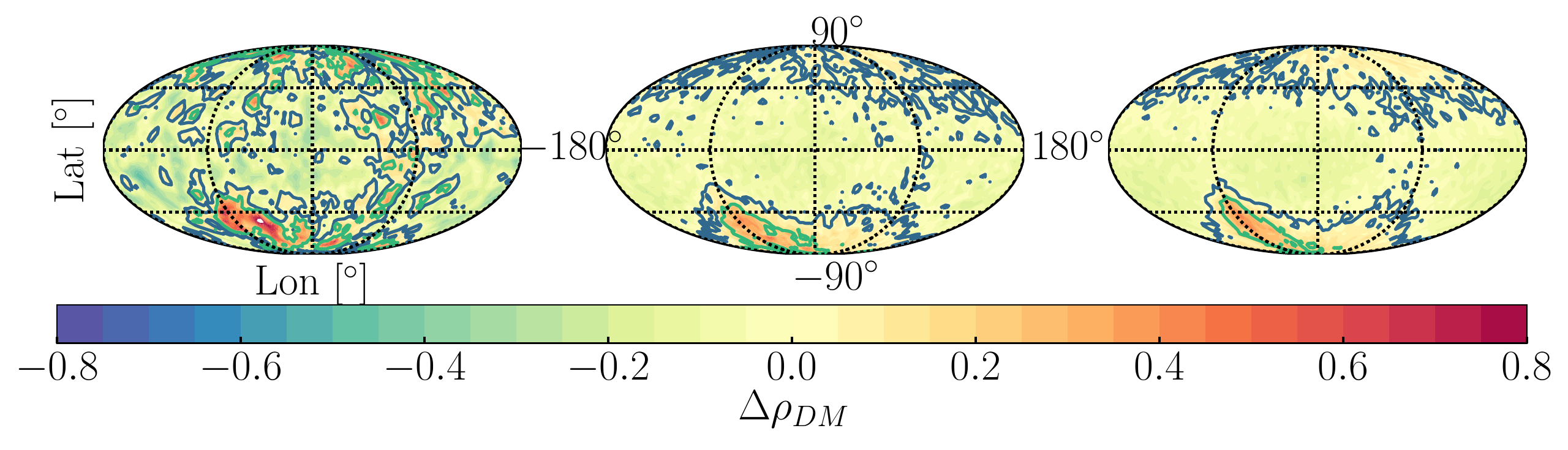}
    \caption{Mollweide all-sky map of the change in DM density in a spherical shell
    (5 kpc in thickness at 45 kpc) using the same set up as Sim. 4 (most massive LMC; Model 1
    halo). The density 
    contrast is measured relative to MW Model 1 in isolation ($\Delta \rho_{DM}$; Equation
    \ref{eq:deltarho}). Contours are defined as in Figure~\ref{fig:wake_yz}. 
    Plotted are results for three versions of Sim 3., using different
    resolutions: $10^{6}$ particles (left), $4\times 10^{7}$ (middle)
    and $10^8$ particles (right). The
    overdensities were computed in a cell of size $(3.6^{\circ})^2$ in all cases. The low
    resolution simulations cannot resolve the details of the DM halo response, while
    the higher resolution simulations show similar morphology, illustrating qualitative 
    convergence.}
    \label{fig:errors_density}
\end{figure*}

Figure \ref{fig:errors_density} illustrates the morphology of the DM wake generated in 
using the same set up as in Sim. 4 (most massive LMC; Model 1 halo) but for
three different resolutions ($10^6$, \textbf{$4 \times 10^7$}, $10^8$ particles) in a
Mollweide projection inside a spherical shell of 5 kpc width at 45 kpc. 
The density contrast is defined relative to the MW modeled in isolation with the same 
resolution (Equation \ref{eq:deltarho}; Figure~\ref{fig:wake_yz}). 
The figure shows that in the lowest-resolution case (left panel) the structure
of the DM wake is barely discernible. In the higher resolution simulations (the middle
and right panels) the structure of the DM wake is clear and is almost identical
in both cases, illustrating qualitative convergence.

\begin{figure}
    \centering
    \includegraphics[scale=0.5]{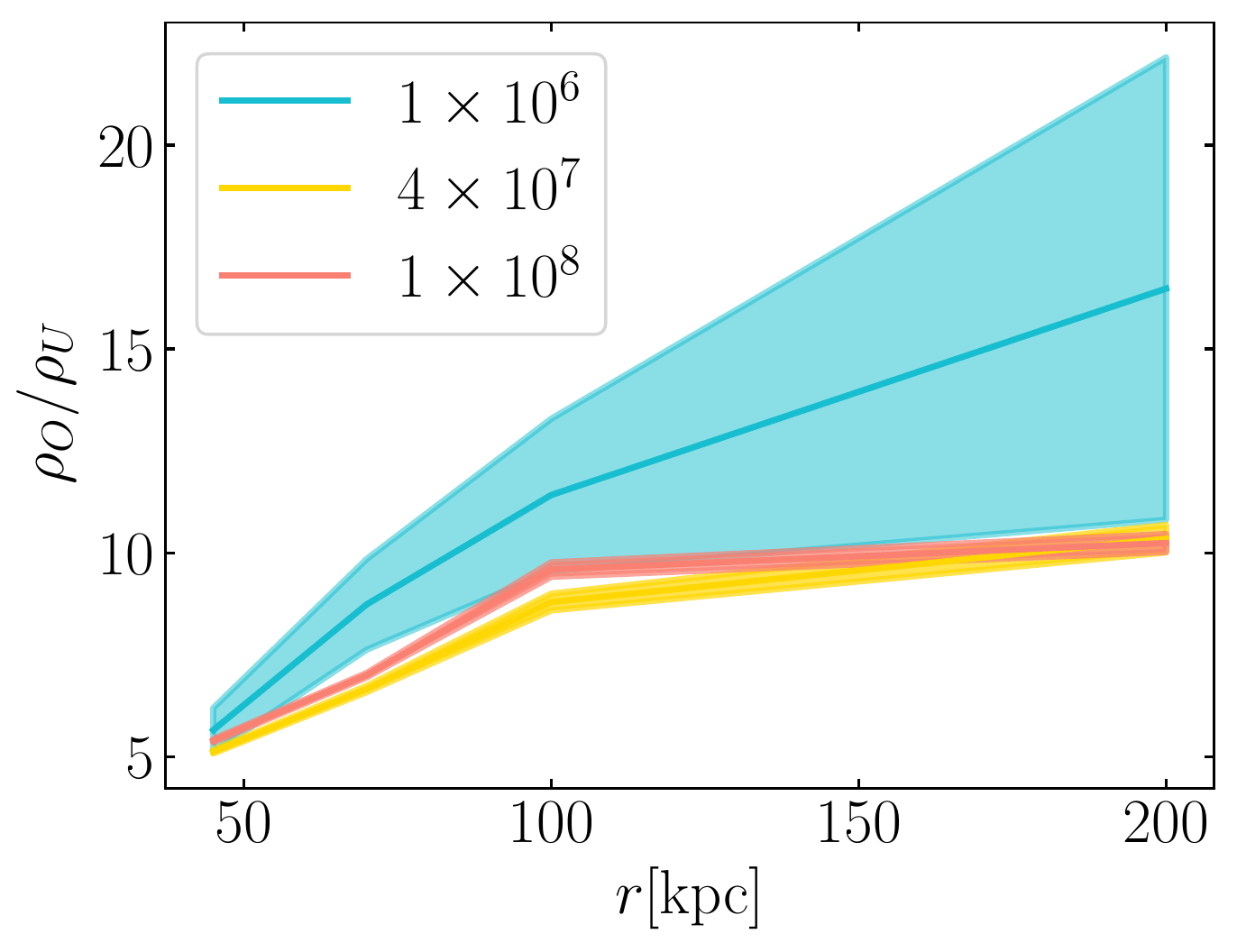}
    \caption{Density ratio between two regions with the same size of those defined in 
    Table \ref{table:densityRegions} for halos of an isolated MW.
    The different lines show the results for three simulations with
    different resolution (number of particles as marked in the legend). 
    The shaded areas represent the 1$\sigma$ error on the
    measurements computed using bootstrapping. The density ratio 
    converges for the two higher resolution simulations (red and yellow lines).}
    \label{fig:errors_wake}
\end{figure}

Figure \ref{fig:errors_wake} shows the stellar number density ratio between 
the underdense and overdense regions associated with the LMC's DM wake as a function of radius. 
The regions chosen to compute the density ratios have the same volumes as those in 
Figure \ref{fig:obs_wake}, whose properties are listed in Table \ref{table:densityRegions}. 
The shaded regions show the errors in the measurements computed using 
the bootstrapping technique. As the 
resolution increases, convergence is achieved within 10\%, indicating that results presented in 
Figure~\ref{fig:obs_wake} are reliable.

\begin{figure}
    \includegraphics[scale=0.5]{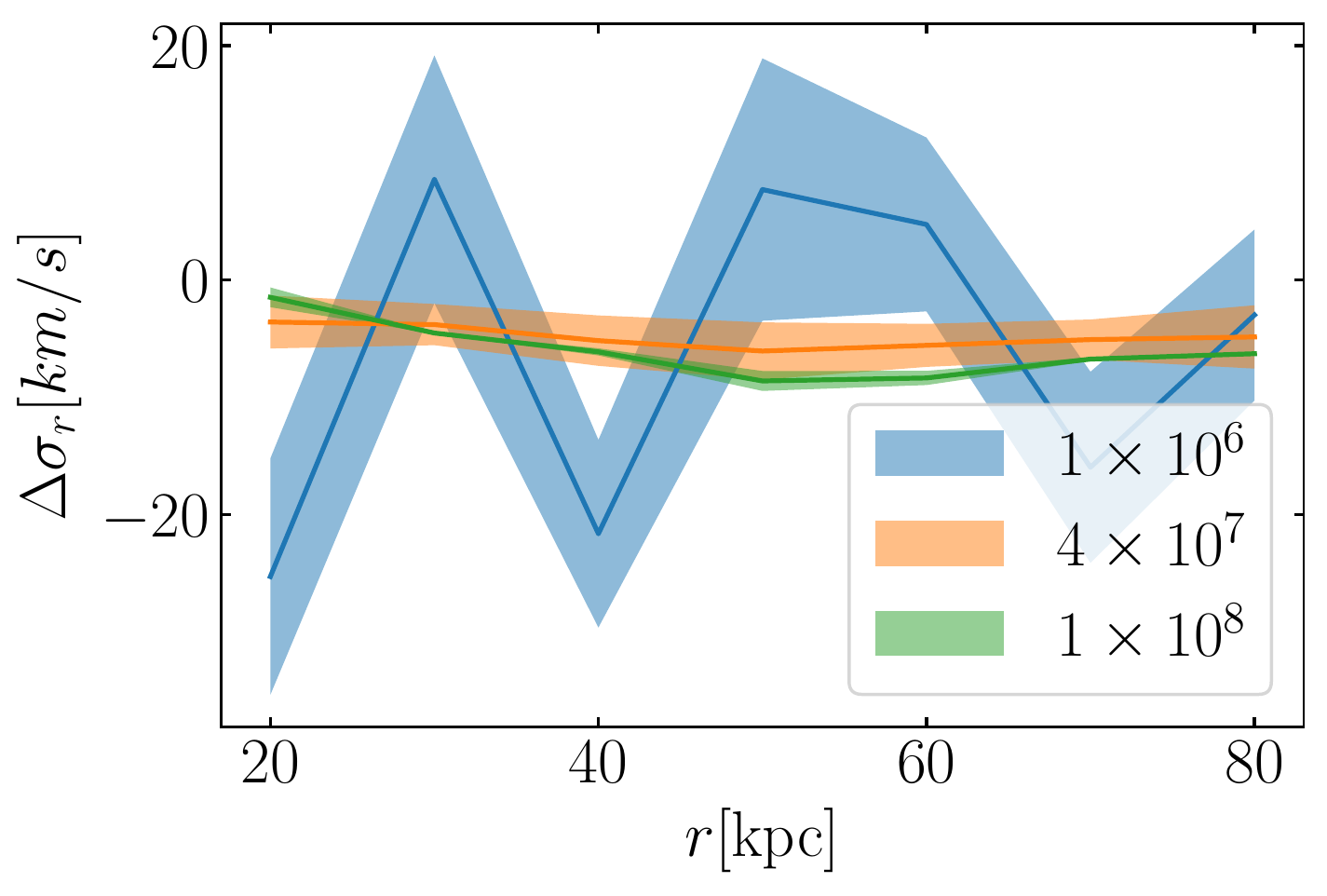}
    \caption{The relative radial velocity dispersion, $\Delta \sigma_r$, profile
    computed in the same regions as those defined in Table \ref{table:densityRegions} for three 
    simulations with different resolution (colored lines). The width of
    the lines show the $1\sigma$ errors in $\sigma_r$ computed using the bootstrapping
    technique. Convergence is achieved within 5 km/s for the two higher resolution
    simulations.}
    \label{fig:errors_sigma}
\end{figure}

In Figure \ref{fig:errors_sigma} we show that the predicted ratio in radial-velocity 
dispersion, $\sigma_r$ (as in the bottom panel of Figure~\ref{fig:observability}),
is similarly converged. 
If one computes the radial-velocity dispersion in 
smaller regions of the halo, the errors will be larger but the mean values are
unchanged.
These results for $\sigma_r$ are also consistent for the other components of the
velocity dispersion: $\sigma_{\theta}$ and $\sigma_{\phi}$.

We conclude that the results for the density and the kinematics of
our simulations with $10^8$ particles are converged and that the high number of
particles allows us to predict the morphological and kinematic properties of the 
DM wake and the halo response with small numerical uncertainties.

\subsection{The Impact of the Mass of the LMC}\label{sec:LMC_mass_wake}

In this section, we study how our results scale with the total halo mass of 
the LMC at infall. So far, our
analysis has focused on the fiducial LMC mass model (LMC3) of $\rm{M}_{vir} = 1.8 \times
10^{11}$M$_{\odot}$. However, as discussed in \S \ref{sec:LMC_mass}, the total
mass of the LMC is unknown within a factor of three.  We created 8
simulations with lighter and heavier LMC masses, see Table \ref{tab:sims}, to
study how the velocity dispersion and the strength of the DM wake are affected by 
the LMC's infall mass.

\begin{figure*}
    \centering
    \includegraphics[scale=0.5]{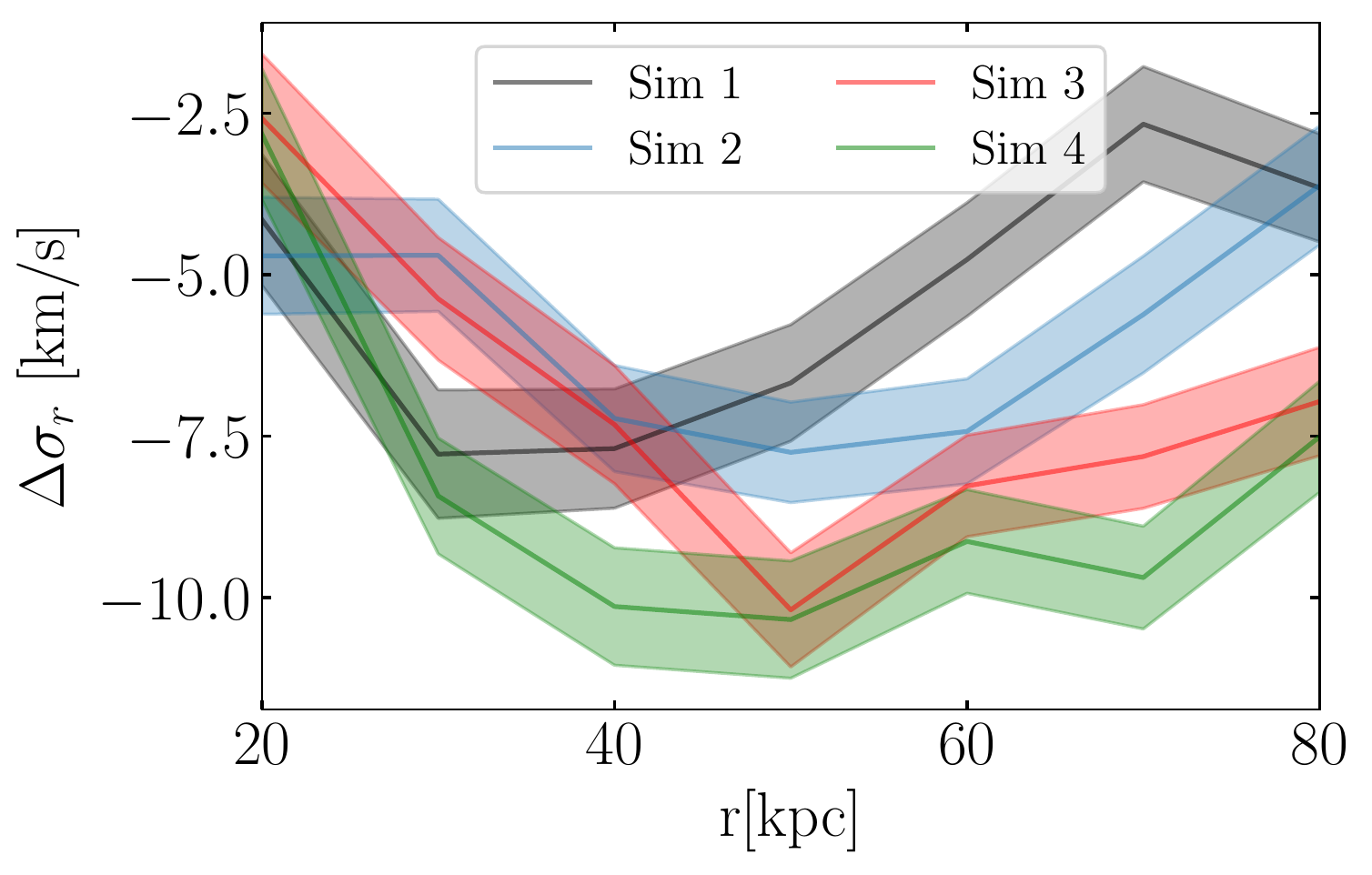}
    \includegraphics[scale=0.5]{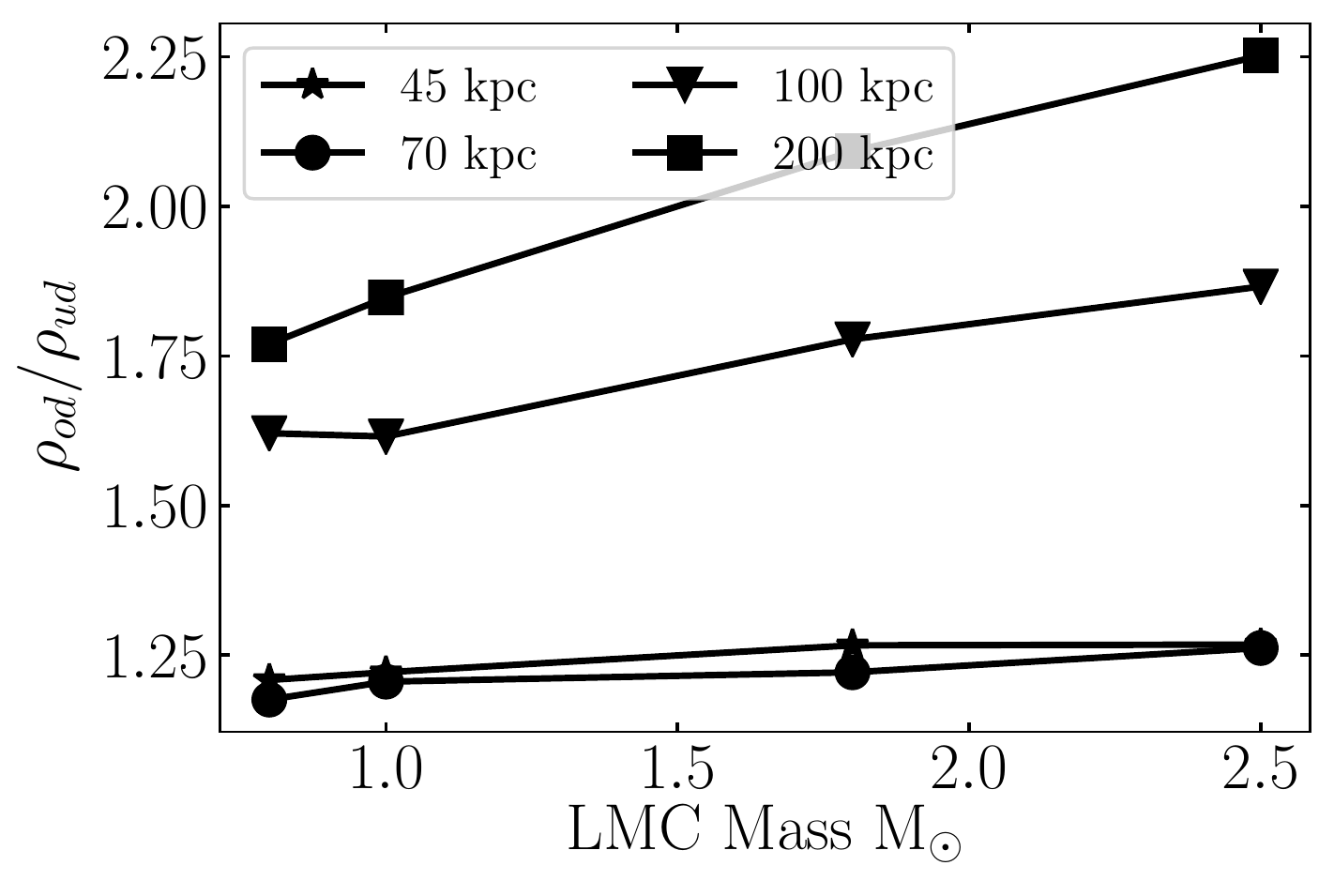}
    \caption{\textit{Left panel:} ratio of the radial-velocity dispersion, $\Delta \sigma_r$, 
    computed in the same magenta regions as marked in Figure \ref{fig:observations_1}. 
    This choice highlights the `cold region' in the north associated with the
    Collective response, which 
    exhibits lower radial dispersion (see Figure \ref{fig:radial_45}). As such, 
    $\Delta \sigma_r$ is negative. The colored lines show the results 
    for simulations with different LMC infall masses (Sim. 1-4; see Table~\ref{tab:sims}).  
    As the mass of the LMC increases, the response grows in 
    strength, becoming increasingly negative. However, the ratio is always decreasing 
    from 20 to 40 kpc, regardless of LMC infall mass. 
    \textit{Right panel:} density ratio between regions on the sky that are
    coincident with the DM wake (overdense; O) and adjacent to the wake 
    (underdense; U), as a function of LMC mass. The selected regions are the 
    same as those in Figure \ref{fig:dm_regions}. Different symbols mark results at 
    different Galactocentric radii. As the mass of the LMC increases, 
    the density contrast grows at all radii. The change is modest for the smallest 
    radii (25 and 45 kpc), suggesting that perturbations in the inner halo 
    will exist irrespective of the LMC's infall mass.}
    \label{fig:LMC_mass_results}
    
\end{figure*}

Figure \ref{fig:LMC_mass_results} shows how key observables associated with the 
LMC's DM wake scale as a function of
the LMC mass. The right panel shows the strength of the Transient response
 (overdense region) as a function of LMC mass and at different distances. 
 Each line shows the ratio in density between the same regions identified in 
 Table \ref{table:densityRegions}
and plotted in Figure~\ref{fig:obs_wake}. One region is always coincident 
with the Transient response (overdense region, $\rho_{O}$) and the other is in an expected 
underdense region adjacent to the wake ($\rho_U$). 

The strength of the Transient response increases as a function of LMC infall mass in the outer
halo, from 15\% at 70 kpc, up to 45\% at 200 kpc. 
Interestingly, at 25 kpc and 45 kpc the strength of the wake is similar for
all LMC mass models. 
These results suggest that
the Transient response created at 45 kpc and the weaker signals in the inner halo should be present regardless of
the assumed LMC mass. Furthermore, we confirm that 
the morphology of the Transient and Collective DM responses are the same for all LMC mass models, 
despite minor differences in their exact orbital trajectories.

The left panel of Figure \ref{fig:LMC_mass_results} illustrates the ratio
in the radial velocity 
dispersion, $\Delta \sigma_r$, between the same two regions on the sky used 
in \S \ref{sec:obs_kinematics} and Figure~\ref{fig:observability} to observe the
`cold region' in the north
associated with the Collective response.
The results presented here are for the
DM particles, but trends are the same using stellar particles.
Each line shows the value of $\Delta \sigma_r$ as a function of Galactocentric
distance for different LMC masses (Sim. 1 through 4). As the mass of the LMC increases, 
$\Delta \sigma_r$ becomes increasingly negative. This region of the sky (the
Collective response)
is impacted by both halo resonances and the COM motion of the disk relative to the outer halo, 
both of which increase in strength with increasing LMC mass.
Similar trends were found by \cite{Laporte18} for the strength of 
the warp in the MW's stellar disk owing to the LMC. 

These results suggest that (1) the strength of the decrease in radial velocity dispersion in 
the ``cold region"; and (2) the magnitude of the bipolar 
radial-velocity signal in the outer halo (redshifts in the north and blue shifts in the 
south; Figure~\ref{fig:radial_70} and \ref{fig:radial_100}), can together constrain
the total mass of the LMC at infall.

In the inner regions of the halo, 20 and 30 kpc, $\Delta \sigma_r$ does not change as 
much with LMC mass as in the outer halo. This is expected since the LMC's
pericenter distance is at $\sim$ 45 kpc; this, its impact is not as strong in the inner
regions of the halo. In addition, the COM motion of the inner halo 
is following that of the disk.
We found similar results for $\sigma_{\theta}$ and $\sigma_{\phi}$.

Note that we have characterized the LMC's wake
ignoring the Small Magellanic Cloud (SMC). However, the SMC is roughly one-tenth of the
stellar mass of the LMC. We expect that the inclusion of the SMC might make the  
structure of the wake
more complicated, since the SMC is modifying the DM halo density
profile
of the LMC as it orbits within it.
Nevertheless, the SMC's orbit largely traces the COM motion of the
LMC \citep{Kallivayalil13}. 
Its impact is most likely captured by increasing the mass of the LMC, which we have 
characterized here. 
We thus do not anticipate that our conclusions about the morphology and kinematics 
of the LMC's wake will change with the inclusion of the SMC.

\subsection{Density Perturbations from both Sgr. \& the LMC}\label{sec:sag+lmc}

We claim that the LMC is currently the strongest perturber
of the MW's DM halo at r$>$45 kpc. The LMC is currently the MW's most massive satellite and
recently passed its first pericenter approach $\sim 50$ Myr ago. However, the
LMC is not the only satellite that has perturbed the MW's DM halo. 
Sgr. has been orbiting the MW for at least the
past 6 Gyr, having made at least 
three pericenter approaches at $\sim$ 20 kpc and apocenter distances of $\sim 100$
kpc \citep{Dierickx17, Fardal18, Laporte18}. 

Fortuitously, the orbital plane of
Sgr is perpendicular to that of the LMC. This means that the
DM wake induced by Sgr is likely in
different regions of the sky than that of the LMC. However,
understanding the complex interplay between these two effects requires
N-body simulations. 

Here, we compare the perturbations to the MW's DM halo from both the LMC and Sgr.
We use the simulations presented in \cite{Laporte18}, a suite of two $N$-body
simulations of the interaction between the MW, LMC, and Sgr and four additional
simulations of the MW-Sgr
interaction alone (no LMC). These simulations were used in \cite{Laporte18} to
quantify the impact of these satellites on the MW's disk. 

The MW and LMC models used to generate these
simulations are the same as those presented in this
work. Four different models for the mass of Sgr. are used \citep[see Table 1 in
][]{Laporte18}. For this study, we use the most massive Sgr. model, with a mass of
M$_{200}=$1$\times$10$^{11}$M$_{\odot}$ and concentration $c_{200}=26$ at
infall, since this massive and concentrated model should generate the strongest DM
wake. In the following, we use the MW-LMC-Sgr and MW-Sgr simulations to
compare the amplitudes and morphology of the DM wakes induced by Sgr. and the LMC at the
present day.   

\begin{figure*}
    \includegraphics{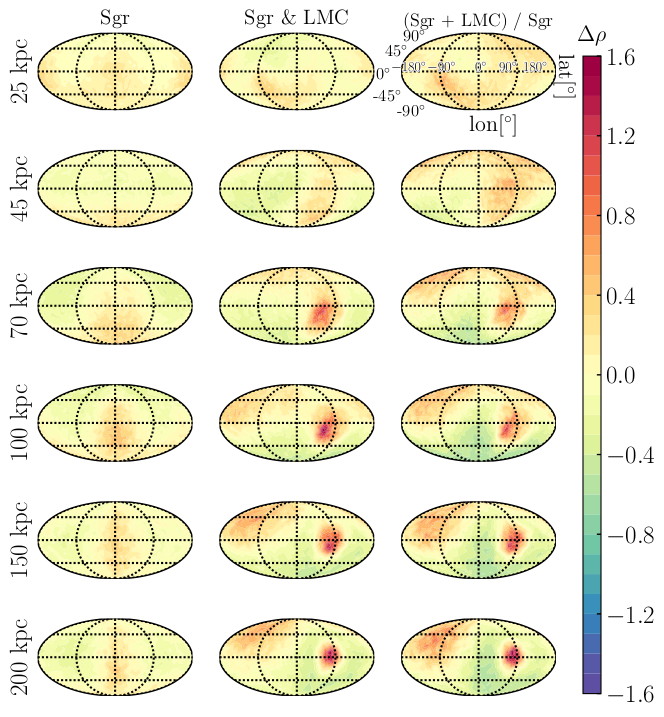}
    \caption{The present-day DM halo response of an isotropic MW DM halo to the orbits 
    of the Sagittarius
    dwarf spheroidal (Sgr.) and the LMC after 6 Gyr of evolution. 
    Sgr.'s infall mass is M$_{200}=1\times 10^{11}$M$_{\odot}$ 
    with a concentration of $c_{200}=26$.
    These results are based on the
    simulations of \cite{Laporte18}. The color bar indicates the ratio of the local DM density to the all-sky average, 
    $\Delta \rho_{DM}$. Only MW DM particles are included in all panels.  Different rows indicate different
    Galactocentric radii. 
    \textit{Left column:} the present day response of the MW's halo to the 
    orbit of Sgr alone. 
    \textit{Middle column:} same as left column, but also including a 
    massive LMC ($2.5 \times 10^{11}$M$_{\odot}$; LMC4), which enters the MW virial 
    radius $\sim$2 Gyr ago. 
    \textit{Right panel:} ratio of the present day wakes from Sgr.+LMC 
    to that of Sgr. alone (middle column/left column). The LMC clearly 
    dominates at all distances, wiping out the response to Sgr. In particular, 
    the orbit of Sgr. does not affect the
    morphology of the LMC's DM wake. We conclude that our results are robust to the presence
    of Sgr.}
    \label{fig:sag+LMC}
\end{figure*}

The left column of Figure \ref{fig:sag+LMC} shows the ratio of the 
local DM overdensities relative to the all-sky average, 
highlighting the DM wake produced by Sgr alone at different
Galactocentric distances. 
The middle column shows the same for the combined DM wakes from both Sgr and the LMC. 
The right column shows the ratio of the DM density perturbations from both
Sgr+LMC to the response to Sgr alone (middle column/left column). 
The present-day halo response to Sgr alone is
predominantly found at $lon=0^{\circ}$ and at $(lat=0^{\circ}, lon=\pm 180^{\circ})$,
as expected given its orbital plane and the location of the Sgr. Stream (Figure
\ref{fig:dm_regions}). However, Sgr Transient response was stronger in the past
and it has decayed over time.

The overdensities produced by the halo response to the motion of Sgr are up to
30\% relative to the mean DM density of the halo and can be observed from 25 to 200
kpc. However, in the presence of the LMC, the halo response to Sgr
is barely discernible. The similarity between the middle (LMC+Sgr)
and right columns (Sgr's contribution removed) illustrates that the LMC's contribution
dominates and that Sgr does not change the morphology of the LMC's DM wake at
$r>25$ kpc, However Sgr could dominate in the inner halo \citep{Laporte18}.

Interestingly, Sgr's DM wake does affect
the density ratios between the underdense and overdense regions created by
the LMC's DM wake. In the most extreme case, the ratio could decrease up to
$\sim$ 12\%, which is not sufficient to significantly modify the expected signal form
the LMC's DM wake. We conclude that our results
presented in \S \ref{sec:results} are still valid in the presence of Sgr.

In addition, as in the case with Sgr, the LMC can erase the
previous signatures in the outer halo of earlier merger events due to its recent and ongoing infall.

\subsection{Distinguishing the Magellanic Stellar Stream from the stellar
Transient response.}\label{sec:MS_wake}

We have discussed the observability of the stellar Transient response in previous sections. However, 
the existence of a stellar component of the Magellanic Stream (MS) has also been
predicted by all tidal models of the Magellanic System \citep[e.g.][]{Gardiner96, Diaz12, Besla13}. 
How might the stellar MS be distinguishable from the stellar Transient response? In this subsection, 
we discuss the differences in the location, density, kinematics, and chemistry between 
the stellar MS and the Transient response.

The Transient response is expected to be well-aligned with the past orbit of the 
LMC on the plane of the sky.  
In contrast, the proper motion of the LMC \citep{Kallivayalil13} indicates the past orbit of the 
LMC is not aligned with the gaseous MS on the plane of the sky
\citep{Besla07}. Note, however, that the gas and stellar MS may also not be spatially coincident. 
The gaseous MS is subject to hydrodynamical forces such as gas drag and ram 
pressure \citep{Mastropietro15}, owing to its motion through the Circumgalactic Medium, which 
may create offsets \citep[e.g.][]{Roediger06}.  Together, this suggests that
the, yet undiscovered, stellar component of the MS is expected to be neither coincident 
with the LMC's orbit \citep{Diaz12, Besla13, Guglielmo14, Pardy18} nor the
Transient response.

\begin{figure}
    \includegraphics[scale=0.6]{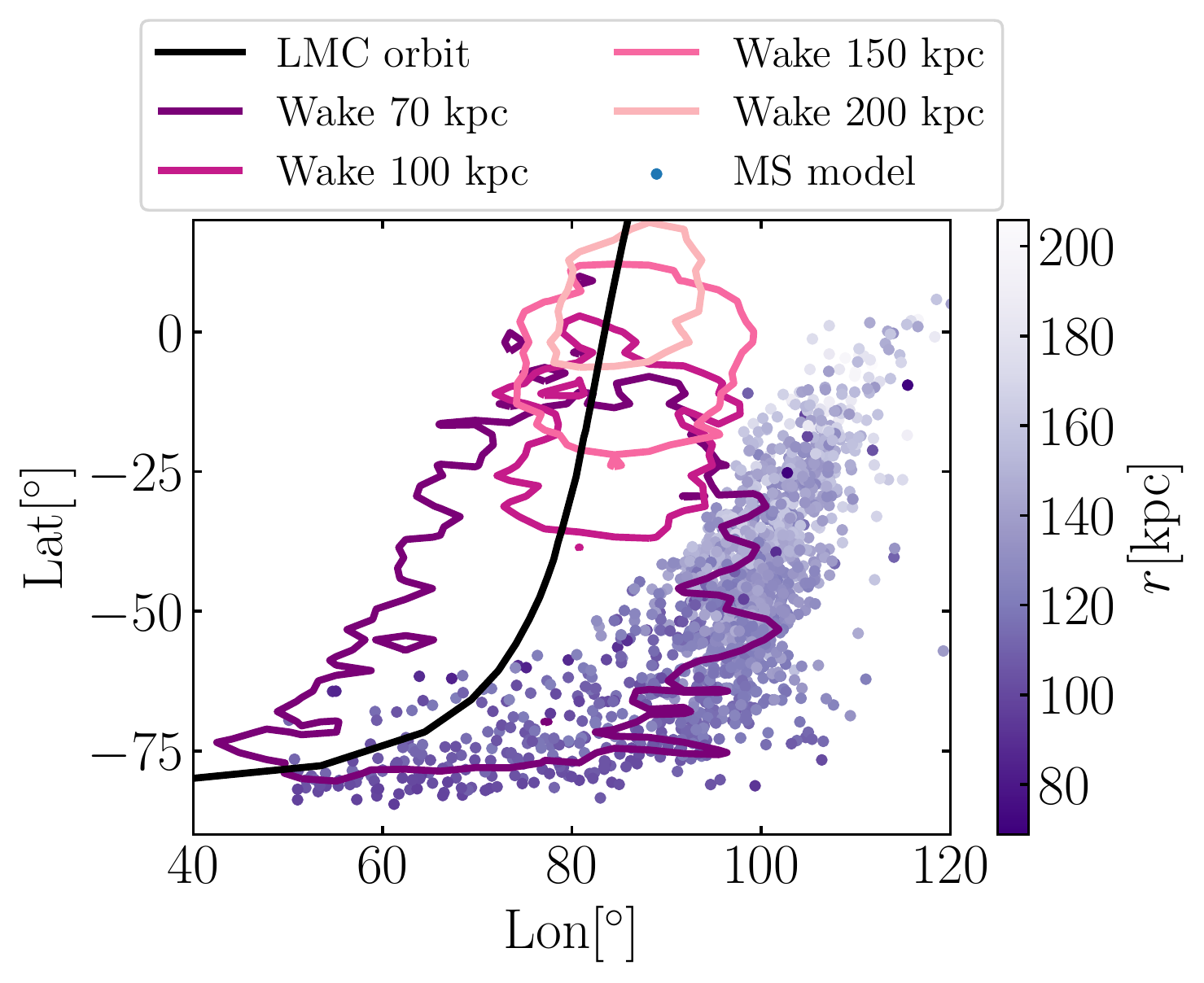}
    \caption{Projection of the location of the stellar Transient response and the
    stellar Magellanic Stream (MS),
    from \cite{Besla13} (their Model 1; purple points), in Galactocentric coordinates.
    The color scale indicates the Galactocentric distance to stellar particles 
    modeled in the MS.
    The contours shows the stellar density enhancement of the MW halo
    ($\Delta \rho$, as defined in Equation
    \ref{eq:deltarho}) for our Sim. 7 (Model 2 halo and most massive LMC model),
    at different Galactocentric distances. The contours are at 0.25, 0.6, 0.75 and 0.7
    for distances 70, 100, 150, and 200 kpc, respectively. 
    The stellar transient response is seen as an
    overdense region that tracks the LMC's past orbit (black solid line). The
    stellar MS is thinner than the stellar Transient response
    and spatially offset from both the LMC's orbit and the Transient response.}\label{fig:MS}
\end{figure}

The predicted locations of the stellar MS, the LMC orbit and the stellar
Transient response are
illustrated in Figure \ref{fig:MS} at Galactocentric distances greater than 70 kpc, where the 
expected deviation of the LMC's orbit from the location of the MS on the sky is more pronounced. 
The plotted stellar MS model is from the Model 1
simulation of \cite{Besla13}, which simulates the interaction history
between the SMC, LMC, and the MW, tracking the tidal stripping of stars and gas from the SMC. 

This figure illustrates that the
stellar Transient response is expected to be much more extended along and across
our line of sight than the stellar MS.  

At every radius, the width of the stellar Transient response is at least five
times the width of the stellar MS. 
These results show that, overall, the stellar MS is expected to have
little overlap spatially with the
stellar Transient response. 

The spatial offset of the MS from the LMC orbit is explainable by 
the MS originating from tidal stripping of the SMC, which is initially modeled as a rotating disk 
whose orbit does not exactly track that of the LMC.
Because the stellar MS is expected to originate from the SMC, the chemistry of any detected stars 
will likely be the most important discriminant between the stellar Transient
response and the stellar MS, the former
being comprised of old halo stars. 

We further find that the density of stars in the modeled MS is higher than the predicted density of the  
stellar Transient response at every Galactocentric distance by at least 1 to 2 orders of
magnitude. This may complicate searches for the stellar Transient response. 
We caution that different authors find strong variations in the predicted density of the
stellar MS \citep[e.g.,][]{Diaz12,Pardy18}. Also, the expected stellar density
of the Transient response is very sensitive to the assumed
stellar halo density profile. 

In addition to the differences in the spatial distribution and 
density of the stellar MS, we also expect differences in the kinematics. 
Stellar streams display kinematically coherent motion in the direction 
of the progenitor (see Sacchi et al.2019, in preparation).
Given the polar orbit of the stellar MS, its kinematic signature is expected
to be the strongest in $\overline{v_r}$, $v_{lat}$ and $\sigma_{lat}$. 
This is the opposite of the stellar Transient response, which is characteristically 
surrounded by converging/diverging flows in $v_{lon}$ and increases in $\sigma_{lon}$ with minimal 
impact on $\sigma_{lon}$ or $v_{lat}$ (\S \ref{sec:results}).

In summary, while the stellar MS is expected to be denser than the stellar
Transient response, 
the latter is expected to be spatially offset, thicker,  
and chemically distinct from the stellar MS. The stellar Transient response is 
also characterized by converging and diverging stellar
motions, whereas stellar streams display motions along the stream toward the progenitor. 
We conclude that it will be possible to distinguish the detection of the stellar
Transient response from the stellar MS.

\subsection{The Transient response as an Indirect Measure of the MW's Total Mass.}\label{sec:MW_mass_wake}

The past orbit of the LMC is strongly influenced by the total mass of the MW. As the mass of
the MW increases, the orbit of the LMC becomes more elliptical, allowing the LMC 
to complete one or more orbits about the MW within a Hubble time \citep{Kallivayalil13}.
Unlike stellar streams, which can deviate substantially from 
the past orbital path of a progenitor disk galaxy, the Transient response thus uniquely traces the 
orbit of the LMC. This provides us with an indirect probe of the underlying DM distribution of the MW. 

In particular, if the virial mass of the MW is of the order of $1.5 \times 10^{12}$ M$_{\odot}$, the LMC will 
have traversed through the northern sky \citep{Patel17a}, leaving a very different signature in 
the stellar halo than that illustrated here. 
If the mass approaches $2 \times 10^{12}$ M$_{\odot}$, the LMC may make multiple orbits, 
in which case, as illustrated in the case of Sgr (\S \ref{sec:sag+lmc}), 
the Transient response will likely be very difficult to identify.
 
As such, by detecting the amplitude, sky location, and distance of the LMC's
Transient response, we can 
constrain the 3D orbital path of the LMC, which in turn will constrain the mass profile 
of the MW's DM halo.

\subsection{Prospects of Studying the Nature of the DM Particle Using the LMC's
DM Wake}\label{sec:DM_nature}

In the era of high-precision astrometry, surveys like LSST and DESI will reveal the
structure and kinematics of the stellar halo out to 300 kpc 
\citep{Ivezic08}. These large-volume data sets have the capability to reveal the
shape and density structure of the DM
halo. But it is less clear how they may inform us about the nature of
the DM particle itself. 

In this work, we have outlined a strategy for observing the signatures of the DM
wake induced by the LMC in the kinematics and density profile of the stellar
halo over a large range of Galactocentric distances. Identifying these
signatures will both constrain the total infall mass of the LMC and provide
proof of dynamical friction in action, which in turn requires the existence of a
DM particle. Furthermore, here we discuss how the properties of the wake itself
may reveal the characteristics of the DM particle. 

We have shown that for the CDM scenario, the morphology of the LMC's wake is largely independent of
both the mass of the LMC and initial MW halo kinematics (Model 1 vs. Model 2). 
However, both of these factors do affect the amplitude of the halo response. 
As such, the amplitude of the wake may allow us to constrain the 
anisotropy profile of the MW's halo, given independent probes of the 
LMC's mass \citep[e.g.,][]{Erkal18b}. The anisotropy profile
is a significant uncertainty in the 
expected velocity distribution of DM particles in the solar neighborhood, 
particularly the high velocity-tail (Besla et al 2019, in preparation), 
which has direct consequences for 
direct detection experiments \citep{Green02a,Green02b}.

In the CDM framework, the location of the LMC's DM wake on the sky is controlled primarily by the orbit of
the LMC over the past 1 Gyr. Each MW+LMC model
explored in this study resulted in final LMC position and velocity vectors that
are not exactly the same, but all agree within 2$\sigma$ with the measurements
of
\cite{Kallivayalil13}, see Figure \ref{fig:LMC_phase_space}. Over the short
timescale considered in this study, uncertainties in MW mass cause only minor
changes to the orbit \citep{Kallivayalil13}. In particular, changes to MW mass
will not change the projected location of the LMC's orbit on the sky \citep{Besla07}.
We thus conclude that our predictions for the general shape and location the
LMC's wake are robust within
the CDM framework owing to the current low uncertainties in the LMC's 6D phase-space properties.
 
It is possible, however, that different DM models may affect the location and
morphology of the DM wake on the sky in a manner distinct from the uncertainties in the
LMC orbit. If the DM is self-interacting (SIDM) and velocity-independent, we
might not expect large differences from the wake in the CDM model. Upper 
limits in the cross-section of SIDM particles already 
suggest that the interaction between particles might not be as strong to create
difference in the wake. However, if the interactions of the DM particles 
are velocity-dependent the relative velocity difference between the LMC and the MW DM
halo could be sufficient to produce a significant deviations in the morphology and 
strength of the wake, relative to the CDM case.

On the other hand, if the DM is Fuzzy (FDM), i.e. composed of light bosons or
axion particles, the properties of the                      
DM wake will also be different than in the CDM model. For example,
\cite{Hui17} showed that the orbital decay times of globular clusters in the
Fornax dwarf galaxy are longer in the FDM                                                                           
model. In FDM, the DM halo is cored, rather than forming a cusped as in CDM, which
changed the density and velocity dispersion of the DM halo. \cite{Read06}
showed that dynamical friction behaves differently within DM cores vs. in cusps. In
DM cores, a satellite initially undergoes a rapid strong dynamical friction
followed by suppression of dynamical friction. In addition, \cite{Hui17} showed
that the dynamical friction is different in FDM, since the de Broglie wavelength of
the FDM particles has to be taken into account. 

Early works by \cite{Furlanetto02} compared the structure of the overdensity associated with DM wakes in CDM
and SIDM, which they approximated as a perfect fluid. They found that the structure of the wake is more complicated in SIDM.
For example, the wake structure can change if the satellite galaxy is moving subsonically
or supersonically. In the subsonic case, the wake is symmetric in front of 
and behind the satellite. In the supersonic case, the wake forms a Mach
cone trailing the satellite. Furthermore, in these simulations, DM was subject to ram-pressure, which
creates an additional DM wake, and a bow shock in the supersonic case. These
studies already show large differences in the structure of the wake in different
DM models. We now have stronger constraints on the nature of the DM particle, and
hence new $N$-body simulations revisiting the structure and morphology of DM wakes
in different DM models are missing. 

In particular, the interaction between the MW and the LMC in such alternative DM models, including 
a live stellar halo, are required in order to
make concrete predictions of the morphology of the DM wake 
and the observable counterpart in the stellar
halo. This will be the subject of future work.

\subsection{Effect of the MW's initial Dark Matter density profile}
In this study, we have modeled the MW's initial DM density distribution 
using a Hernquist profile. The kinematics of the DM halo will be
different if the density profile changes - as such, the resonant response of the
DM halo can also change. The minor differences exhibited between our
adopted Model 1 and Model 2 kinematic profiles suggest that such differences
will not significantly modify the morphology of the halo response, but may 
impact the amplitude of the perturbations. Preliminary work studying the halo
response to the LMC using an NFW profile suggests that the overall
morphology and kinematics of the wake are qualitatively similar to the
results shown in this study (M. Weinberg 2019, private communication). 
As such, we expect that uncertainties in the mass of the LMC are the dominant source of uncertainty in the predictions presented in this study.  

%% file: conclusions.tex
\section{Conclusions}\label{sec:conclusion}

Since its first entry within the MW's virial radius, $\sim 2$ Gyr ago, the LMC 
has been the most massive perturber of the MW's DM and stellar halo. 
Yet, the strength, nature, and location of these perturbations has remained unknown. 
In this work, we quantified the density and kinematic perturbations induced within the MW's
DM and stellar halo using high-resolution $N$-body simulations of
the recent pericentric approach of a massive LMC ($8-25$ $\times 10^{10}$ M$_\odot$).

The ultimate goal of this study is to characterize global patterns in the stellar halo 
kinematics and densities that are correlated with the orbital motion of the LMC.
When taken together, these signals can confirm the identification of the LMC's DM wake. 
Given the rarity of LMC analogs at pericentric approach about MW analogs in cosmological simulations, 
studying this process in detail necessitates the use of controlled $N$-body simulations. 
In this work, we presented a suite of 8 $N$-body simulations, with
four different LMC masses (8, 10, 18, 25 $\times 10^{10}$ M$_\odot$) and two MW models
with the same mass ($1.2 \times 10^{12}$ M$_\odot$), but with different kinematic structure. 
MW Model 1 has an
isotropic anisotropy parameter ($\beta=0$) while Model 2 has a radially biased
anisotropy parameter (Eq. \ref{eq:beta}). In all simulations, 
the final position and velocity of the 
simulated LMC is ensured to be within 2$\sigma$ of the current 
measured values \citep{Kallivayalil13}.
In order to describe the wake structure as observables, we include a MW stellar halo 
that was constructed by assigning weights to the MW DM particles'
masses and velocities in order to reproduce the observed MW density profiles.

We identify global changes in the density and kinematics of the stellar halo
that persist over large Galactocentric distances (40-200 kpc). 
Specifically, we identified three main components of the wake generated by the
LMC. \textbf{The Transient response}, trailing the LMC and following the orbital history of the LMC out to the virial
radius of the MW. \textbf{The Collective response}, in the Galactic north, leading the
LMC; and the \textbf{The Global Underdensity}, an underdense region
surrounding the Transient response mainly in the south and 
extending out to the virial radius of the MW. 

We find that the density enhancement associated with 
the Transient response is stronger in Model 2 at every distance, reflecting the 
resonant nature of the halo response. The Collective response
on the other hand, has a different morphology in both models and its associated
density enhancement is stronger in Model 1. We conclude from these results that 
the initial kinematic state of the MW affects the strength and the 
morphology of the wake, and that deviations from isotropic models can strengthen
the wake.

The kinematics and density distribution 
of the MW's stellar halo are found to be perturbed by the LMC and correlated with 
the location and properties of the DM wake. 
Below, we summarize our main results for the stellar counterpart to the 
Transient and Collective DM responses generated by the LMC:

\begin{enumerate}
    
    \item{\bf Overdensities Associated with the Transient response} are stronger at distances 
    larger than 45 kpc. The Transient response should be detectable as an overdense region with 
    respect to underdense regions if at least 20 stars are identifiable 
    within a 20 square degree area, 5 kpc in thickness (corresponding to stellar number densities
    of $\sim$0.01-0.03 kpc$^{-1}$) (Figure \ref{fig:obs_wake}). Such number densities are consistent 
    with measurements of the number density profile of RR Lyrae and K-Giants. 
    At distances greater than 70 kpc, the strength of the Transient response increases with increasing 
    LMC infall mass, but is largely independent of LMC mass at smaller radii.
    The Transient response is not found to be coincident with known halo substructures - in 
    particular, regions of the Transient response can be found that avoid the Sagittarius Stream.
    
  \item {\bf Radial velocities:} The mean radial stellar velocities 
  show a bipolar behavior.
    Overall, stars in the Galactic north are moving away from the disk (redshifts), while 
    stars in the south are
    moving toward the disk (blueshifts). This is most clearly predicted at distances 
    larger than 45 kpc (see Figure \ref{fig:radial_70}). This bipolar behavior is indicative of the
    barycenter movement of the disk due to the gravitational acceleration from
    the LMC \citep{Gomez15}. Correspondingly, the strength of this velocity shift will
    scale with the infall mass of the 
    LMC. This behavior is associated with the Collective response and should be
    observable in ongoing or upcoming radial velocity surveys.

  \item {\bf Radial velocity dispersion:} In the north, there is a decrease in the radial velocity dispersion of $\sim$10 km/s that
    we call the ``Cold region". This region is also located within 
    the Collective response and corresponds to a region of the sky where the LMC has not yet 
    passed through, but will in the future. 
    Stars in the ``Cold region" have smaller mean radial velocities than
    stars in other regions in the north. The ``Cold region" persists over a large distance range, 
    from 45 kpc to the virial radius, and large area on the sky.  This structure 
    should also be observable in radial velocity surveys (see Figure~\ref{fig:observability}).

    \item {\bf Tangential motions in the vicinity of the LMC:} There are flows of stars
    converging towards the LMC. This is apparent in both latitudinal and 
    longitudinal velocity components in the vicinity of the LMC. This motion is associated 
    with an increase in the velocity dispersion in both components. However, the observability 
    of such motions will be complicated owing to the extended stellar populations associated  with the Magellanic Clouds.
    
    \item {\bf Motions along the Transient response:} In the latitudinal tangential direction,
    stars within the Transient response follow the orbital direction of the LMC's COM 
    (Figures \ref{fig:theta_45} and \ref{fig:theta_70}).
    In contrast, in the longitudinal tangential direction, we find diverging flows, where 
    stars that were once converging toward the LMC (see previous point)
    have already crossed each other and now are
    going in opposite directions (Figures \ref{fig:phi_45} and \ref{fig:phi_70}). 
    Consequently, the latitudinal velocity dispersion is enhanced in the regions
    surrounding the Transient response, and decreased within it. 
    In contrast, the longitudinal velocity dispersion is unaffected. 
    This opposite behavior is expected to be 
    observable from 70 to 100 kpc, provided that the tangential 
    motions of at least $10^3$ RR Lyrae or K-Giant stars 
    are measured within a 20 square degree region, 5 kpc in thickness, and
    tangential velocity errors are less than 100 km/s. Such accuracies are expected to be plausible with LSST. Furthermore, within the 
    Transient response, the radial velocity dispersion is 
    expected to be decreased relative to the average. The Transient response can thus be identified
    by these correlated, opposite kinematic signals in velocity dispersion.

    \item {\bf The anisotropy parameter, $\beta$:} is affected in all regions of
    the sky, as a result of the perturbations in all components of the velocity dispersion.
    We found that changes in $\beta$ are largest in the
    isotropic MW model (Model 1) but are still present in the radially anisotropic Model 2.
    The strongest changes in $\beta$ are found to be
    from -0.4 up to 0.35 at 45 kpc. At larger distances, the Transient response is discernible in 
    the $\beta$ maps (see Figure \ref{fig:beta_mollweide}).

    \item {\bf The stellar counterpart of the Magellanic Stream} is expected to be
    distinguishable from the Transient response, through its differing kinematics,
    spatial offset from the LMC orbit (which the Transient response tracks) and its chemical
    composition. The stellar stream will be comprised of SMC stars, whereas the
    Transient stellar response will be comprised of older halo stars.

  \item {\bf The halo response to the orbit of a massive 
  Sagittarius dwarf spheroidal is insufficient to wipe out the perturbations induced by the LMC} 
  (see Figure \ref{fig:sag+LMC}). This suggests that our results are robust to perturbations 
  from cosmological substructure, but this must be tested in a cosmological setting and is the subject of future work.

\end{enumerate}

    Given the imminence of the 
    era of all-sky photometric and kinematic surveys of the stellar halo to 
    large distances, we are optimistic that the Transient and Collective
    response induced by the recent passage of the LMC through the stellar 
    and DM halos will be detected. Indeed there are few regions of the sky where the density and/or kinematics
    of the stellar halo are not expected to be impacted by the LMC.
    
    Ultimately, the detection of the Transient response will track the past orbit of the LMC and
    constrain the eccentricity of that orbit, which is an indirect measure of the total
    mass of the MW. Owing to the expected
    dependence of dynamical friction to the properties of the 
    DM particle, the kinematics and density signatures of the
    halo response will provide an indirect measurement and a new test bed 
    of the nature of the DM particle.

%% file: appendix.tex
\section{Simulation details}
\label{appendix:sim}

Here, we present details of our simulations. Table \ref{tab:ICs}
lists present-day positions and velocities of the LMC in our simulations
with respect to the observed values from \cite{Kallivayalil13} (Columns 4-9). 
We also show the initial condition coordinates in Column 10. 
All of the simulations are within 2$\sigma$ of the total position and
velocity vectors, as shown in the right-most columns of ($\Delta v$ and $\Delta
r$). 
However, three simulations exceed 2$\sigma$ in their present-day $\hat{z}$
coordinates, and all
of the simulations have difficulty exactly matching the $\hat{v}_y$ component.

\begin{rotatetable*}
\begin{deluxetable*}{c c c c c c c c c c}
\tablecaption{Summary of LMC-MW simulations.  Initial conditions list the position and velocity vectors
of the LMC at infall, $\sim$2 Gyr ago. $\Delta x, y, z, v_x, v_y, v_z$, 
denote the difference in each component of the present-day 3-D position and
velocity vectors with respect to the measured values found in
\citep{Kallivayalil13}.\label{tab:ICs}}
\tabletypesize{\scriptsize}
\tablehead{ \colhead{Sim.} & \colhead{MW model} &  \colhead{LMC model} & 
\colhead{$\Delta x$ [kpc]} & \colhead{$\Delta y$[kpc]} & 
\colhead{$\Delta z$ [kpc]} & \colhead{$\Delta v_x$ [km/s]} & 
\colhead{$\Delta v_y$ [km/s]} & \colhead{$\Delta v_z$ [km/s]} & 
\colhead{Initial Conditions:}
}

\startdata
1 &  Model 1 & LMC1 & 1.75 & 2.68 & 2.56 &  -5.04& -55.17& 7.65 & $\vec{r} = (27, 276, 57)$ , $\vec{v}=(7.22,-61,-78)$  \\
2 &  Model 1 & LMC2 & 1.94 & 1.69 & 3.89 & -20 & -47 & 15  & $\vec{r} = (28, 310, 90)$  $\vec{v}=(7, -35, -70) $ \\
3 &  Model 1 & LMC3 & 2.12 & 0.61 & 1.99 & -4.47 & -42.65 & 1.53 & $\vec{r}=$ (12, 238, 130)  $\vec{r}=$(12, 13, -77) \\
4 &  Model 1 & LMC4 & -0.16 & 1.19 & 4.92 & -5.99 & -38.49 & 19.61 &  $\vec{r}=$(11.58, 248, 130)  $\vec{v}=$(12, 11, -77) \\
5 &  Model 2 & LMC1 & -0.13 & 2.68 & -1.21  & -0.19 & -55.74 & 1.03 & $\vec{r}=$(24, 278, 58), $\vec{v}=$(4, -59, -77)  \\
6 &  Model 2 & LMC2 & 2.31& 2.54& -0.5 & -12.07& -50.89& 7.86 &  $\vec{r}=$(28, 306, 90)  $\vec{v}=$(7, -35, -70) \\
7 & Model 2 & LMC3 & 2.66 &-0.4& 0.07 & 1.66 & -36.59& 0.81 &  $\vec{r}=$(12, 238, 130) , $\vec{v}=$(12, 13, -77)  \\
8 &  Model 2 & LMC4 & 2.69& 0.46& 3.21 & -2.42 & -35.56& 19.34 &  $\vec{r}=$(11, 238, 130) , $\vec{v}=$(12, 11, -77) \\
\enddata
\end{deluxetable*}
\end{rotatetable*}

\begin{figure}
  \centering
  \includegraphics[scale=0.6]{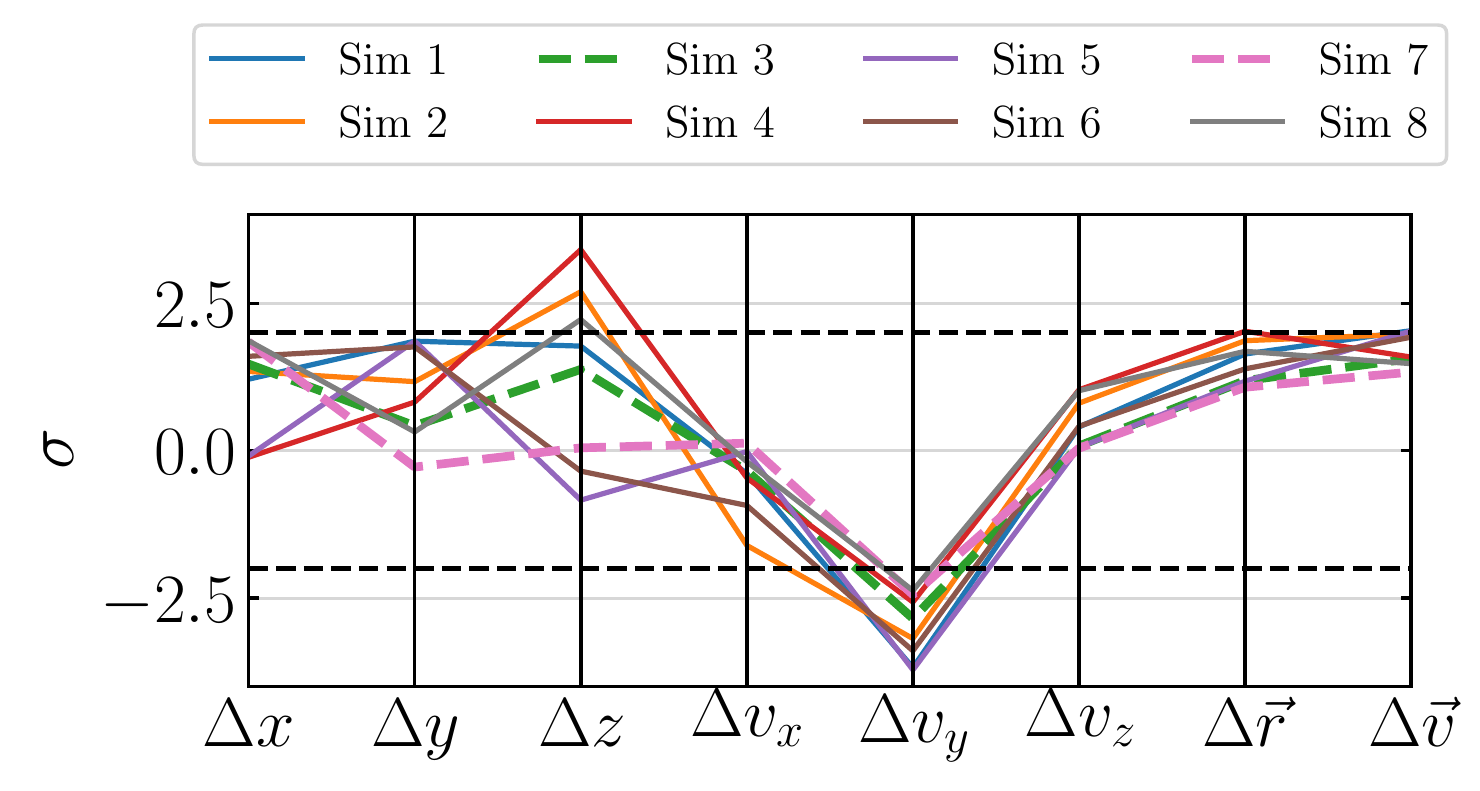}
  \caption{Final phase space coordinates of the LMC in all simulations. The
  y-axis shows the deviation of each simulated LMC position and velocity 
  component at the present day 
  from that observed by \cite{Kallivayalil13}. 
  The black dashed horizontal lines indicate a 2$\sigma$
  deviation from the observations.} 
  \label{fig:LMC_phase_space}
\end{figure}

Figure \ref{fig:LMC_phase_space} shows the final phase-space  coordinates of the
eight LMC+MW simulations described in Table~\ref{tab:sims}. 
The magnitude of the position and velocity vectors are all within 2$\sigma$ of 
the observations. Most of the individual velocity and position components 
are also within 2$\sigma$ of the observations (dashed lines). However, three simulations 
exceed 2$\sigma$ in the $\hat{z}$ component and all 
exceed 2$\sigma$ in the $\hat{v}_y$ component. The simulations cover a range of 
final positions and coordinates that generally span the allowable error space. 
The good agreement between the resulting structure and properties of the DM wake
in all simulations 
indicate that our conclusions are not significantly affected by the 
uncertainties in the LMC's orbital parameters. 

\section{Observational Surveys: Tangential Velocities Accuracies}\label{app:errors}

\begin{figure}
  \centering
    \includegraphics[scale=0.65]{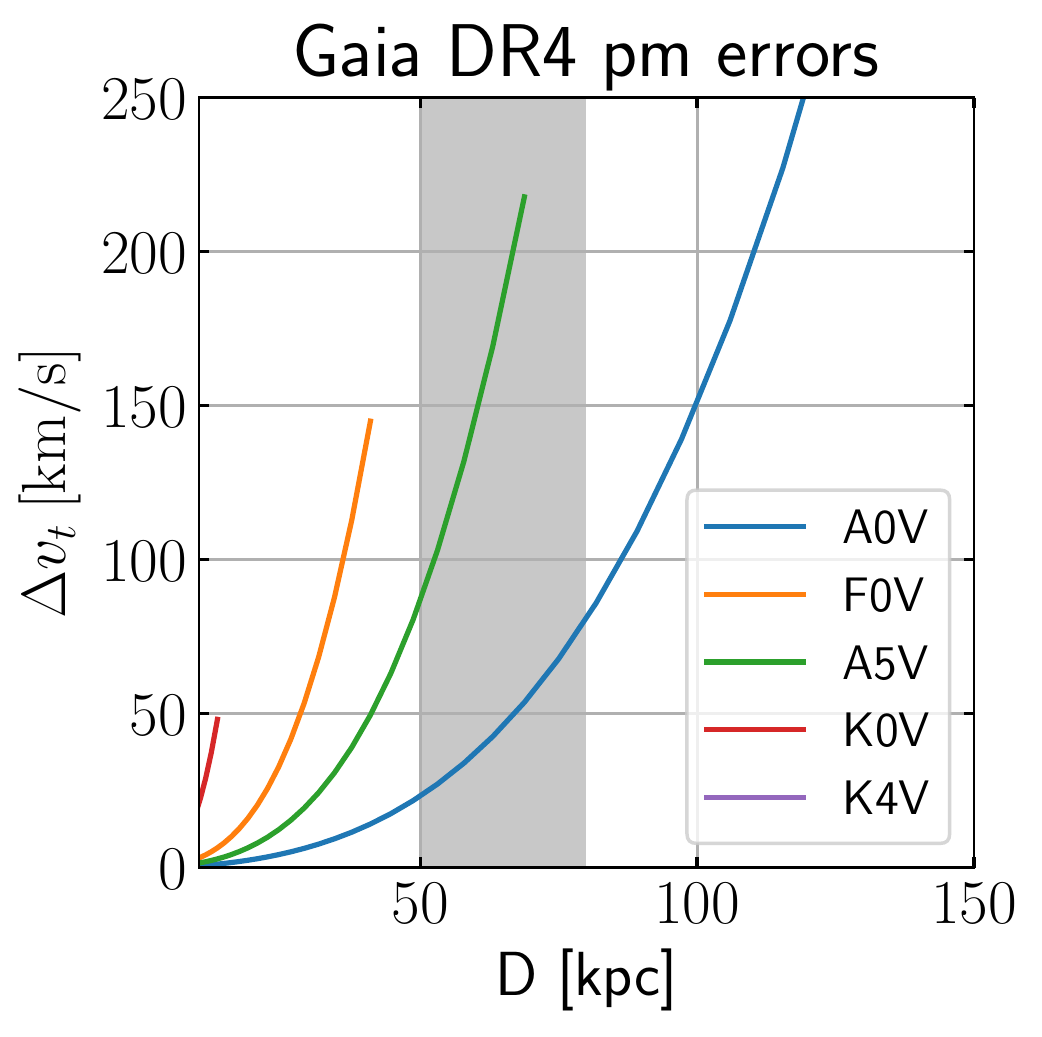}
    \includegraphics[scale=0.65]{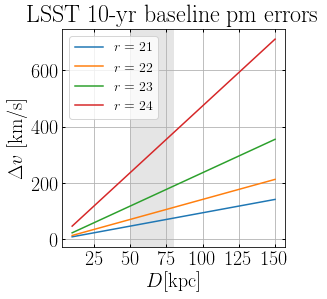}
    \caption{\textit{Top Panel:} Accuracies in the tangential velocities as a  
    function of Galactocentric distances for different
    stellar tracers for Gaia data release 4. Each colored solid line ends at the Gaia
    sensitivity limit. Beyond 50 kpc, A-type stars are the only observable stars in Gaia DR4
    with errors in tangential velocities in the range of 50-100 km/s. These estimates
    where computed using the PyGaia library. \textit{Bottom Panel:} LSST
    10-year long baseline accuracies in the tangential 
    velocities as a function of Galactocentric distances for different magnitudes.
    The shaded grey regions in both panels illustrates the regions of interest for 
    our analysis. In those regions, LSST accuracies range from 30 up to 400 km/s for the
    faintest objects.}\label{fig:gaia_errors}    
\end{figure}

In this section, we compute the accuracies in the tangential velocities used in 
our analysis of Section \S \ref{sec:obs_kinematics}. The accuracies are computed 
for both \textit{Gaia} and LSST. For \textit{Gaia} we compute the accuracies as a function of
Galactocentric distance ($D$) using the \verb+Pygaia+ package.
We compute the accuracies for five spectral-type stars: A0V,
F0V, A5V, K0V, and K4V. These spectral types are of stars commonly found in the 
MW stellar halo e.g: K-giants, BHB stars, and RR Lyrae. Figure 
\ref{fig:gaia_errors} shows the expected accuracy in the tangential velocity for 
\textit{Gaia's} data release 4 (DR4). At distances larger than 50 kpc,
\textit{Gaia} will observe
only A-type stars with errors of $\sim 100$ km/s. In our analysis in Section 
\S \ref{sec:obs_kinematics}, we are interested in distances within the range of 50-80 kpc. 
At those distances, the errors in the A0V-type stars range from 20 up to 100 km/s.
We thus decide to use errors in the tangential velocity of 50 and 100 km/s.

For LSST, the expected accuracies after a 10 yr long baseline will be 
similar to \textit{Gaia's} accuracies, but for fainter sources up to r$\sim$21, see Figure 21 in \cite{Ivezic13} and
Table 3 in \cite{Ivezic08}. LSST's sensitivity will allow us to observe fainter
objects with r$\sim$24. Figure \ref{fig:gaia_errors} shows the accuracies in
tangential velocities as a function of Galactocentric distances for different magnitudes.
Between 50 and 80 kpc (our region of interest), LSST accuracies range from 
30 km/s up to 400 km/s for the faintest objects $r=24$.